\pdfoutput=1
\documentclass{article}
\PassOptionsToPackage{numbers, compress}{natbib}
\usepackage[preprint]{neurips_2020}
\usepackage[utf8]{inputenc} 
\usepackage[T1]{fontenc}    
\usepackage{setspace}            
\usepackage{booktabs}       
\usepackage{amsfonts,amsthm,amsmath,amssymb,stmaryrd}       
\usepackage{nicefrac}       
\usepackage{microtype,color}      
\usepackage[linesnumbered,ruled,vlined]{algorithm2e}
\usepackage{graphicx,enumitem,url}
\def\bs{\mathbf{s}}
\def\bg{\mathbf{g}}
\def\bG{\mathbf{G}}
\def\bP{\mathbf{P}}
\def\bA{\mathbf{A}}

\def\bx{\mathbf{x}}
\def\bX{\mathbf{X}}

\def\Yg{Y_{\bg}}

\def\nmax{n_{\max}}
\def\bz{\mathbf{z}}
\def\by{\mathbf{y}}
\def\bY{\mathbf{Y}}
\def\YG{\bY_{\bG}}

\def\Sen{s}

\def\Spe{\sigma}
\def\Seg{\Sen_g}
\def\Spg{\Spe_g}

\def\Spgi{\Spe_{g_i}}

\def\cG{\mathcal{G}}
\def\Gg{\mathcal{G}_g}
\def\cI{\mathcal{I}}
\def\cP{\mathcal{P}}
\def\cV{\mathcal{V}}
\def\RR{\mathbb{R}}
\def\NN{\mathbb{N}}
\def\E{\mathbb{E}}
\newcommand{\interv}[1]{\llbracket #1\rrbracket}
\def\P{\mathbb{P}}
\def\Prob{\P}
\newcommand{\Pt}[1]{\pi_{#1}}
\newcommand{\hPt}[1]{\hat{\pi}_{#1}}
\DeclareMathOperator{\lse}{lse}
\DeclareMathOperator{\PO}{Pos}
\DeclareMathOperator{\NE}{Neg}

\DeclareMathOperator{\argmax}{argmax}
\DeclareMathOperator{\argmin}{argmin}

\DeclareMathOperator{\MI}{MI}
\def\one{\mathbf{1}}

\newtheorem{lemma}{Lemma}
\newtheorem{remark}{Remark}
\newcommand{\jp}[1]{\textbf{[JP: #1]}}

\title{Noisy Adaptive Group Testing \\ using Bayesian Sequential Experimental Design}

\author{%
  Marco Cuturi \quad Olivier Teboul \quad Quentin Berthet \quad Arnaud Doucet \quad Jean-Philippe Vert\\
  \texttt{\{cuturi,oliviert,qberthet,arnauddoucet,jpvert\}@google.com} \\
}

\begin{document}

\maketitle

\begin{abstract}
When the infection prevalence of a disease is low, Dorfman showed 80 years ago that testing groups of people can prove more efficient than testing people individually. 
Our goal in this paper is to propose new group testing algorithms that can operate in a \textit{noisy} setting (tests can be mistaken) to decide \textit{adaptively} (looking at past results) which groups to test next, with the goal to converge to a good detection, as quickly, and with as few tests as possible.
We cast this problem as a Bayesian sequential experimental design problem. Using the posterior distribution of infection status vectors for $n$ patients, given observed tests carried out so far, we seek to form groups that have a maximal utility. We consider utilities
such as mutual information, but also quantities that have a more direct relevance to testing, such as the AUC of the ROC curve of the test.  Practically, the posterior distributions on $\{0,1\}^n$ are approximated by sequential Monte Carlo (SMC) samplers and the utility maximized by a greedy optimizer. Our procedures show in simulations significant improvements over both adaptive and non-adaptive baselines, and are far more efficient than individual tests when disease prevalence is low.
Additionally, we show empirically that loopy belief propagation (LBP), widely regarded as the SoTA decoder to decide whether an individual is infected or not given previous tests, can be unreliable and exhibit oscillatory behavior. Our SMC decoder is more reliable, and can improve the performance of other group testing algorithms.
\end{abstract}

\section{Introduction}
Singling out infected individuals in a population that has little immunity to a pathogen is of paramount importance to control the propagation of an epidemic. 
When tests are expensive and the base infection rate is low, an approach first pioneered by~\citet{dorfman1943detection} consists in pooling individuals in disjoint groups (e.g. by pooling 5 nasal swabs) and test only those pooled samples first (e.g. to detect traces of virus RNA in each pool). In a second stage, only samples that belonged to positive groups are re-tested, one-by-one, to single out positives. \citeauthor{dorfman1943detection} showed that this two-stage group testing procedure was optimal in an idealized setup, by choosing a group size that is a (decreasing) function of the disease prevalence.  \citeauthor{dorfman1943detection}'s procedure is therefore well motivated mathematically, and reportedly in use to test for SARC-CoV-2 infection at scale~\citep{Seifried,yelin2020evaluation}. Since Dorfman's seminal work on medical testing, the field of group testing at large has significantly grown, with applications considered in quality control~\citep{Sobel1959Group}, communications~\citep{Berger1984Random,Wolf1985Born}, molecular biology~\citep{Balding1996comparative,Ngo2000survey}, pattern matching~\citep{Macula2004group,Clifford2010Pattern}, database systems~\citep{Cormode2005Whats}, traitor tracing~\citep{Meerwald2011Group,Laarhoven2015Asymptotics}, or machine learning~\citep{Zhou2014Parallel}; see \citep{aldridge2019group} for a recent review.

\textbf{Group testing regimes: adaptiveness and noise.}
Group testing strategies can be \emph{non-adaptive}, when every group to be tested is decided beforehand, or \emph{adaptive}, when the tests are performed in several stages, and when groups to be tested at the next stage are decided using results from all tests performed previously~\citep{Scarlett2019Noisy}. For example, Dorfman's strategy is adaptive and has two stages. Group testing strategies can be also be designed to handle \textit{noisy} tests, \textit{i.e.} account for the fact that tests can make mistakes, or, like Dorfman's, expect on the contrary that tests are \textit{noiseless}. There exists a large body of work on adaptive and non-adaptive group testing in the noiseless setting~\citep{mitchell1987computer,Du2000Combinatorial}, where adaptive strategies tend to have better theoretical guarantees and result in more practical algorithms than non-adaptive ones~\citep{Scarlett2016Phase,Aldridge2017capacity,Scarlett2019Noisy}. For instance, \citet{Hwang1972method} proposed a multi-stage adaptive binary splitting algorithm, which achieves the information-theoretical asymptotic lower bound  on the number of tests needed to identify all infected individuals when the population size increases, and the proportion of infected individuals vanishes~\citep{Baldassini2013capacity}. Additionally, it is also known, in the noiseless case, that non-adaptive designs can be suboptimal compared to adaptive strategies in some regimes~\citep{Agarwal2018Novel}, while optimal two-stage adaptive algorithms are also well understood~\citep{Mezard2011Group,DeBonis2005Optimal}. 

\textbf{Noisy, adaptive group testing.\;} With Covid-19 as a backdrop, where RT-PCR tests are known to be both in short supply and noisy~\citep{Wikramaratna,yelin2020evaluation}, the \textit{noisy adaptive} setting is relevant: As noise increases, the possibly contradictory results of noisy tests can put a spoke in the wheel of combinatorial approaches. In the noisy regime, information-theoretic limits of group testing are well understood~\citep{Malyutov1978separating,Malyutov1980Planning,Atia2012Boolean,Baldassini2013capacity,Scarlett2016Phase,Aldridge2017capacity} but most existing group testing strategies are non-adaptative~\citep{Malyutov1980Planning,Chan2011Non,Chan2014Non,Scarlett2018Optimal}, with the exception of \citet{Cai2013GROTESQUE} and \citet{Scarlett2019Noisy}. These two algorithms have various optimality properties in an asymptotic regime, when the population size increases and the fraction of infected individuals vanishes. However, little is known about the quality of these methods in a non-asymptotic regime, with a finite horizon, and a small but non-vanishing proportion of infections in the population.

\textbf{Our contributions.} In this work, we depart from the standard asymptotic analysis, to propose a sequential Bayesian optimal experimental design (BOED) approach to group testing, consisting of:

\begin{itemize}[noitemsep,nosep,wide,partopsep=-4pt]
\item We derive a BOED~\citep{Chaloner1995Bayesian} approach for group testing, in which groups are sequentially selected to maximize the \textit{expected} utility of their hypothetical test results at the next stage. We consider two utility functions: the information gain (or mutual information) provided by a new wave of tests, or, closer to health professionals' requirements, the AUC of the ROC curve given by the marginal posterior distribution. Both are optimized using greedy forward-backward selection.

\item Evaluating these utilities requires having access to an approximation of the posterior distribution of infection states of all $n$ individuals, given all group tests observed up to that stage, with known priors on infection and noise. We use SMC samplers~\citep{del2006sequential} at \textit{each} testing stage, to approximate that posterior as a cloud of particles in $\{0,1\}^n$.
SMC was used previously for other Bayesian design problems~\cite{ryan2016review}; ours builds upon~\citep{schafer2013sequential}, using a Gibbs sampler as Markov chain Monte Carlo (MCMC) kernel.

\item Noisy group testing approaches use a \textit{decoder}, an algorithm tasked with outputting an infection probability vector from test results. We show that the marginal distribution produced by our SMC samplers outperform those produced by LBP, the SoTA decoder \cite[\S3.3]{sejdinovic2010note,aldridge2019group}. Open source code:
\end{itemize}
\url{https://github.com/google-research/google-research/tree/master/grouptesting}

\section{Background on Group Testing}
\textbf{Prior on infection.\;} We consider a population of $n$ individuals, who can be either infected or not. The infection status of the $i$-th individual is modeled with a binary random variable (r.v.) $X_i$, where $X_i=1$ if that individual is infected and $X_i=0$ otherwise. We write $\bX=(X_1,\ldots,X_n) \in\{0,1\}^n$ for the infection status of the whole population. We assume that a prior probability distribution for $\bX$ is given.
For example, each infection may be modelled as an independent Bernoulli r.v. $X_i \sim \mathbb{B}(q_i)$. Here, $q_i$ is a prior infection rate, either shared across individuals, or estimated for each using other covariates.
Under this model, the probability mass function (pmf) of the prior probability would satisfy, for any $\bx=(x_1,\ldots,x_n)\in\{0,1\}^n$,
$\Pt{0}(\bx):=\P_0(\bX=\bx) = \prod_{i=1}^n q_i^{x_i} (1-q_i)^{1-x_i}$.
~More informed and non-independent priors (relating for instance two people in the same household) may be considered; as discussed later in \S\ref{sec:SMC}, we approximate the prior $\Pt{0}$ with a weighted cloud of particles in $\{0,1\}^n$, giving us the flexibility to consider any sort of initial prior.

\textbf{Group \textit{vs.} individual testing in the presence of testing noise.\;} Our goal is to infer which individuals are infected and which ones are not. A straightforward approach to do so would be to test each individual one-by-one. However, this raises two issues: \textit{(i)} this requires $n$ tests, which is costly if $n$ is large, and inefficient if infection prevalence is low; \textit{(ii)} tests can be noisy (e.g., nose swabs tested with RT-PCR create false negatives and, to a lesser extent, false positives), so by testing only once each individual, there is a risk of error. In this paper we solve both issues by relying on group tests. We assume that for any given group $\bg\subset \{1,\dots,n\}$, we can pool samples from that group and test that ``mixture of samples'' to reveal the group's binary status: it is either \textit{negative}, when none of the individuals in the group is infected, or \textit{positive}, when one or more individuals are infected.


\textbf{Probabilistic inference from a single group.\,} For an integer $n$, we write $\interv{n}:=\{1,\dots,n\}$. $\cG$ is the set of all non-empty groups, i.e. non-empty subsets of $\interv{n}$. With a slight overload of notations, a group $\bg\in\cG$ is also equivalently represented as a binary vector $\bg\in\{0,1\}^n$, where the $i$-th element of $\bg$ is 1 if and only if $i\in\bg$. We write $g$ for $\bg^T\mathbf{1}_n$, the size of group $\bg$. We write 
\begin{equation}\label{eq:binsum}
\text{for all } \bg,\bx \in\{0,1\}^{n},\;   [\bg,\bx]:=\bigvee_{i\in\bg}x_i = 1 - \prod_{i\in\bg}(1-x_i) = \max(1,\bg^T\bx)\;\in\{0,1\}\;,
\end{equation}
the binary status of group $\bg$ given the binary status of individuals $\bx$. Indeed, $[\bg,\bx]$ is equal to $0$ if and only if all entries in $\bx$ indexed by $\bg$ are equal to $0$. 
Given a group $\bg\in\cG$, the output of a \emph{group test} associated to $\bg$ is a binary r.v. $Y_\bg$ that assesses the group status $[\bg,\bX]$. We assume that conditioned on $\bX$, all considered group tests are independent from each other, and that each group test suffers from noise due to the specificity $\Spg$ and sensitivity $\Seg$ parameters of the testing device, which both depend on the size $g$ of $\bg$.
For any group $\bg\in\cG$, writing $\rho_g:=\Seg+\Spg-1$, we have
\begin{gather}
\P(Y_\bg = 1\,|\,[\bg,\bX]=1)=\Seg\,,\quad \P(Y_\bg = 0\,|\,[\bg,\bX]=0)=\Spg\,, \label{eq:senspe}\\
\!\!\!\forall\, \bx\in\{0,1\}^n, y\in\{0,1\}, \P(Y_{\bg}=y\,|\,\bX = \bx ) 
= \left( \Spg - \rho_g[\bg,\bx]\right)^{(1-y)} \, \left(1-\Spg + \rho_g [\bg,\bx]\right)^{y} .\label{eq:testlikelihood}
\end{gather}

\textbf{Inference for batches of groups.\;} We assume that up to $k$ tests can be run simultaneously, in parallel, on a testing device. Consequently, we tailor our strategies so that they propose a \textit{batch} of $k$ groups $\bG = (\bg_1,\dots,\bg_k) \in \cG^k$, equivalently represented as a $n\times k$ binary membership matrix. 
Given a batch $\bG$ of $k$ groups, we define the random vector $\YG:=\left(Y_{\bg_1},\dots, Y_{\bg_k}\right)$ of its $k$ independent test outcomes. 
The probability of $\bY_\bG$ taking values $\by\in\{0,1\}^k$ conditionally on $\bX$ is, using~\eqref{eq:testlikelihood}:
\begin{equation}\label{eq:grouptestlikelihood}
\P(\YG = \by\, | \bX= \bx) = \prod_{i=1}^k \left( \Spgi - \rho_{g_i}[\bg_i,\bx]\right)^{(1-y_i)} \,\left(1-\Spgi + \rho_{g_i}[\bg_i,\bx]\right)^{y_i}\,.
\end{equation}


\section{Bayesian Optimal Experimental Design to Select Useful Groups}\label{sec:bayesexp}
In $T$-stage adaptive group design, given a finite horizon $T\in\NN$, our goal is to select sequentially batches of groups $\bG^t \in \cG^k$ at each stage $1\leq t\leq T$. At the end of stage $t\geq 1$, batches $\bG^1,\dots,\bG^t$ were selected previously, and tested with observed outcomes $\bY_{\bG^{1}}=\by^1,\dots,\bY_{\bG^{t}}=\by^t$. Let us denote by $\P_t$ the probability conditioned to all tests seen up to stage $t$, i.e., for any new batch $\bG$,
\begin{equation*}
    \begin{split}
    \P_t(\bX=\bx\,,\bY_\bG = \by_\bG) 
    &:= \P\left(\bX=\mathbf{x}\,,\bY_\bG = \by_\bG\,|\,\bY_{\bG^1}=\by^1,\ldots,\bY_{\bG^t}=\by^t \right) \\
    &= \pi_t(\bx)\times \P\left(\bY_\bG = \by_\bG\,|\,\bX=\mathbf{x}\right)\,,
    \end{split}
\end{equation*}
where $\P\left(\bY_\bG = \by_\bG\,|\,\bX=\mathbf{x}\right)$ is given by (\ref{eq:grouptestlikelihood}) and $\Pt{t}$ is the posterior pmf of the vector $\bX$ of infection states, given all those test results revealed up to stage $t$, i.e.:
\begin{equation}\label{eq:posterior}
\Pt{t}(\mathbf{x}) := \P_t(\bX=\bx) = \P\left(\bX=\mathbf{x}\,|\,\bY_{\bG^1}=\by^1,\ldots,\bY_{\bG^t}=\by^t \right)\,.
\end{equation}
From $\Pt{t}$, we propose to follow a myopic approach~\citep{Chaloner1995Bayesian} by choosing for stage $t+1$ a new batch $\bG$ that has largest \emph{utility} $U(\bG,\Pt{t})$ w.r.t. $\Pt{t}$. We introduce first a simple utility grounded on information theory, before presenting a more general and flexible formulation for utilities $U$ in \eqref{eq:maxutil} below.

\textbf{Maximizing mutual information.\;} 
%
Ideally, a batch of $k$ tests to be tested at time $t+1$ should be such that $\YG$ reveals as much information as possible on $\bX$, at time $t$. Since $(\bX,\bY_\bG)$ are both r.v. under $\P_t$, this can be achieved by maximizing their mutual information (MI) utility in $\bG$~\citep{Cover1990Elements}: 
$$
U_{\text{MI}}(\bG, \Pt{t}) := I_{\P_t}(\bX;\bY_{\bG}) = H_{\P_t}\!(\bY_{\bG}) - H_{\P_t}\!(\bY_{\bG}|\bX) = H_{\P_t}\!(\bY_{\bG}) - \sum_\bx \!\Pt{t}(\bx)H_{\P}(\bY_{\bG}|\bX=\bx)\,, 
$$
where, for any r.v. $\mathbf{Z}$ with distribution $\P_{\mathbf{Z}}$ and pmf $\eta(\bz)$, $H_{\P_\mathbf{Z}}(\mathbf{Z})=-\E_{\P_\mathbf{Z}} \left[\log \eta(\mathbf{Z})\right]$ is the entropy.
The MI is a standard utility function in Bayesian experimental design \citep{Lindley1956Measure,Chaloner1995Bayesian,foster2019variational}. 
In our particular setting, $U_{\text{MI}}$ can be evaluated thanks to this lemma (proof in \S\ref{subsec:prooflemma1}):
\begin{lemma}\label{thm:MI}
For a group $\bg$, define $f_{\Pt{t}}(\bg):=\sum_\bx \Pt{t}(\bx)\, [\bg,\bx]$. For $\bG=(\bg_1,\ldots,\bg_k) \in \cG^k$, one has
\begin{equation}\label{eq:thmMIgroup}
    I_{\P_t}(\bX\,;\,\YG) = H_{\P_t}(\YG) - \sum_{i=1}^k \left( h_{\Spe_{g_i}} + \gamma_{g_i} f_{\Pt{t}}(\bg_i) \right)\,,
\end{equation}
where $h(u)= -u\log u - (1-u)\log(1-u)$ is the binary entropy, and for any group size $g$, $h_{\Spg}=h(\Spg)$, $h_{\Seg}=h(\Seg)$, and $\gamma_g = h_{\Seg} - h_{\Spg}$. In the case of a single group $\bg\in\cG$, this reduces to
\begin{equation}\label{eq:thmMI}
    I_{\P_t}(\bX;Y_\bg) = h\left(\rho_g \, f_{\Pt{t}}(\bg) +1 - \Spg\right) - \gamma_g f_{\Pt{t}}(\bg) -h_{\Spg}\,.
\end{equation}
\end{lemma}
When selecting one and only one group, choosing $\bg$ boils down to maximizing~\eqref{eq:thmMI}. The MI utility of a group is directly evaluated from $f_{\Pt{t}}(\bg)$, the expected value of its negative/positive status. Therefore, to maximize its MI utility, $f_{\Pt{t}}(\bg)$ should be close of the real argmax of $z\mapsto h(\rho_g z + 1 -\Spg) -\gamma_g z -h_{\Spg}$. If $\Spg=\Seg=1$, that map reduces to $h(z)$, and is maximized at 1/2. Therefore, in a noiseless setting, a group $\bg$ is deemed useful, from a MI viewpoint, if its test r.v. $Y_{\bg}$ is almost an unbiased coin flip.

\textbf{Other utilities.} Instead of relying on information theoretic quantities, we may want to handle more specific criteria. Because the posterior $\Pt{t}$ defines what test results $\bY_{\bG}$ are likely to be at time $t$, we can define a random probability for $\bX$, by conditioning on a test outcome $\bY_{\bG}$ for $\bG$:
\begin{equation}\label{eq:marginaly}
\pi_{t}^{\bG}(\bx) := \P_t(\bX=\bx | \bY_{\bG})\,.
\end{equation}
$\pi_{t}^{\bG}$ can be interpreted as a family of $2^k$ hypothetical posteriors at time $t+1$ for $\bX$, one for each possible outcome for $\bY_{\bG}$. In BOED~\cite{Chaloner1995Bayesian} a guiding principle is to score each of these hypothetical posteriors using a functional $\Phi$, and to choose groups that maximize that score in expectation:
\begin{equation}\label{eq:maxutil}
\bG^{t+1} \in \argmax_{\bG\in\cG^k} U_{\Phi}\left(\bG,\Pt{t}\right):= \E_{\P_t} \Phi(\pi_{t}^{\bG})\,.
\end{equation}
If we set $\Phi$ to be the negative entropy, choosing groups $\bG$ that decrease maximally the expected conditional entropy of $\bX$ at $t+1$, we recover the MI criterion (see \S\ref{sec:MIcomputation}). 
Although we expect lower entropy to correlate with improved testing performance, we use the flexibility of the BOED framework to optimize directly more relevant utilities. We propose to maximize the expected area under the ROC curve (AUC) of the marginal decoder: For a pmf $\pi$ on $\{0,1\}^n$, the marginal decoder $m(\pi) \in[0,1]^n$ is the vector of marginal probabilities that each individual is infected under $\pi$. Given an infection status $\bx\in\{0,1\}^n$, where $\PO(\bx) = \sum_{i=1}^n x_i$ and $\NE(\bx) = \sum_{i=1}^n (1-x_i)$ are the total number of infected and non-infected, we write $\psi_{\text{AUC}}(\bs,\bx)$ the AUC of a predictor $\bs\in\RR^n$: 
\begin{equation}\label{eq:psiutility}
    \psi_{\text{AUC}}(\bs,\bx) = \frac{\sum_{i,j=1}^{n} x_i (1-x_j)\left( \one(s_i > s_j) + \frac{1}{2}\one(s_i = s_j)\right)}{\PO(\bx)\NE(\bx)}\,.
\end{equation}
where, if either $\PO$ or $\NE$ is 0, the AUC is discarded from our computations. 
The expected AUC of the marginal decoder is therefore $\Phi(\pi) = \sum_{\bx}\pi(\bx)\psi_{\text{AUC}}(m(\pi),\bx)$, which we plug directly in~\eqref{eq:maxutil}.

\section{Algorithms}\label{sec:algo}
In order to implement the sequential BOED procedure described in \S\ref{sec:bayesexp}, we now describe in more details the algorithmic components needed at each stage to \textit{(i)} maintain a computationally tractable description of the posterior distribution \eqref{eq:posterior} after each stage, \textit{(ii)} compute the utility of a batch of groups (r.h.s. of \ref{eq:maxutil}), \textit{(iii)} find a batch that solves (\ref{eq:maxutil}), and \textit{(iv)} compute individual infection probabilities.

\textbf{Algorithm to store and update the posterior.}
At every stage $t$, we need the posterior (\ref{eq:posterior}) in order to evaluate and optimize utilities to select groups. This posterior is then updated by observing the results from tests carried out at stage $t$, before moving to stage $t+1$. One way to proceed would be to store the posterior as a $2^n$-dimensional vector, keeping track of the probability of each possible population infection status vector, and update it using Bayes' rule, given test results $\by^{t} \in\{0,1\}^{k}$ for group $\bG^{t}$: 
 \begin{equation}\label{eq:postupdate}
 \Pt{t}(\bx) \propto \Pt{t-1}(\bx) \P(\mathbf{Y}_{\mathbf{G}^{t}}=\by^{t}\,|\,\bX=\bx)\propto \Pt{0}(\bx)\prod_{i=1}^t 
 \P(\mathbf{Y}_{\mathbf{G}^i}=\by^i\,|\,\bX=\bx)\,.
 \end{equation}
While this approach is tractable for up to $n\approx 25$ (leading to $2^{25}\approx 33$ million probabilities to store), it does not scale further. We instead use a SMC sampler~\citep{del2006sequential} to approximate $\Pt{t}$. This provides an approximation by a cloud of weighted particles of the form $\hPt{t} = \sum_{i=1}^N \omega_i \delta_{\bx_i}$, where $N\ll 2^n$. 
We follow closely the algorithmic approach outlined by~\citet[Algo. 2]{schafer2013sequential}, with two modifications: we sample initially from $\Pt{0}$ rather than the uniform prior, and as $t$ grows, use $\hPt{t}$ to produce $\hPt{t+1}$; we consider a few variants for the MCMC kernel used within SMC~\cite[Proc. 4]{schafer2013sequential}, including theirs, which all appear to provide similar results. We pick the modified Gibbs kernel for discrete spaces introduced by \citet{liu1996peskun}, where we loop on the $n$ coordinates of all particles, 4 times by default at each kernel application (see \S\ref{sec:SMC} for algorithmic details and comparisons).


\textbf{Algorithms to Evaluate Utilities.\;} 
Given a functional $\Phi$, Algo.~\ref{algo:computeutility} shows how to estimate the utility $U_{\Phi}$~\eqref{eq:maxutil} for a group $\bG$ and posterior pmf $\hPt{t} = \sum_{i=1}^N \omega_i \delta_{\bx_i}$ obtained at any time $t$ through a SMC sampler. The space complexity of Algo.~\ref{algo:computeutility} is $O(N\times \max(2^k,n))$, where $n$ is the number of individuals, $k$ is the number of groups allowed per stage, and $N$ is the size of the support of pmf $\hPt{t}$, e.g. the number of particles. The time complexity is dictated by line 6, where we repeat $2^k$ times a call to the utility function $\Phi$ where $\hPt{t}$ has a support of size $N$ in $\{0,1\}^n$. If this operation has complexity $C(N,n)$, then the time complexity of  Algo.~\ref{algo:computeutility} is $O(2^k C(N,n))$. For example, for utilities based on marginals such as AUC, we need to compute the marginal first in $O(N n)$; then sort the marginal itself in $O(n\ln(n))$; then compute the AUC on each particle in $O(N n)$ in total, resulting in $C(N,n) = O(n \max(N, \ln(n))$. If we set $\Phi$ to be the neg-entropy, this results in $C(N,n)=O(N)$, but in that case we can use the equivalent formulation of entropy minimization as MI maximization to derive a specific $O(2^k + N)$ algorithm (instead of $O(2^k N)$ with Algo.~\ref{algo:computeutility}), as detailed in Algo.~\ref{algo:computeMI}. 

\begin{algorithm}
\DontPrintSemicolon 
\KwIn{$\hPt{t}(\bx)=\sum_{i=1}^N \omega_i \delta_{\bx_i}(\bx) = \hat{\P}_t(\bX=\bx)$; $\bG = (\bg_1,\ldots,\bg_k)\in\{0,1\}^{n\times k}$ a set of groups; $\Spe,\Sen\in[0,1]^{k}$ the specificities and sensitivities of the test for each group in $\bG$; $\Phi:[0,1]^N \times \{0,1\}^{n\times N}$ a utility function evaluated on a weighted cloud of $N$ particles.}
\KwOut{The utility $U_{\Phi}(\bG,\hPt{t})$}
$A_{ij} \gets 1-\Spe_i + (\Spe_i+\Sen_i -1) [\bg_i,\bx_j]$  for $(i,j)\in\interv{k} \times \interv{N}$ \tcp*{$\P(Y_{\bg_i}=1\,|\,\bX=\bx_j)$}
$B_{ij} \gets \prod_{t=1}^k  A_{tj}^{b_{it}} (1-A_{tj})^{1-b_{it}}$ for $(i,j)\in\interv{2^k}\times \interv{N}$, where $b_{it}$ is the $t$-th bit from the right in the binary expansion of $i$ \tcp*{$\P(\YG=i\,|\,\bX=\bx_j)$}
$C_{ij} \gets B_{ij}\times \omega_j$ for $(i,j)\in\interv{2^k}\times \interv{N}$ \tcp*{$\hat{\P}_t(\YG=i\,,\,\bX=\bx_j)$}
$D_i \gets \sum_{j=1}^N C_{ij}$ for $i\in\interv{2^k}$ \tcp*{$\hat{\P}_t(\YG=i)$}
$E_{ij} \gets C_{ij} / D_i$ for $(i,j)\in\interv{2^k}\times \interv{N}$ \tcp*{$\hat{\P}_t(\bX=\bx_j \,|\,\YG=i)$}
$F_i \gets \Phi( E_{i\cdot} , \bX)$  for $i\in\interv{2^k}$\tcp*{$\Phi(\hat{\P}_t(\bX\,|\,\YG=i))$}
$U \gets \sum_{i=1}^{2^k} D_i F_i$ \tcp*{$\E_{Y_\bG}\Phi(\hPt{t}^{\bG})$}
\Return{U}
\caption{Compute utility of a set of groups given posterior $U_{\Phi}(\bG,\hPt{t})$}
\label{algo:computeutility}
\end{algorithm}

\textbf{Algorithms to Maximize Utility.\;}
Taking for granted that we can evaluate $U_{\Phi}(\bG,\hPt{t})$ given a candidate batch $\bG$ and posterior pmf $\hPt{t}$ using Algo.~\ref{algo:computeMI}, the question of finding a batch $\bG$ that maximizes $U_\Phi$ in~\eqref{eq:maxutil} is a difficult discrete optimization problem, with costly evaluations. Any standard algorithm for discrete optimization can in principle be used to find suboptimal solutions, such as greedy forward/backward optimization, simulated annealing or genetic algorithms. We implemented a greedy approach that adds incrementally groups one by one, to build up a batch. Each group itself grows in a greedy manner, using standard forward/backward steps: we first greedily add the best $F$ individuals one by one (forward step), and then delete the $B$ worst ones (backward step) until we either stop improving utility, or reach the maximal group size $\nmax$. In the case of MI maximization, we detail a way to carry out such greedy optimization more efficiently than by applying repeatedly Algo.~\ref{algo:computeMI} (in the appendix).

\textbf{Algorithms for Marginal Inference and Decoding.\;}
At any point in the testing campaign, one may want to compute the marginal probability for each individual to be infected, a step known as \textit{decoding} test results. This marginal may be used to make informed decisions at any stage of the campaign, but could also be used to design new groups to test (to illustrate this, we propose in our experiments to use the informative Dorfman (ID) procedure~\citep{mcmahan2012informative} after a first round of test has been carried out, see \S\ref{sec:policies}). To estimate the marginal, we implemented two approaches: \textit{(i)} marginalizing the approximate posterior $\hPt{t}$ maintained by the SMC sampler (\S\ref{sec:SMC}), which consists simply in computing the ``mean'' particle in $\hPt{t}$, or \textit{(ii)} computing the marginal with a LBP algorithm \citep{pearl1982reverend}, as detailed in \S\ref{sec:LBP}. The LBP is a fast and popular decoding algorithm in group testing \cite{sejdinovic2010note,aldridge2019group}, which is not, however, guaranteed to converge to the correct marginal. On the other hand, although the SMC-based estimator may be inaccurate if the particle approximation of the posterior is poor, it does not suffer from convergence issues. We show in experiments (see Fig.~\ref{fig:lbpisbad}) that SMC outperforms LBP decoding. More problematically, we find that LBP can display oscillatory behavior, notably for certain group testing strategies that re-test several times the same individuals. However, because LBP is significantly faster, we propose a mixed approach, which tests whether LBP has converged within a maximal number of iterations, and, if not, switches to a SMC. This strategy seems to be almost as effective in our simulations to using a SMC by default, and we adopt it for \textit{all} group testing strategies.

\section{Simulations}\label{sec:simuls}

\begin{algorithm}
\DontPrintSemicolon 
\KwIn{horizon $T\geq 1$, maximum number of tests per cycle $k$, maximum group size $\nmax$.\;
Ground truth infection probability $\Prob_0$ of $n$ patients on state space $\{0,1\}^{n}$.\;
Ground truth noise parameters $\Spe,\Sen\in [0,1]^{\nmax}\times[0,1]^{\nmax}$ depending on group size.\; 
Prior on infection probability $\hat{\Prob}_0$ used by algorithms.\;
Prior on noise parameters $\hat{\Spe},\hat{\Sen}\in [0,1]^{\nmax}\times[0,1]^{\nmax} $  used by algorithms.\; 
\texttt{Test}$(\bG,\Spe,\Sen)$ returns $\texttt{col}(\bG)$ noisy tests using $\Spe,\Sen$ by pooling samples according to $\bG$.\; 
\texttt{Policy}$(t, d , \nmax, \by^t, \bG^t, \hat{\Spe},\hat{\Sen}, \hPt{t-1} \text{ or } \bar{\bx}_{t-1})$
calls $t$-th selector to produce up to $d$ new groups of size at most $\nmax$. May use: past test results $\by^t, \bG^t$; priors $\hat{\Spe},\hat{\Sen}$; posterior or marginal approx.\; 
\texttt{Sampler}$(\by^t, \bG^t, \hat{\Spe}, \hat{\Sen}, \hPt{t-1})$ produces $N$ approx. weighted samples from $\P_{t}$ using new tests.\; 
\texttt{MarginalSampler}$(\by^t, \bG^t, \hat{\Spe}, \hat{\Sen}, \hPt{t-1})$ produces only approx. marginal distribution.\;
}
\KwOut{ground-truth vector, $T$ marginal predictions of infection.}
$\bx_{\mathrm{truth}} \sim \Prob_0$\tcp*[f]{sample the ground truth status}\; 
$\hPt{0} \overset{N \text{i.i.d.}}{\sim} \hat{\Prob}_0$\tcp*[f]{sample $N$ i.i.d particles from prior}\; 
$\bG^{\mathrm{totest}} \gets  \mathbf{0}_{n\times 0}$\tcp*[f]{initialize groups} \;
\For{$t\gets 1$ \KwTo $T$}{
\If(\tcp*[f]{produce new groups if needed}) {$\mathtt{col}(\bG^{\mathrm{totest}})<k$}{
$\bG^{\mathrm{add}} \gets \texttt{Policy}(t, k-\texttt{col}(\bG^{\mathrm{totest}}),\by^{t-1}, \bG^{t-1}, \hat{\Spe}, \hat{\Sen}, \hPt{t-1} \text{ or } \bar{\bx}_{t-1})$.\;
$\bG^{\mathrm{totest}} \gets [\bG^{\mathrm{totest}}, \bG^{\mathrm{add}}]$\tcp*[f]{add groups to stack}
}
$r\gets\min(k,\texttt{col}(\bG^{\mathrm{totest}})), \bG^{t}\leftarrow \bG^{\mathrm{totest}}_{:r}, \bG^{\mathrm{totest}}\leftarrow \bG^{\mathrm{totest}}_{r:}$ \tcp*[f]{set new tests}\;
$\by^{t}\gets\texttt{Test}\left(\bG^{t}, \Spe, \Sen\right)$\tcp*[f]{receive lab tests}\;
$\hPt{t}\leftarrow \texttt{Sampler}(\by^t,\bG^t, \hPt{t-1})$\tcp*[f]{sample particles using test results}\;
$\bar{\bx}_t \leftarrow \texttt{MarginalSampler}(\by^t,\bG^t,\hPt{t-1})$\tcp*[f]{compute marginal using tests}\;
}
\Return{$\bx_{\mathrm{truth}},(\bar{\bx}_1,\dots,\bar{\bx}_T)$\tcp*[f]{ground truths + marginal predictions}\;}
\caption{Simulator to evaluate the performance of a group testing \texttt{Policy} }
\label{algo:simul}
\end{algorithm}

\textbf{Policies: Ours and baselines.\;} We call a \textit{selector} any algorithm, adaptive or not, that is able to choose groups at any stage, using possibly the knowledge of past tests. A group testing \textit{policy} is a \textit{sequence} of group selectors to be used at each stage. In the group testing literature, it is common that a policy sticks to a single selector throughout all stages. We propose here several new baseline policies: some use a single selector throughout, some use different selectors. For instance, we consider policies that may start with a non-adaptive selector in the first stage, followed next by an adaptive selector.

Our BOED selectors maximize, using greedy forward-3/backward-2 selection, either the mutual information (\textbf{G-MIMAX}) or the expected AUC utility (\textbf{G-AUCMAX}). We consider them as single-selector policies, and compare them to the following baseline policies. On the one hand, we consider the standard 2-stage \textbf{Dorfman} policy \cite{dorfman1943detection}, which first splits the population in groups of size $\approx \min(\nmax,1+\lceil 1/\sqrt{q}\rceil)$, and then tests all individuals in positive groups, and the multi-stage \textbf{Binary Dorfman} policy that implements \citeauthor{Hwang1972method}'s hierarchical binary splitting approach \cite{Hwang1972method} instead of the second stage of Dorfman's strategy. On the other hand, we test two non-adaptive selectors where groups are either uniformly \textbf{Random} (composed of $g$ patients, where $g$ is chosen so that the probability of a test being positive is close to $1/2$, which is asymptotically optimal in the absence of noise \cite{Mezard2011Group}), or fixed using the predefined \texttt{Origami M3} (\textbf{OM3}) assay matrix~\citep{kainkaryam2008poolhits} containing 22 groups of maximal size 10 for 70 individuals, which was optimized to deal with an infection rate of $\approx5\%$. We consider the \textbf{Random} selector as a policy in itself, and consider 3 mixed policies: \textit{(i)} \textbf{Random-ID}, where a first batch of \textbf{Random} groups are formed, which is used to form a first guess for the marginal distribution, which can be used in the second stage by a variant of \textbf{Dorfmann}'s splitting known as \textbf{Informative Dorfman (ID)} \cite{mcmahan2012informative}, where the first uniform split of groups in the \textbf{Dorfman} strategy is replaced by an optimized strategy; \textit{(ii)} \textbf{Origami-Random}, which first tests the 22 groups of \textbf{(OM3)}, and then switches to \textbf{Random} groups; 
(iii) \textbf{Origami-ID}, which switches instead to an \textbf{ID} strategy, using the posterior marginal computed from observing tests from \textbf{(OM3)}. These strategies are described in more detail in \S\ref{sec:policies}.

\textbf{Group testing simulations: algorithm and parameters.\,}
Each simulation runs for a predefined $T$ test cycles, during which we can carry out up to $k$ tests simultaneously. We consider settings where $Tk <n$. The testing simulator is described in Alg.~\ref{algo:simul} using the following notations: $\texttt{col}(\bA)$ is the number of columns of a matrix $\bA$; $\bA_{:i}$, the first $i$ columns of a matrix; $\bA_{i:}$, the matrix $\bA$ stripped of those $i$ first columns. 
We use the following parameters in our simulations:
\begin{itemize}[noitemsep,nosep,wide,partopsep=-4pt]
    \item population size $n = 70$; infection rate of $q = 2\%$ or $5\%$ (see \S\ref{subsec:IR10} for $10\%$); constant specificity $\sigma=97\%$ and sensitivity $s = 85\%$ (see \S\ref{subsec:varying} for results with varying sensitivity);
    \item $k=8$ tests per cycle, horizon of $T = 5$ cycles (total 40 tests), maximal group size: $\nmax=10$.
    \item 5,000 simulation runs for each policy.
\end{itemize}    

\begin{figure}
\hskip-.8cm
    \includegraphics[width=.6\textwidth]{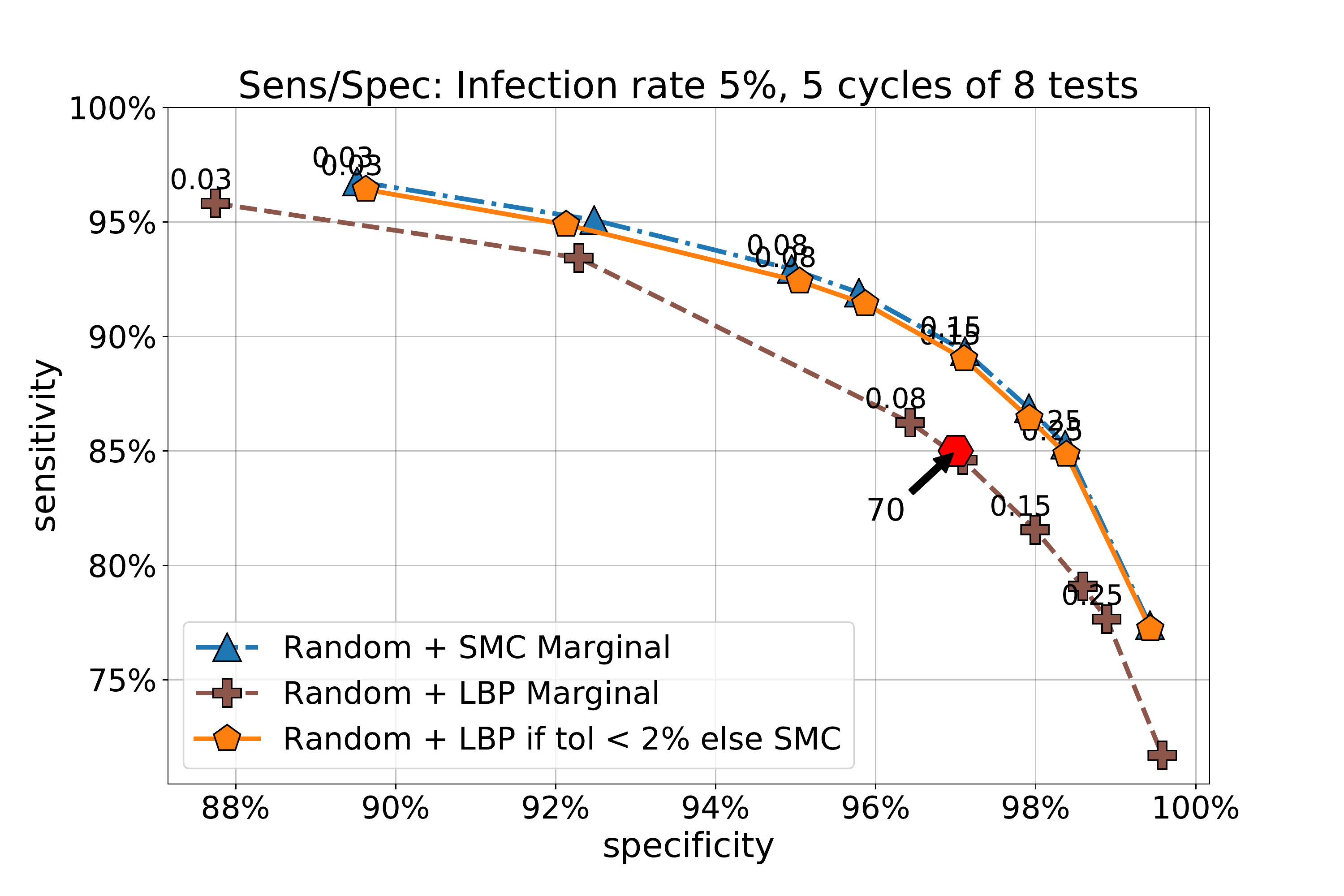}\hskip-.7cm
    \includegraphics[width=.6\textwidth]{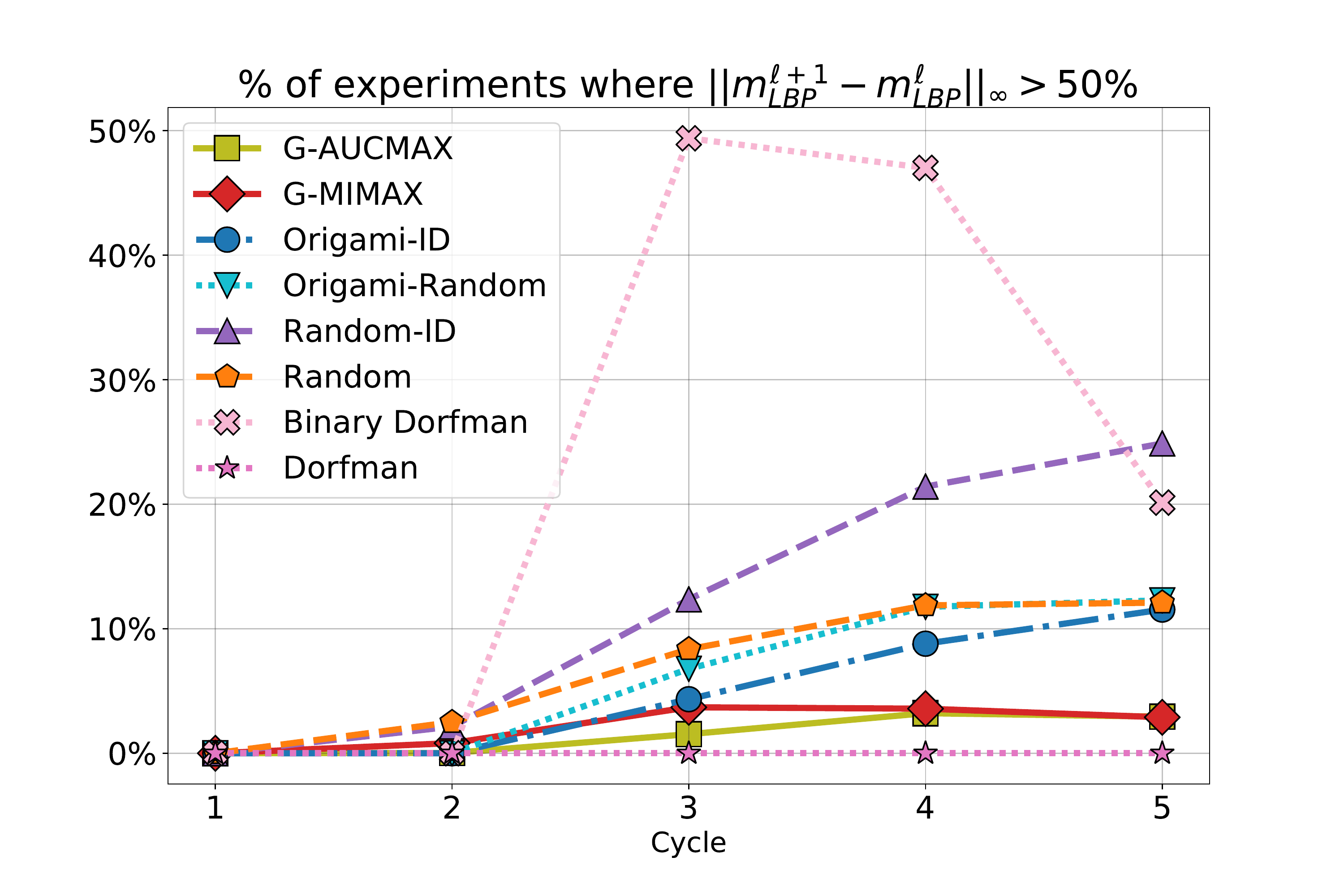}
\caption{Issues with LBP as a decoder are highlighted in these figures. The left plot reveals that using LBP significantly degrades the performance of the \textbf{Random} policy, as measured by its average sensitivity/specificity after 40 tests. In the right plot, we count the proportion of simulations in which, at each given cycle (each cycle corresponds to $k=8$ tests), the LBP marginal oscillates significantly, in the sense that even after $\ell=1000$ iterations, the difference between two iterates is bigger, in at least one coordinate, by more than 0.5, which is a significant contradiction for at least one individual.}
    \label{fig:lbpisbad}
\end{figure}

\begin{figure}
    \hskip-.6cm
    \includegraphics[width=.57\textwidth]{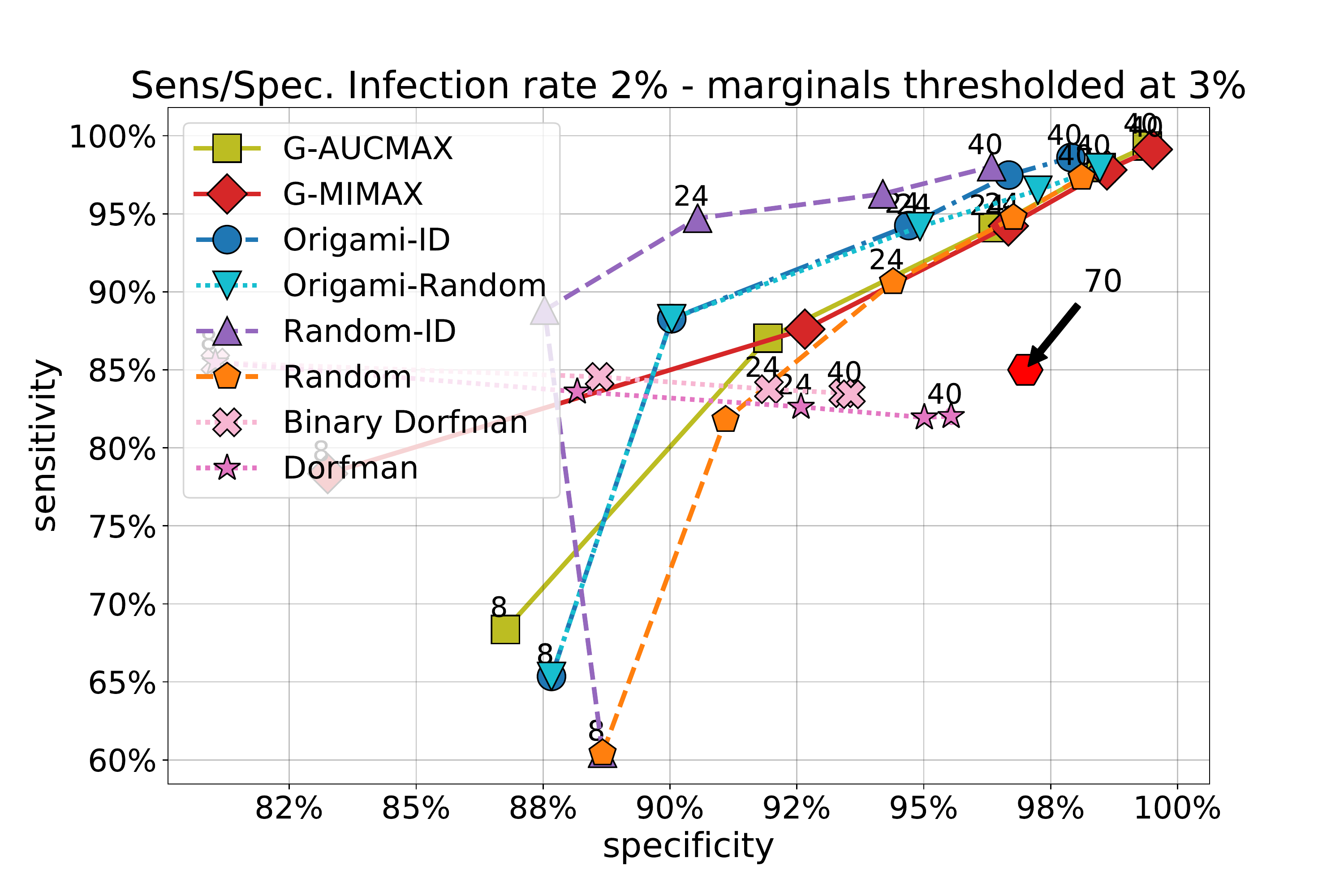}\hskip-.7cm
    \includegraphics[width=.57\textwidth]{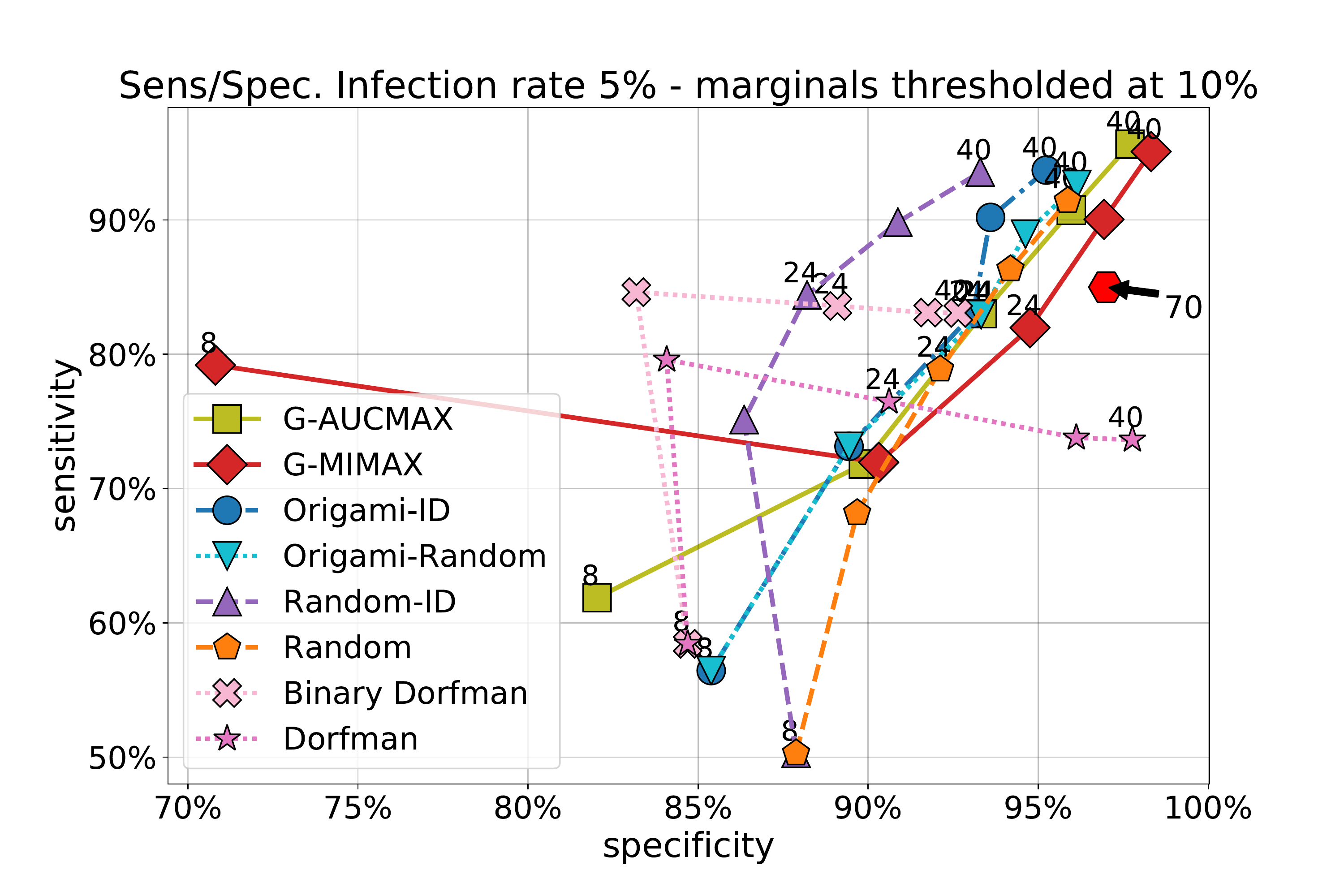}
\caption{We provide in this plot average specificity/sensitivity for each policy, for two infection rates. For each policy, and at each step $t$, we use its marginal approximation $\bar{\bx}_t$ from Algo.\ref{algo:simul} and threshold its coordinates at levels 3\% and 10\% respectively for 2\% and 5\% base infection rates, to make a binary decision. Comparing it to the corresponding $\bx_{\textrm{truth}}$, we compute that simulation's specificity/sensitivity, and average them over 5000 simulations. The specificity/sensitivity of individual tests (requiring 70 tests) is plotted in red, and is significantly outperformed with our approaches.}
    \label{fig:perf_prog}
\end{figure}

\begin{figure}
    \hskip-.6cm
    \includegraphics[width=.57\textwidth]{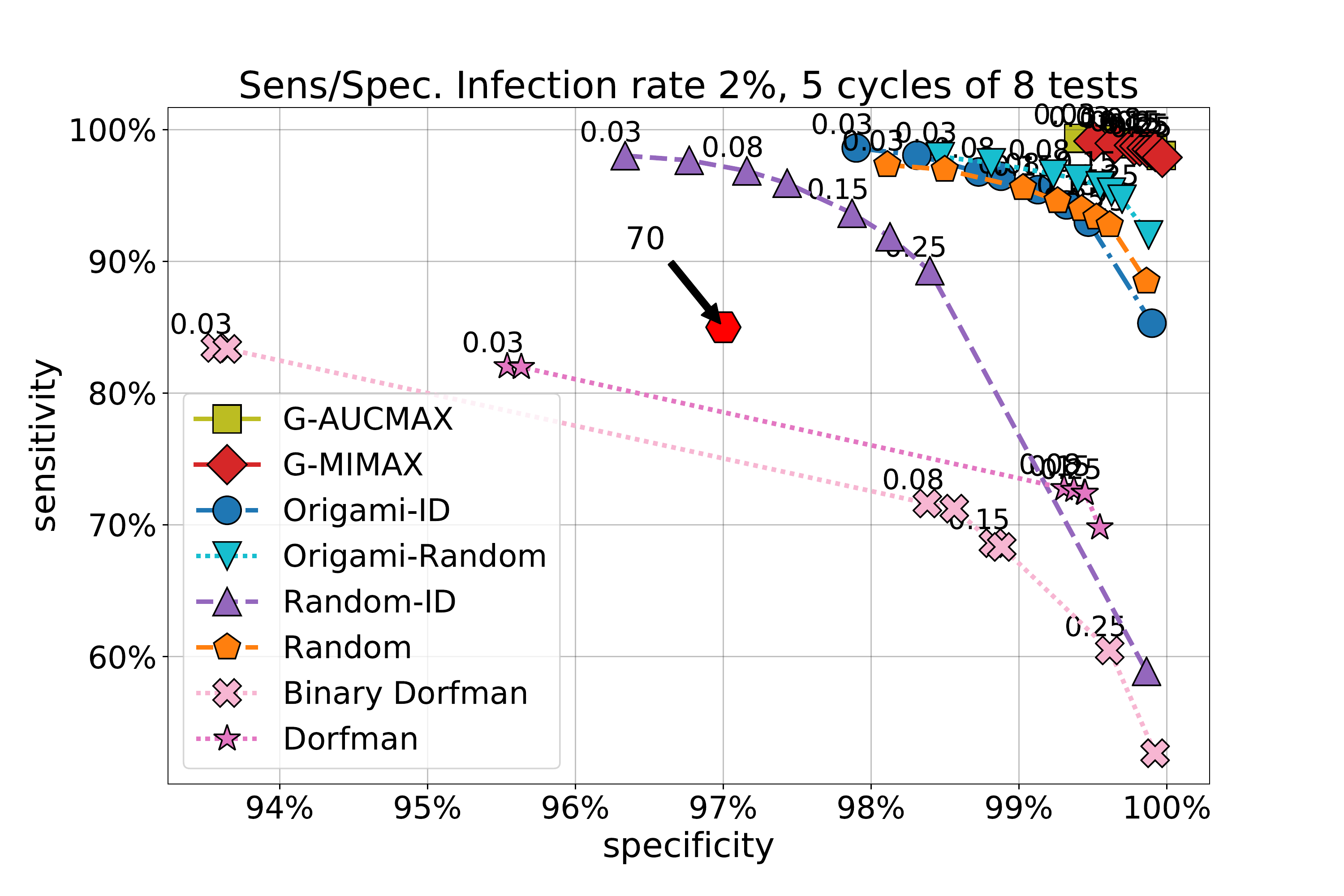}\hskip-.7cm
    \includegraphics[width=.57\textwidth]{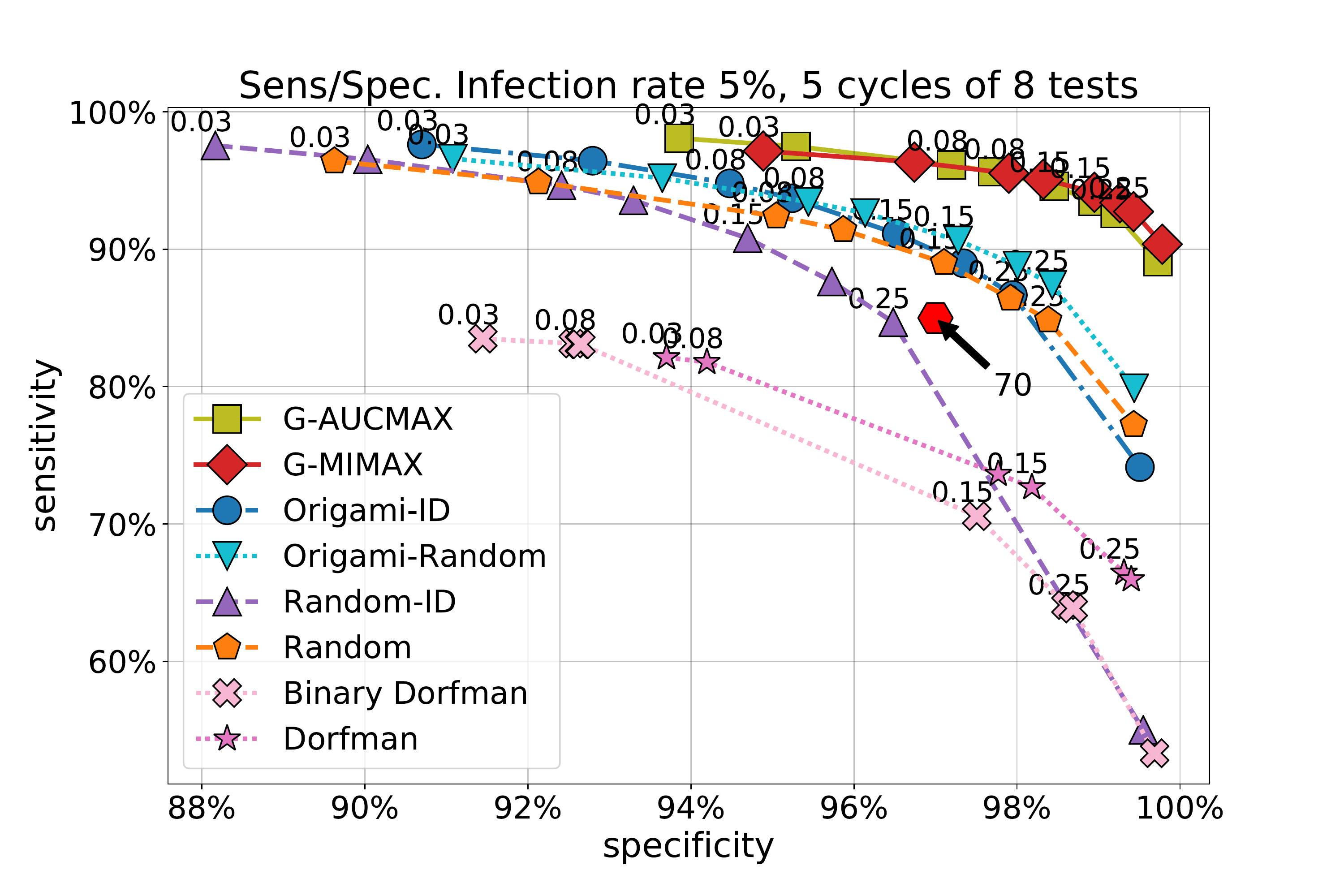}
\caption{As in Figure~\ref{fig:perf_prog} we report average specificity/sensitivity for various policies, but we focus on the final cycle ($t=5$) and vary the threshold used to make decisions, which produces a global specificity/sensitivity curve for all 5,000 experiments.}
    \label{fig:perf_curve}
\end{figure}

\textbf{Decoder and discussion on LBP's convergence.\;} To define the \texttt{MarginalSampler} referred to in Algo.~\ref{algo:simul}, we considered two choices: LBP (\S\ref{sec:LBP}) and the marginal of a posterior sample produced from SMC (\S\ref{sec:algo},\S\ref{sec:SMC}). We compare their performance in Fig.~\ref{fig:lbpisbad}, using the setup of Fig.~\ref{fig:perf_prog}~\&~\ref{fig:perf_curve}, to decode 40 tests generated with the \textbf{Random} policy. Fig.~\ref{fig:lbpisbad} (left) reveals that using the SMC marginal as a decoder, rather than LBP, significantly improves performances (we observed similar results for all other policies). However, because LBP is orders of magnitude faster than SMC, we propose a practical compromise, using a hybrid approach: we run LBP and check whether its iterates have stabilized after at most 1000 iterations. If the marginals produced on the two final successive iterations differ by more than 2\% on any coordinate, we conclude that LBP has not stabilized and is possibly oscillating; in that case we run an SMC, and use its marginal instead. The performance of that approach is comparable to that of SMC. Notice, in Fig.~\ref{fig:lbpisbad} (right), that the number of times LBP is \textit{significantly} unstable is far from negligible. We believe that LBP failures arise because of its inability to handle contradictory tests due to noise, notably for small groups, as can for instance happen in the Binary-Dorfman approach.

\textbf{Performance in terms of sensitivity/specificity.\;} 
In Fig.~\ref{fig:perf_prog} we apply the \textit{same} threshold on the marginal sampler's output $\bar{\bx}_t$ (see Algo.\ref{algo:simul}) at all steps, of all simulations, of all policies, to decide which individuals are classified as positive (marginal above threshold) or negative (below). We record the resulting sensitivity/specificity by comparing it to the corresponding $\bx_{\mathtt{truth}}$. We then obtain 5,000 pairs per policy, and at each cycle, on 5,000 simulations. For those simulations with entirely negative ground-truth state vector, i.e. $\bx_{\mathtt{truth}}=\mathbf{0}$, which happens regularly when $q=2\%$, the sensitivity cannot be evaluated, and those simulations are therefore only used to record specificity. Although we have considered previously the AUC of the marginal in an earlier version of this paper\footnote{https://arxiv.org/abs/2004.12508v1}, we argue that computing average specificity/sensitivity for a fixed threshold results in a more realistic performance assessment: If these policies were to be deployed, one would need to ``ship'' them set with a threshold set beforehand. We report the dynamic progress of average specificity/sensitivity as a function of $t$, here labelled next to markers as total tests carried out. In Figure~\ref{fig:perf_curve}, we plot the average specificity/sensitivity of each policy, obtained this time by varying the threshold (labelled next to markers) after 5 cycles of 8 tests(\textit{i.e.} 40 in total). This recovers a ``frontier'' curve of average sensitivity/specificity levels using all experiments (more plots in \S\ref{subsec:earliercycles} at earlier cycles).

\paragraph{Conclusion.}
Our goal in this work was to maximize the efficiency of group testing in a noisy adaptive setting. We proposed a general framework to do so using Bayesian optimal sequential experimental design. By relying on a particle representation of the posterior, we formulate the problem of designing groups as a combinatorial maximization problem, solved with a greedy forward-backward approach. We have benchmarked our proposals against several baselines (some of our own design), and have shown a substantial improvement in performance. Results obtained with our G-MIMAX and G-AUCMAX approaches beat all other approaches by a wide margin. This work suggests several directions for improvement: quality of posterior sampling, alternative utility functions $\Phi$, improvement of the combinatorial solver tasked to produce groups out of posterior samples. Since our method currently scales exponentially with the number $k$ of requested groups (which we equate in this work with the number of tests available per cycle, for a pool of $n$ patients), an extension of our work that carries out resampling at each group optimization iteration might be required for larger $k$.

\paragraph{Statement of contributions.}
MC, OT and JPV produced the first version of this work. MC proposed to optimize MI using SMC samplers, wrote the first draft of the paper, coded the simulator and ran all experiments contained in the paper. JPV proposed and coded the LBP and G-AUCMAX, carried out the bibliographic survey, and re-wrote extensively the draft. OT prepared all figures and later took the lead on all aspects related to the code, refactoring and open-sourcing it. From v.5, QB and AD were added to the authors' list. QB proposed and coded the simulated annealing approach mentioned in \S\ref{sec:algoMImax}. AD provided guidance on SMC for binary spaces early on in the project, designed the dynamic re-sampling scheme and re-wrote the paper with JPV and MC. 

\paragraph{Acknowledgements} 
We would like to thank Kevin Murphy, Olivier Bousquet, Dan Popovici, Peter Bartlett, Phil Long (Google) for their feedback at various stages of this project; Ignacio Anegon, Jeremie Poschmann and Laurent Tesson (INSERM) for providing us background information on RT-PCR tests, and Nicolas Chopin (CREST) for giving guidance on his work to define SMCs for binary spaces.

\bibliographystyle{plainnat}
\bibliography{biblio.bib}

\newpage
\appendix
\section{Proofs and Algorithms}
We provide in this section more details on the mathematics of our paper. We start in \S\ref{subsec:prooflemma1} with a proof for Lemma \ref{thm:MI}. This in turn motivates the link we make between maximizing mutual information and maximizing the expected neg-entropy of the conditional distribution of $\bX$ given $\YG$, which can be evaluated more quickly than by applying directly Algo.~\ref{algo:computeMI} and even maximized more efficiently using a greedy F/B algorithm as presented in \S\ref{sec:algoMImax}. We conclude this section by providing in \S\ref{sec:LBP} the details of the message passing algorithm run to compute the LBP decoder, as well as, in \S\ref{sec:SMC}, details on the SMC implementation we have considered. 

\subsection{Proof of Lemma~\ref{thm:MI}}\label{subsec:prooflemma1}
Let us start with a single group $\bg\in\cG$. We use the fact that $I_\P(X;\Yg)$ can also be written as 
\begin{equation}\label{eq:MIdecomp}
I_\P(\bX;Y_\bg) := H_\P(\Yg) -\E_{\P_\bX}[H_\P(\Yg|\bX)]\,,
\end{equation}
and compute each term in turn. $H_\P(\Yg)$ can be computed easily from the law of $\bX$ since, by \eqref{eq:testlikelihood},
$$
\P(\Yg=1)= \E_\bX \P(\Yg=1\,|\,\bX) = \E_\bX \left( 1-\Spg + \rho_g [\bg,\bX]\right) = 1-\Spg + \rho_g f_{\P_\bX}(\bg)\,,
$$
from which we obtain
\begin{equation}\label{eq:H1}
H_\P(\Yg) = h\left(\rho_g f_{\P_\bX}(\bg) + 1 - \Spg\right)\,.
\end{equation}
For the second term, we notice that, conditionally to $\bX=\bx$, $Y_\bg$ is a Bernoulli random variable whose expectation only depends on $[\bg,\bx]$, which itself can only take two values $0$ and $1$. By (\ref{eq:senspe}) we deduce:
$$
H_\P(Y_{\bg}\,|\,\bX=\bx) = \begin{cases}h_{\Seg} &\text{if } [\bg,\bx] = 1\,, \\ h_{\Spg} &\text{if } [\bg,\bx] =0\,,\end{cases}
$$
which we can summarize as
\begin{equation}\label{eq:entcond}
H_\P(Y_{\bg}\,|\,\bX=\bx) = h_{\Spg} + \gamma_g[\bg,\bx] \,.
\end{equation}
We deduce that
\begin{equation}\label{eq:H2}
\E_{\P_\bX}[H(\Yg|X)] = \E_{\P_\bX}\left( h_{\Spg} + \gamma_g[\bg,\bX] \right) = h_{\Spg} + \gamma_g f_{\P_\bX}(\bg)\,.
\end{equation}
Plugging (\ref{eq:H1}) and (\ref{eq:H2}) into (\ref{eq:MIdecomp}) gives (\ref{eq:thmMI}). Moving now to the case of a batch $\bG=(\bg_1,\ldots,\bg_k) \in \cG^*$, we use the fact that the entries of $Y_\bG$ are independent from each other given $\bX$ to write, for any $\bx\in\{0,1\}^n$, and using (\ref{eq:entcond}),
$$
H_\P(Y_\bG\,|\,\bX=\bx) = \sum_{j=1}^k H_\P(Y_{\bg_j}\,|\,\bX=\bx) = \sum_{j=1}^k \left( h_{\Spe_{g_j}} + \gamma_{g_j}[\bg_j,\bx] \right)\,.
$$
As a result,
$$
\E_{\P_\bX}[H(Y_\bG\,|\,\bX)] = \sum_{j=1}^k \left( h_{\Spe_{g_j}} + \gamma_{g_j}f_{\P_\bX}(\bg_j) \right)\,.
$$
which gives (\ref{eq:thmMIgroup}).

\subsection{Neg-Entropy and Mutual Information}\label{sec:MIcomputation}
We notice first that, using the identity that defines the mutual information in \S\ref{sec:bayesexp}:
$$\begin{aligned}
U_{\text{MI}}(\bG, \Pt{t}) :=& I_{\P_t}(\bX;\bY_{\bG}) = H_{\P_t}\!(\bY_{\bG}) - H_{\P_t}\!(\bY_{\bG}|\bX) = H_{\P_t}\!(\bY_{\bG}) - \sum_\bx \!\Pt{t}(\bx)H_{\P}(\bY_{\bG}|\bX=\bx)\\
=& H_{\P_t}\!(\bX) - H_{\P_t}\!(\bX|\bY_{\bG}) = H_{\P_t}\!(\bX) - H_{\P_t}\!(\bX|\bY_{\bG})\\
=& H_{\P_t}\!(\bX) - \sum_\by \!\Pt{t}(\by)H_{\P_t}(\bX=\bx|\bY_{\bG}=\by) = H_{\P_t}\!(\bX) + \E_{\Prob_t} [\Phi_{\text{NegEnt}}(\Pt{t}^{\bG})]
\end{aligned}$$
where 
$$
\Phi_{\text{NegEnt}}(\hat{\pi}) = \sum_{i=1} \omega_i \log \omega_i\,, \text{ where } \hat{\pi} = \sum \omega_i \delta_{\bx_i},
$$
is the negative entropy utility. Since $\bG$ has no influence on $H_{\P_t}\!(\bX)$ in the r.h.s of the last line above, maximizing mutual information is equivalent to maximizing the expected neg-entropy of $\bX$ when conditioned on hypothetical test results for group $\bG$. 

From the equality 
$$
I_{\Prob_t}(\bX;\YG) = H_{\Prob_t}(\YG)-\E_\bX[H(\YG|\bX)]\,,
$$
we propose an algorithm that is able to directly evaluate the first term from the vector $D$ and the second term from the matrix $A$, with notations from Algo.\ref{algo:computeutility}. Since $\YG$ is a product distribution conditioned to $\bX$ we have
$$
H(\YG|\bX=\bx) = \sum_{i=1}^k H(Y_{\bg_i}|\bX=\bx).
$$
The resulting algorithm is shown in Algorithm~\ref{algo:computeMI}. Compared to using Algorithm~\ref{algo:computeutility} with $\Phi=\Phi_{\text{NegEnt}}$, the computation of $F$ in $O(N\times 2^k)$ operations to compute $2^k$ entropies over a space of cardinality $N$ in Algorithm~\ref{algo:computeutility}, line 6, is replaced by the computation of $H_2$ in $O(N\times k)$  (Algorithm~\ref{algo:computeMI}, line 2) and of $H_1$ in $O(2^k)$ to compute a single entropy over a space of cardinality $2^k$ (Algorithm~\ref{algo:computeMI}, line 7).

\begin{algorithm}
\DontPrintSemicolon 
\KwIn{$\hPt{t}(\bx)=\sum_{i=1}^N \omega_i \delta_{\bx_i}(\bx) = \hat{\P}_t(\bX=\bx)$; 
$\bG = (\bg_1,\ldots,\bg_k)\in\{0,1\}^{n\times k}$ a set of groups; $\Spe,\Sen\in[0,1]^k$ the specificities and sensitivities of the test for each group in $\bG$.}
\KwOut{The MI utility of the groups $U(\bG) = \MI(\bX;\YG)$.}
$L_{ij} \gets [g_i,\bx_j]$  for $(i,j)\in\interv{k}\times \interv{N}$\;
$h_2 \gets \sum_{i=1}^k  \left[\left(\sum_{j=1}^N \omega_j L_{ij}\right)\left(h(\Sen_i)-h(\Spe_i) \right)+ h(\Spe_i)\right]$ \tcp*{$\E_X[H(\YG|\bX)]$}
$A_{ij} \gets 1-\Spe_i + (\Spe_i+\Sen_i -1) L_{ij}$  for $(i,j)\in\interv{k}\times \interv{N}$ \tcp*{$\P(Y_{\bg_i}=1\,|\,\bX=\bx_j)$}
$B_{ij} \gets \prod_{t=1}^k  A_{tj}^{b_{it}} (1-A_{tj})^{1-b_{it}}$ for $(i,j)\in\interv{2^k}\times \interv{N}$, where $b_{it}$ is the $t$-th bit from the right in the binary expansion of $i$ \tcp*{$\P(\YG=i\,|\,\bX=\bx_j)$}
$C_{ij} \gets B_{ij}\times \omega_j$ for $(i,j)\in\interv{2^k}\times \interv{N}$ \tcp*{$\P(\YG=i\,,\,\bX=\bx_j)$}
$D_i \gets \sum_{j=1}^N C_{ij}$ for $i\in\interv{2^k}$ \tcp*{$\P(\YG=i)$}
$h_1 \gets -\sum_{i=1}^{2^k} D_j \log(D_j)$ \tcp*{$H(\YG)$}
\Return{$h_1 - h_2$}
\caption{Compute MI utility of a set of groups}
\label{algo:computeMI}
\end{algorithm}

\subsection{Algorithm to Maximize MI}\label{sec:algoMImax}
In this section we describe an algorithm to maximize the mutual information utility, subject to the constraint that each group should have at most $\nmax$ individuals, and that the batch should contain any number $m\leq k$ of groups. Simply put, we greedily create groups one by one, until we have $m$ groups. Once we have created groups $\bG^{j}=(\bg_1,\ldots,\bg_{j})$, we create a new group $\bg_{j+1}$ by starting from an empty group $\bg=\emptyset$ (line 3) and growing iteratively the group by selecting the individual that adds the most mutual information
$$
\bg \leftarrow \bg\cup\{i\}\text{ where }i \in \arg\max_{u} I_\P( \bX \,;\, Y_{(\bG^{j}, \bg\cup\{u\})}),
$$
until either we stop making progress in terms of mutual information, or when the group has already reached size $\nmax$. We consider additionally a variant in which we do not only consider greedy \textit{addition} of individuals to form a group, but also \textit{removal}, resulting in Forward-Backward iterations. Algorithm~\ref{algo:recursiveMIOptim} describes an efficient way to carry out such forward passes more efficiently than by evaluating repeatedly Algorithm~\ref{algo:computeMI}, because it leverages the fact that $\bG$ is built sequentially, column by column. We omit the backward pass which only consists in changing line 5 (by setting $\bg_\omega = 1$ instead) and removing (rather than adding) $\boldsymbol{\iota}_{u^*}$ from $\bg$ in line 16. At each loop index in $i$ (line 4), having a number $F$ of forward passes and $B$ of backward passes, with $F>B$, requires executing the body of the loop (lines 5 to 14) $F$ times in forward mode, and $B$ times in backward mode.

We use the following notations in Algorithm~\ref{algo:recursiveMIOptim}: small letters denote constants, small bold letters denote vectors, bold capital letters are matrices and bold greek letters are 3D tensors.

\begin{algorithm}
\DontPrintSemicolon 
\KwIn{$\hPt{t}(\bx)=\sum_{i=1}^N \omega_i \delta_{\bx_i}(\bx) = \hat{\P}_t(\bX=\bx)$\; 
Number of $m$ groups to add, $\nmax$ upperbound on group size \; 
$\rho_i = \sigma_i + s_i - 1, \gamma_i = h_{s_i} - h_{\sigma_i}, \,i\in\interv{\nmax}.$
}
\KwOut{Approximate maximizer $\bG$ of $U(\bG) = \MI(\bX;\YG)$.}
$\bG \gets \mathbf{0}_{n\times 0},\, \bP \gets \mathbf{1}_{n\times 1},\, h \gets 0$\;
\For{$j\gets 1$ \KwTo $m$}{
    $\bg \gets \mathbf{0}_n, \, f_0\gets 0, \mathbf{p}=\mathbf{0}_{N} $\tcp*{initialize group, objective, positive in group across particles indicator}
    \For{$i\gets 1$ \KwTo $\nmax$}{
        $\boldsymbol{\iota} \gets (w \in \interv{n} : \mathbf{g}_w = 0), r \gets |\boldsymbol{\iota}|$\tcp*{indices that can be added}
        $T_{uv} \gets \bx_v[\boldsymbol{\iota}_u] \vee \mathbf{p}_v , (u,v)\in\interv{r}\times\interv{N}$\tcp*{detect positive in candidates}
        $\mathbf{h}_u^2 \gets h_{\sigma_i} + \gamma_{i} \sum_{v} T_{uv}\, \omega_v + h, \,u\in\interv{r}$\tcp*{conditional entropies}
        $\boldsymbol{\Gamma}_{u,v,0} \gets 1- \sigma_i + \rho T_{uv},\; \boldsymbol{\Gamma}_{u,v,1} \gets \sigma_i - \rho T_{uv}, u,v),\;(u,v)\in\interv{r}\times\interv{N}$\tcp*{probabilities of 2 possible test results, tensorized}
        $\boldsymbol{\Xi}_{u,v,b} \gets \boldsymbol{\Gamma}_{u,v,0} \bP_{v,b},\, \boldsymbol{\Pi}_{u,v,b+2^{j-1}} \gets \boldsymbol{\Gamma}_{u,v,1} \bP_{v,b},\; (u,v,b)\in\interv{r}\times\interv{N}\times\interv{2^{j-1}}$ \tcp*{probability tensor across all possible candidate groups $\times$ particles $\times$ $2^j$ hypothetical test results across $j$ groups. }
        $Q_{u,b} \gets  \sum_v \boldsymbol{\Pi}_{u,v,b}\, \omega_v, (u,b)\in\interv{r}\times\interv{2^{j-1}}$\tcp*{marginalization / particles}
        $\mathbf{h}^1 \gets -\sum_{u,b} Q_{u,b} \log(Q_{u,b})$ \tcp*{unconditional entropy}
        $\mathbf{m} \gets \mathbf{h}^1 - \mathbf{h}^2$ \tcp*{MI objective function}     
        $u^* \gets \argmax_u \mathbf{m}_u, f_i\gets \mathbf{m}_{u^*}, \bP^{\text{new}}_{v,b}=\boldsymbol{\Pi}_{u^*,v,b}$ \tcp*{greedy selection}
        $h^{\text{new}} \gets \mathbf{h}^2_{u^*}$ \tcp*{record conditional entropies of all tests so far}
        \If{$f_i>f_{i-1}$}{
            $\bg  = \bg \cup  \{\boldsymbol{\iota}_{u^*}\}$\tcp*{incorporate candidate}
            $\mathbf{p} = T_{u^*,\cdot}$ \tcp*{update vector of positive in group across particles} 
         }
        \Else{
        $\bG = [\bG, \bg]$\tcp*{incorporate $\bg$}
        $\bP=\bP^{\text{new}},\; h = h^{\text{new}}$\tcp*{update probability \& entropy after adding $\bg$}
        break\;
        }
    } 
}
\caption{G-MIMAX: Optimize MI of $m$ prospective group tests with greedy search}
\label{algo:recursiveMIOptim}
\end{algorithm}

As an alternative to greedy approaches, we have also considered stochastic optimization approaches based on simulated annealing, with a constant temperature. These approaches can also be combined with our greedy algorithm (or any algorithm), by choosing its output as initialization, rather than a random set of groups. A simple implementation did not yield significant improvement in performance for a comparable running time.

\subsection{Approximate posterior estimation by loopy belief propagation}\label{sec:LBP}
A standard way to compute an approximation of the posterior marginals is to run loopy belief propagation (LBP) until convergence. Here we detail the LBP equations for our setting. Given $n$ individuals and $m$ tests performed with groups $\bg_i, \ldots, \bg_m \in\mathcal{G}$, LBP alternates  passing messages $\mu_{i\rightarrow j} = (\mu_{i\rightarrow j}(0), \mu_{i\rightarrow j}(1)) \in \RR^2$ from individuals $i\in\interv{n}$ to groups $j\in\interv{m}$ with $i\in\bg_j$, and $\tilde{\mu}_{j\rightarrow i} = (\tilde{\mu}_{j\rightarrow i}(0), \tilde{\mu}_{j\rightarrow i}(1)) \in \RR^2$ from groups $j$ with $i\in\bg_j$ to individuals $i$, respectively. 

Adding a superscript $(t)$ to clarify the messages sent at the $t$-th iteration of LBP, the messages from an individual $i\in\interv{n}$ to a group $j\in\interv{m}$ with $i\in\bg_j$
follow the standard equations: 
\begin{equation}\label{eq:forward}
    \begin{cases}
      \mu_{i\rightarrow j}^{(t+1)}(0) &= (1-q_i) \prod\limits_{j'\neq j\,:\,i \in \bg_{j'}} \tilde{\mu}_{j'\rightarrow i}^{(t)}(0) \,, \\
      \mu_{i\rightarrow j}^{(t+1)}(1) &= q_i \prod\limits_{j'\neq j\,:\,i \in \bg_{j'}} \tilde{\mu}_{j'\rightarrow i}^{(t)}(1) \,. 
    \end{cases}
\end{equation}

The messages from a group $j\in\interv{m}$ to an individual $i\in\interv{n}$ with $i \in \bg_j$ depend on the result of the test $Y_{\bg_j}$: if $Y_{\bg_j} = 0$ (negative test), then
\begin{equation}\label{eq:backward0}
    \begin{cases}
      \tilde{\mu}_{j\rightarrow i}^{(t)}(0) &= \Spe_{\bg_j}\!\!\!\!\! \prod\limits_{i'\neq i\,:\,i' \in \bg_{j}}\!\!\!\!\! \mu_{i'\rightarrow j}^{(t)}(0) + (1-\Sen_{\bg_j})\left( \prod\limits_{i'\neq i\,:\,i' \in \bg_{j}}\!\!\!\!\! (\mu_{i'\rightarrow j}^{(t)}(0) + \mu_{i'\rightarrow j}^{(t)}(1)) - \!\!\!\!\! \prod\limits_{i'\neq i\,:\,i' \in \bg_{j}}\!\!\!\!\! \mu_{i'\rightarrow j}^{(t)}(0) \right) \\
      &= (1 - \Sen_{\bg_j})\!\!\!\!\! \prod\limits_{i'\neq i\,:\,i' \in \bg_{j}}\!\!\!\!\! (\mu_{i'\rightarrow j}^{(t)}(0) + \mu_{i'\rightarrow j}^{(t)}(1)) + (\Spe_{\bg_j} + \Sen_{\bg_j} - 1) \prod\limits_{i'\neq i\,:\,i' \in \bg_{j}} \mu_{i'\rightarrow j}^{(t)}(0)\,, \\
      \tilde{\mu}_{j\rightarrow i}^{(t)}(1) &= (1 - \Sen_{\bg_j})\!\!\!\!\! \prod\limits_{i'\neq i\,:\,i' \in \bg_{j}} \!\!\!\!\!(\mu_{i'\rightarrow j}^{(t)}(0) + \mu_{i'\rightarrow j}^{(t)}(1)) \,,
    \end{cases}
\end{equation}
while if $Y_{\bg_j} = 1$ (positive test), then
\begin{equation}\label{eq:backward1}
    \begin{cases}
      \tilde{\mu}_{j\rightarrow i}^{(t)}(0) &= (1 - \Spe_{\bg_j})\!\!\!\!\! \prod\limits_{i'\neq i\,:\,i' \in \bg_{j}}\!\!\!\!\! \mu_{i'\rightarrow j}^{(t)}(0) + \Sen_{\bg_j} \left( \prod\limits_{i'\neq i\,:\,i' \in \bg_{j}}\!\!\!\!\! (\mu_{i'\rightarrow j}^{(t)}(0) + \mu_{i'\rightarrow j}^{(t)}(1)) - \!\!\!\!\! \prod\limits_{i'\neq i\,:\,i' \in \bg_{j}} \!\!\!\!\!\mu_{i'\rightarrow j}^{(t)}(0) \right) \\
      &= \Sen_{\bg_j} \prod\limits_{i'\neq i\,:\,i' \in \bg_{j}} (\mu_{i'\rightarrow j}^{(t)}(0) + \mu_{i'\rightarrow j}^{(t)}(1)) - (\Spe_{\bg_j} + \Sen_{\bg_j} - 1) \prod\limits_{i'\neq i\,:\,i' \in \bg_{j}} \mu_{i'\rightarrow j}^{(t)}(0)\,, \\
      \tilde{\mu}_{j\rightarrow i}^{(t)}(1) &= \Sen_{\bg_j} \prod\limits_{i'\neq i\,:\,i' \in \bg_{j}} (\mu_{i'\rightarrow j}^{(t)}(0) + \mu_{i'\rightarrow j}^{(t)}(1)) \,.
    \end{cases}
\end{equation}
To simplify these equations let us introduce some notations:
\begin{equation*}
    \begin{split}
        e^{-\mu_i} &= \frac{q_i}{1-q_i} \text{ for }i\in \interv{n}\,,\\
        e^{\gamma_j^0} &= \frac{\Spe_{\bg_j} + \Sen_{\bg_j} - 1}{1 - \Sen_{\bg_j}} \text{ for }j\in \interv{m}\,,\\
        e^{\gamma_j^1} &= \frac{\Spe_{\bg_j} + \Sen_{\bg_j} - 1}{ \Sen_{\bg_j}} \text{ for }j\in \interv{m}\,.\\
    \end{split}
\end{equation*}
Furthermore, let us make the change of variables, for any $(i,j,t)\in\interv{n}\times \interv{m} \times \NN$, 
\begin{equation}
    \begin{split}
        \alpha_{ij}^{(t)} &= \ln\left( \frac{\mu_{i\rightarrow j}^{(t)}(0)}{\mu_{i\rightarrow j}^{(t)}(0) + \mu_{i\rightarrow j}^{(t)}(1)}\right) \,, \\
        \beta_{ij}^{(t)} &= \ln\left( \frac{\tilde{\mu}_{j\rightarrow i}^{(t)}(0)}{\tilde{\mu}_{j\rightarrow i}^{(t)}(1)}\right) \,.
    \end{split}
\end{equation}
Then (\ref{eq:forward}) can be rewritten as:
\begin{equation}
    \begin{split}
        \alpha_{ij}^{(t)} &= - \ln \left( 1 + \frac{q_i}{1-q_i} \prod\limits_{j'\neq j\,:\,i \in \bg_{j'}} \frac{ \tilde{\mu}_{j'\rightarrow i}^{(t)}(1)}{ \tilde{\mu}_{j'\rightarrow i}^{(t)}(0)} \right) \\
        &= - \ln \left( 1 + e^{-\mu_i - \sum_{j'\neq j\,:\,i \in \bg_{j'}} \beta_{ij'}^{(t)}} \right) \\
        &= - \ln \left( 1 + e^{-\mu_i - \bar{\beta}_i^{(t)} + \beta_{ij}^{(t)}} \right) \,,
    \end{split}
\end{equation}
where 
$$
\bar{\beta}_i^{(t)} = \sum_{ j\,:\,i \in \bg_{j}} \beta_{ij}^{(t)} \,.
$$
Similarly, denoting
$$
\bar{\alpha}_j^{(t)} = \sum_{ i\,:\,i \in \bg_{j}} \alpha_{ij}^{(t)} \,,
$$
we can rewrite (\ref{eq:backward0}) and (\ref{eq:backward1}) as follows: if $Y_{\bg_j}=0$,
\begin{equation}
    \begin{split}
        \beta_{ij}^{(t)} &=  \ln \left( 1 + \frac{\Spe_{\bg_j} + \Sen_{\bg_j} - 1}{1-\Sen_{\bg_j}} \prod\limits_{i'\neq i\,:\,i' \in \bg_{j}} \frac{ \mu_{i'\rightarrow j}^{(t)}(0)}{ \mu_{i'\rightarrow j}^{(t)}(0) + \mu_{i'\rightarrow j}^{(t)}(1)} \right) \\
        &= \ln \left( 1 + e^{ \gamma_j^0 + \sum_{i'\neq i\,:\,i' \in \bg_{j}} \alpha_{i'j}^{(t)}} \right) \\
        &= \ln \left( 1 + e^{\gamma_j^0 + \bar{\alpha}_j^{(t)} - \alpha_{ij}^{(t)}} \,, \right)\,,
    \end{split}
\end{equation}
and if $Y_{\bg_j}=1$,
\begin{equation}
    \begin{split}
        \beta_{ij}^{(t)} &=  \ln \left( 1 - \frac{\Spe_{\bg_j} + \Sen_{\bg_j} - 1}{\Sen_{\bg_j}} \prod\limits_{i'\neq i\,:\,i' \in \bg_{j}} \frac{ \mu_{i'\rightarrow j}^{(t)}(0)}{ \mu_{i'\rightarrow j}^{(t)}(0) + \mu_{i'\rightarrow j}^{(t)}(1)} \right) \\
        &= \ln \left( 1 - e^{ \gamma_j^1 + \sum_{i'\neq i\,:\,i' \in \bg_{j}} \alpha_{i'j}^{(t)}} \right) \\
        &= \ln \left( 1 - e^{\gamma_j^1 + \bar{\alpha}_j^{(t)} - \alpha_{ij}^{(t)}} \,, \right)\,.
    \end{split}
\end{equation}
After convergence of the messages (denoted as $t=\infty$), we estimate the posterior marginal of the $i$-th individual as
\begin{equation}
    \begin{split}
        \ln \frac{P_{\text{LBP}}(D_i = 1\,|\,Y_{\bg_1},\ldots,Y_{\bg_m})}{P_{\text{LBP}}(D_i = 0\,|\,Y_{\bg_1},\ldots,Y_{\bg_m})} &= \ln \frac{q_i}{1-q_i} \prod\limits_{j\,:\,i\in\bg_j} \frac{\tilde{\mu}_{j\rightarrow i}^{(\infty)} (1)}{\tilde{\mu}_{j\rightarrow i}^{(\infty)} (0)} \\
        &= -\mu_i - \sum_{j\,:\,i\in\bg_j} \beta_{ij}^{(\infty)} \,,
    \end{split}
\end{equation}
that is,
\begin{equation}
        P_{\text{LBP}}(D_i = 1\,|\,Y_{\bg_1},\ldots,Y_{\bg_m}) = \frac{1}{1 + e^{\mu_i + \sum_{j\,:\,i\in\bg_j} \beta_{ij}^{(\infty)}}} \,.
\end{equation}

\subsection{Approximate posterior estimation by sequential Monte Carlo sampler}\label{sec:SMC}

We detail here the SMC sampler algorithm used to provide a Monte Carlo approximation 
$$\hPt{t} = \sum_{i=1}^N \omega^t_i \delta_{\bx^t_i},$$ 
of $\Pt{t}$ given an approximation $\hPt{t-1} = \sum_{i=1}^N \omega^{t-1}_i \delta_{\bx^{t-1}_i}$  of $\Pt{t-1}$. The main idea is to introduce intermediate pmfs $\Pt{t}^{\gamma_k}$ of the form 
\begin{equation}
    \Pt{t}^{\gamma}(\bx) \propto \Pt{t-1}(\bx) \{\P(\mathbf{Y}_{\mathbf{G}^{t}}=\by^{t}\,|\,\bX=\bx)\}^{\gamma},
\end{equation}
 bridging smoothly  $\Pt{t-1}$ to  $\Pt{t}$ using a real sequence $\gamma_k$ increasing from $0$ to $1$ so that $\Pt{t}^{0}=\Pt{t-1}$ and $\Pt{t}^{1}=\Pt{t}$.
We then approximate sequentially these pmfs using a combination of importance sampling, resampling and MCMC steps \cite{del2006sequential,schafer2013sequential}. This method is detailed in Algorithm \ref{algo:SMC}.

Practically given an approximation of $\hPt{t}^{\gamma_{k-1}} = \sum_{i=1}^N \omega_i \delta_{\bx_i}$  of $\Pt{t}^{\gamma_{k-1}}$, an importance sampling approximation of $\Pt{t}^{\gamma_{k}}$ is given by $\hPt{t}^{\gamma_{k}}=\sum_{i=1}^N \omega'_i \delta_{\bx_i}$ where
\begin{equation}\label{eq:updateimportanceweights}
    \omega'_i \propto \omega_i~\{\P(\mathbf{Y}_{\mathbf{G}^{t}}=\by^{t}\,|\,\bX=\bx_i)\}^{\gamma_k-\gamma_{k-1}},\quad \sum_{i=1}^N \omega'_i=1,
\end{equation}
and a proxy measuring the ``quality'' of this approximation is the Effective Sample Size (ESS):
\begin{equation}\label{eq:ESS}
    \text{ESS}= \frac{1}{N\sum_{i=1}^N (\omega'_i)^2}\in [1/N,1].
\end{equation}
Simply put, the higher the ESS, the better the approximation. For equally weighted particles, one has $\text{ESS}=1$. We select here $\gamma_k$ such that the ESS is equal to a pre-specified value in $[1/N,1)$ (set to 0.9 in our experiments) and, if this yields $\gamma_k>1$, we set $\gamma_k=1$ . Practically, this is achieved using a bisection search as described in~\citep[Proc.2]{schafer2013sequential}. 
Once we have determined $\gamma_k$, we then compute the new importance weights using~\eqref{eq:updateimportanceweights}, and then use a resampling procedure to replicate particles with high weights and discard particles with low weights; i.e. we approximate $\hPt{t}^{\gamma_{k}}$ by 
\begin{equation}\label{eq:resampledapproximation}
    \tilde{\pi}_t^{\gamma_k}=\frac{1}{N}\sum_{i=1}^N n_i \delta_{\bx_i}= \frac{1}{N}\sum_{i=1}^N \delta_{\tilde{\bx}_i}.
\end{equation}
Each particle $\bx_i$ is copied $n_i\in\interv{N}$ times with $\sum_{i=1}^N n_i=N$. This can be achieved by sampling $N$ times from $\tilde{\bx}_i\sim \hPt{t}^{\gamma_{k}}$ so that $(n_1,...,n_N)$ follow a multinomial distribution. However, we use here instead the systematic resampling scheme described in \cite[Proc.3]{schafer2013sequential} which is faster to implement and enjoys better theoretical properties. 

To improve the particle approximation \eqref{eq:resampledapproximation} of $\Pt{t}^{\gamma_{k}}$, the particles $\tilde{\bx}_i$ are then evolved according to a MCMC kernel of invariant pmf $\Pt{t}^{\gamma_{k}}$. The simplest scheme consists in using the Gibbs sampler \cite{geman1984stochastic} which cycles through the $n$ components of $\bx\in\{0,1\}^n$ 
\begin{equation}\label{eq:Gibbskernel}
    \mathbb{P}_{\text{G}}(\bX'=\bx'|\bX=\bx)=\prod_{j=1}^n \frac{\Pt{t}^{\gamma_{k}}(\bx'_{:{j-1}},x'_j,\bx_{j+1:})}{\Pt{t}^{\gamma_{k}}(\bx'_{:{j-1}},0,\bx_{j+1:})+\Pt{t}^{\gamma_{k}}(\bx'_{:{j-1}},1,\bx_{j+1:})},
\end{equation}
where $\bx'_{:0}=\emptyset$, $\bx'_{:{j-1}}=(x'_1,...,x'_{j-1})$ for $j\geq1$, $\bx_{:k+1}=\emptyset$, $\bx_{{j+1}:}=(x_{j+1},...,x_{k})$ for $j<k$. We used in the paper a modified variant of that Gibbs sampler proposed in~\cite{liu1996peskun}, i.e.
\begin{equation}\label{eq:ModifiedGibbskernel}
    \mathbb{P}_{\text{MG}}(\bX'=\bx'|\bX=\bx)=\prod_{j=1}^n \{\alpha_j(\bx'_{:{j-1}},\bx_{j:})\delta_{\neg x_{j}}(x'_j)+(1-\alpha_j(\bx'_{:{j-1}},\bx_{j:}))\delta_{x_{j}}(x'_j)\},
\end{equation}
for 
\begin{equation}
\alpha_j(\bx'_{:{j-1}},\bx_{j:})=\min\left(1,\frac{\Pt{t}^{\gamma_{k}}(\bx'_{:{j-1}},\neg x_j,\bx_{j+1:})}{\Pt{t}^{\gamma_{k}}(\bx'_{:{j-1}},x_j,\bx_{j+1:})}\right).
\end{equation}
This boils down to proposing to flip sequentially each coordinate $j$, this flip being accepted with probability $\alpha_j(\bx'_{:{j-1}},\bx_{j:})$.

We also tried the independent Metropolis-Hastings sampler described in \cite{schafer2013sequential} that uses all the particles to build a proposal on $\{0,1\}^n$. 
We iterate these steps - schedule calculation, importance sampling, resampling and MCMC moves - until $\gamma_k=1$.

\begin{algorithm}\label{alg:SMCsampler}
\DontPrintSemicolon 
\KwIn{
Approximation $\hPt{t-1} = \sum_{i=1}^N \omega^{t-1}_i \delta_{\bx^{t-1}_i}$ of $\Pt{t-1}$\; 
}
\KwOut{Approximation $\hPt{t} = \sum_{i=1}^N \omega^{t}_i \delta_{\bx^{t}_i}$ of $\Pt{t}$}
$ \boldsymbol{\omega}  \gets \boldsymbol{\omega}_{t-1}, \mathbf{X} \gets \mathbf{X}_{t-1}$.  \;  
$\gamma \gets\texttt{AdaptiveSchedule}(0,\boldsymbol{\omega},\mathbf{X},\by^t)$.   \tcp*[f]{determine first $\gamma$}\; 
$\boldsymbol{\omega} \gets \texttt{ImportanceWeights}(\gamma,\boldsymbol{\omega},\mathbf{X})$.   \tcp*[f]{compute importance weights}\; 
\While{$\gamma < 1$}{
$\widetilde{\mathbf{X}} \gets \texttt{Resample}(\boldsymbol{\omega},\mathbf{X}).$\tcp*[f]{discard/multiply particles with low/high weights}\;

$ \boldsymbol{\omega} \gets \mathbf{1}_{N}/N$.\;

$\mathbf{X} \gets \texttt{MCMC}(\gamma, \widetilde{\mathbf{X}},\by^t).$ \tcp*[f]{MCMC moves targeting $\Pt{t}^{\gamma}$}\;

$ \gamma_{\text{old}} \gets \gamma$.\;

$\gamma \gets\texttt{AdaptiveSchedule}(\gamma_{\text{old}},\boldsymbol{\omega},\mathbf{X},\by^t).$    \tcp*[f]{determine next $\gamma$}\; 
$\boldsymbol{\omega} \gets \texttt{ImportanceWeights}(\gamma-\gamma_{\text{old}},\boldsymbol{\omega},\mathbf{X})$.   \tcp*[f]{compute importance weights}\; 
}
$\boldsymbol{\omega}_{t} \gets \boldsymbol{\omega}, \mathbf{X}_{t} \gets \mathbf{X}$.  \; 
\Return{$\hPt{t} = \sum_{i=1}^N \omega^{t}_i \delta_{\bx^{t}_i}$}

\caption{\texttt{Sampler}$(\by^t, \bG, \hPt{t-1})$ returns $N$ approximate samples from $\P_{t}$ given $\by^t$ and $\hPt{t-1}$.}
\label{algo:SMC}
\end{algorithm}

\section{Policies and group selectors}\label{sec:policies}
We provide in this section more details on the various baselines we have considered in \S\ref{sec:simuls} in the main body of the paper.
\subsection{Group Selectors}
We start with selectors that require no knowledge other than the base infection rate; introduce the informative Dorfman procedure that builds on marginal information; and conclude with our selectors, G-MIMAX and G-AUCMAX. All selectors are constrained by a maximal size for groups $\nmax$.

\begin{itemize}[noitemsep,nosep,wide,partopsep=-2pt]
\item \textbf{\citeauthor{dorfman1943detection} Splitting (D).} It splits all $n$ patients into subgroups of size $\approx \min(\nmax,1+\lceil 1/\sqrt{q}\rceil)$.

\item \textbf{Split only positives (SplitPos).} The second stage of \citeauthor{dorfman1943detection} tests, focusing exclusively on those groups that tested positive after \textbf{(D)}. \textbf{(SplitPos:0)} tests individually all samples that have appeared in a positive group; \textbf{(SplitPos:2)}  uses~\citeauthor{Hwang1972method}'s hierarchical binary splitting approach~\citeyearpar{Hwang1972method}.

\item \textbf{\citeauthor{Mezard2011Group} (MT).} It selects randomly groups of the same size $g$ across all $n$ possible patients. Given a prior $q$, the group is chosen to get an acceptance probability of $1/2$, yielding $g =\min(\nmax,\log((\Seg-1/2)/\rho_g)/\log(1-q)).$
\citep{Mezard2011Group} proves that in the absence of noise this choice is asymptotically optimal.

\item \textbf{Origami fixed design (OM3).} We also consider predefined groups, as enumerated in the \texttt{Origami M3} assay matrix~\citep{kainkaryam2008poolhits} of size $70\times 22$, with 22 groups whose size is equal to or smaller than $10$. This matrix was proposed with a deterministic decoder that operates assuming an infection rate lower than $\approx5\%$, in a noiseless setting. We therefore expect that assay to be the most useful when $q\leq 5\%$.

\item \textbf{Informative Dorfman (ID).} Given results from the first exploitable wave of tests, we plug the marginal distribution produced by a sampler in the informative Dorfman rule \citep{mcmahan2012informative}, a generalization to a noisy setting of an approach proposed by \citep{hwang1975generalized}. The rule proceeds by sorting patients by increasing marginal infection probability, and group them with groups that are initially large (to clear large subsets of unlikely infected patients) to small (to test individuals likelier to be infected in smaller groups). More precisely, given a sorted list of individuals with increasing infection probability $\mathbf{p}= (p_1,\dots,p_n)$, \citep{mcmahan2012informative} propose in their pool specific optimal Dorfman (PSOD) algorithm to group together the first $c^*$ individuals, where $c^*$ is defined as
\begin{equation}\label{eq:id}c^* = \argmin_c \frac{1}{c}\left(1+ \mathbf{1}_{c>1} c \left(s + (1-s-\sigma)\prod_{u=1}^c(1-p_{u})\right)\right),
\end{equation}
remove them from the queue and proceed until all individuals are grouped. We constrain $c$ to be smaller than $\nmax$. In this work, because the infection prior is uniform, we first run a first wave of tests (using either \textbf{Random} or \textbf{Origami}) and use the resulting marginal as an estimate for $\mathbf{p}$. When appropriate, we also use group size specific sensitivites and specificities in~\eqref{eq:id}.


\item \textbf{Greedy Maximization of Mutual Information (G-MIMAX) and AUC (G-AUCMAX).} We optimize the MI and AUC utilities using $k$ groups as described in Algorithm~\ref{algo:computeutility}. The greedy approach for G-MIMAX is detailed in Algo.~\ref{algo:recursiveMIOptim}, G-AUCMAX uses calls to Algo.~\ref{algo:computeutility} along with a greedy forward/backward solver. We use 3 Forward steps and 2 Backward steps in all our experiments. We have experiemnted with a larger number of forward / backward steps, but we find that the resulting computational overhead is not worth the small variation in performance that is obtained.
\end{itemize}

\subsection{Policies}
We consider the following policies, all composed by one or at most two group selectors.
\begin{itemize}[noitemsep,nosep,wide,partopsep=-4pt]
    \item \textbf{Dorfman}: starts with Dorfman splitting first \textbf{(D)} followed by \textbf{(SplitPos:0)}
    \item \textbf{Binary Dorfman}: starts with Dorfman splitting first \textbf{(D)} followed repeatedly by \textbf{(SplitPos:2)}.
    \item \textbf{Random} : generates random groups at each stage with the \textbf{(MT)} selector. 
    \item \textbf{Random-ID} : starts with a random group \textbf{(MT)}, follows with \textbf{(ID)}.
    \item \textbf{Origami-ID}: uses \textbf{(OM3)} for 22 tests, and then switches to \textbf{(ID)}.
    \item \textbf{Origami-Random}: uses \textbf{(OM3)} for 22 tests, and then switches to \textbf{(ID)}.
    \item \textbf{G-MIMAX} and \textbf{G-AUCMAX}: optimize first utilities on a sample with $N=10^4$ particles from the prior, and then from the posterior distribution using a SMC sampler with the same size $N$.
\end{itemize}
All policies are decoded using the same decoded, a hybrid rule that runs an LBP, checks if it has converged (tolerance of 2\% for any coordinate) and, if not, runs a SMC (see discussion in \S\ref{sec:simuls}).
\section{Additional Experiments}
We provide in this section many more results that validate further the good performance of our approaches and illustrate their robustness. 

\begin{itemize}
\item \textbf{10\% infection rate}: we list in \S\ref{subsec:IR10} new plots, comparable to those already included in the main body of the paper, for a higher infection prevalence prior of $q=\hat{q}=10\%$. 
\item \textbf{Dynamics of testing performance}: other plots are listed in \S\ref{subsec:earliercycles} to provide a more detailed assessment of the performance of each policy as the number of tests that is revealed grows. 
\item \textbf{Robustness to Mis-specification}: we list plots in \S\ref{subsec:misspecification} in which we use different values for $q,\Sen$ (parameters used by the simulator) and $\hat{q},\hat{\Sen}$ (parameters used by the policies to produce groups and marginal decoders to interpret them). Obviously some policies are more sensitive to these gaps: For instance, Dorfman splitting does not consider sensitivities to form groups (but uses them to decode test results). Although one may have expected a substantial decrease in performance of our proposals G-AUCMAX and G-MIMAX, neither seems to materialize, as their performance stays clearly above that of all other baselines.
\item \textbf{$k=1$ and single test steps}: In \S\ref{subsec:1cycle} we report fine grained results for \textbf{G-MIMAX} and other baselines that show the evolution of the performance of this method in the most favorable setting on paper, that in which they are free to choose a new test based on the result of the previous test. While this setup would be unrealistic, because it would involve waiting hours required to output a single test result before choosing a new group, it showcases the speed at which our method reaches better performance than other baselines.
\item \textbf{More challenging setup: $n=96$ and varying $\Seg$}: we end this section with a last setup that is more challenging ($n=96$ is larger, and so one may worry about the ability of our posteriors to cover a space of size $2^96$) and which also factors a degraded sensitivity as a function of group size. In that setting, we still observe a substantial gap between G-MIMAX and the other techniques. Additionally, we show that $N$ (the number of particles) and F/B (number of forward/backward iterations) also seem to have a small impact on the performance of G-MIMAX.
\end{itemize}
\subsection{$q=10\%$ infection rates}\label{subsec:IR10}
\begin{figure}
    \includegraphics[width=.57\textwidth]{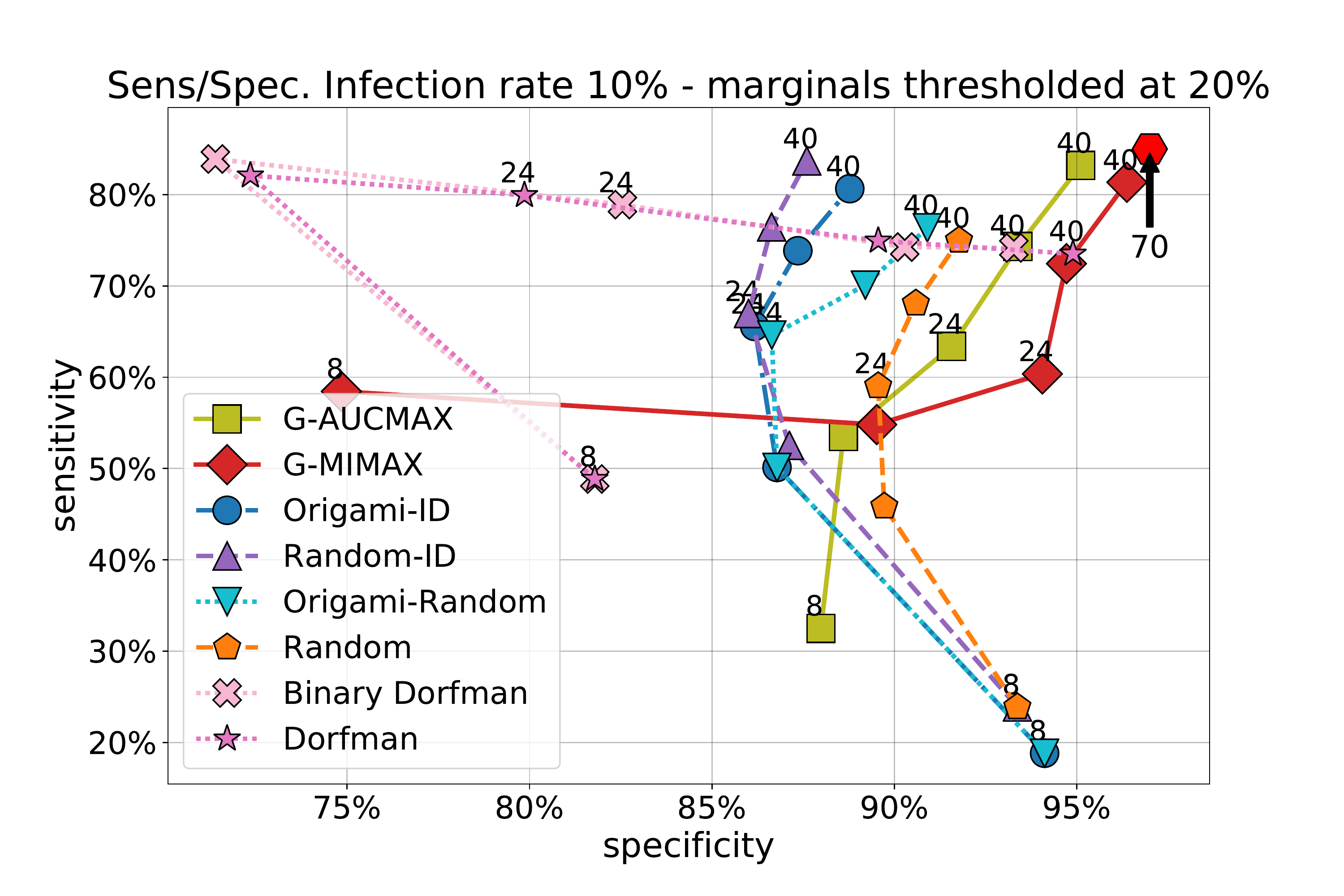}\hskip-.2cm
    \includegraphics[width=.57\textwidth]{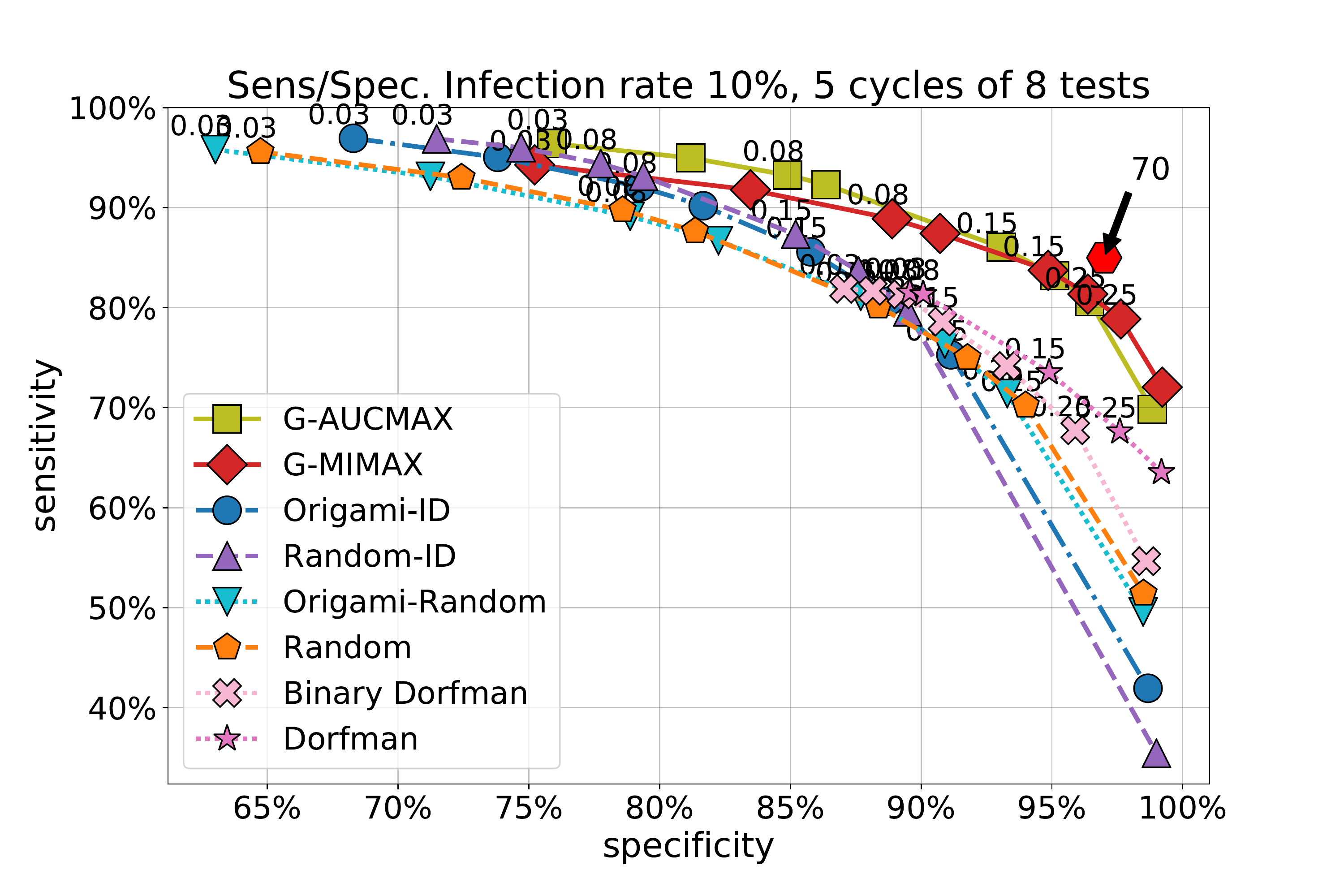}\\
\caption{Using the same setup as in Fig.~\ref{fig:perf_curve} and \ref{fig:perf_prog}, we report results for an infection base rate of $q=10\%$. The red hexagon depicts the sensitivity/specificity of 70 individual tests.}
    \label{fig:10percent}
\end{figure}

We provide in Fig.~\ref{fig:10percent} additional results for a base infection rate of 10\%.

\subsection{Specificity / Sensitivity curves as function of cycles}\label{subsec:earliercycles}
\begin{figure}
    \includegraphics[width=.57\textwidth]{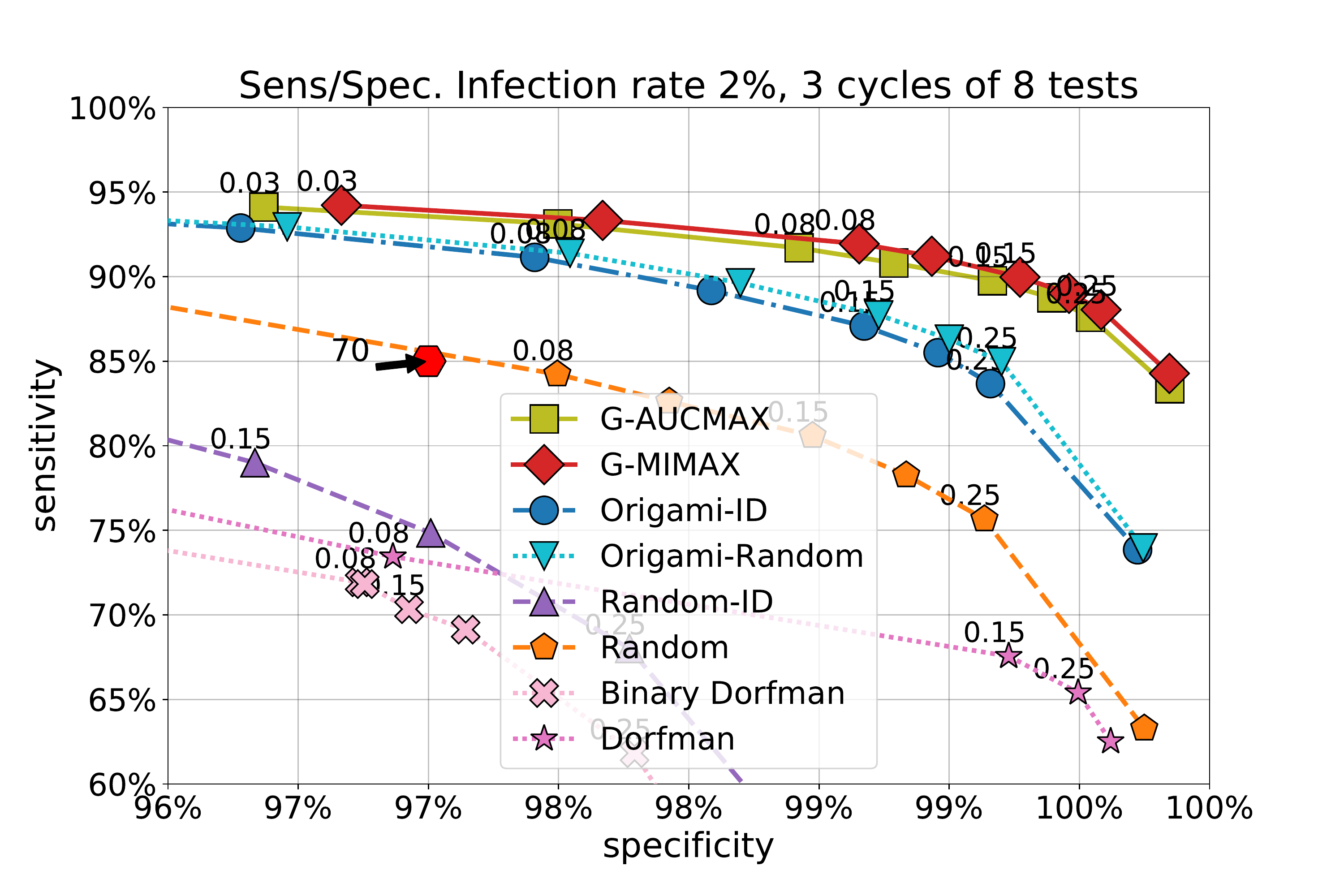}\hskip-.2cm
    \includegraphics[width=.57\textwidth]{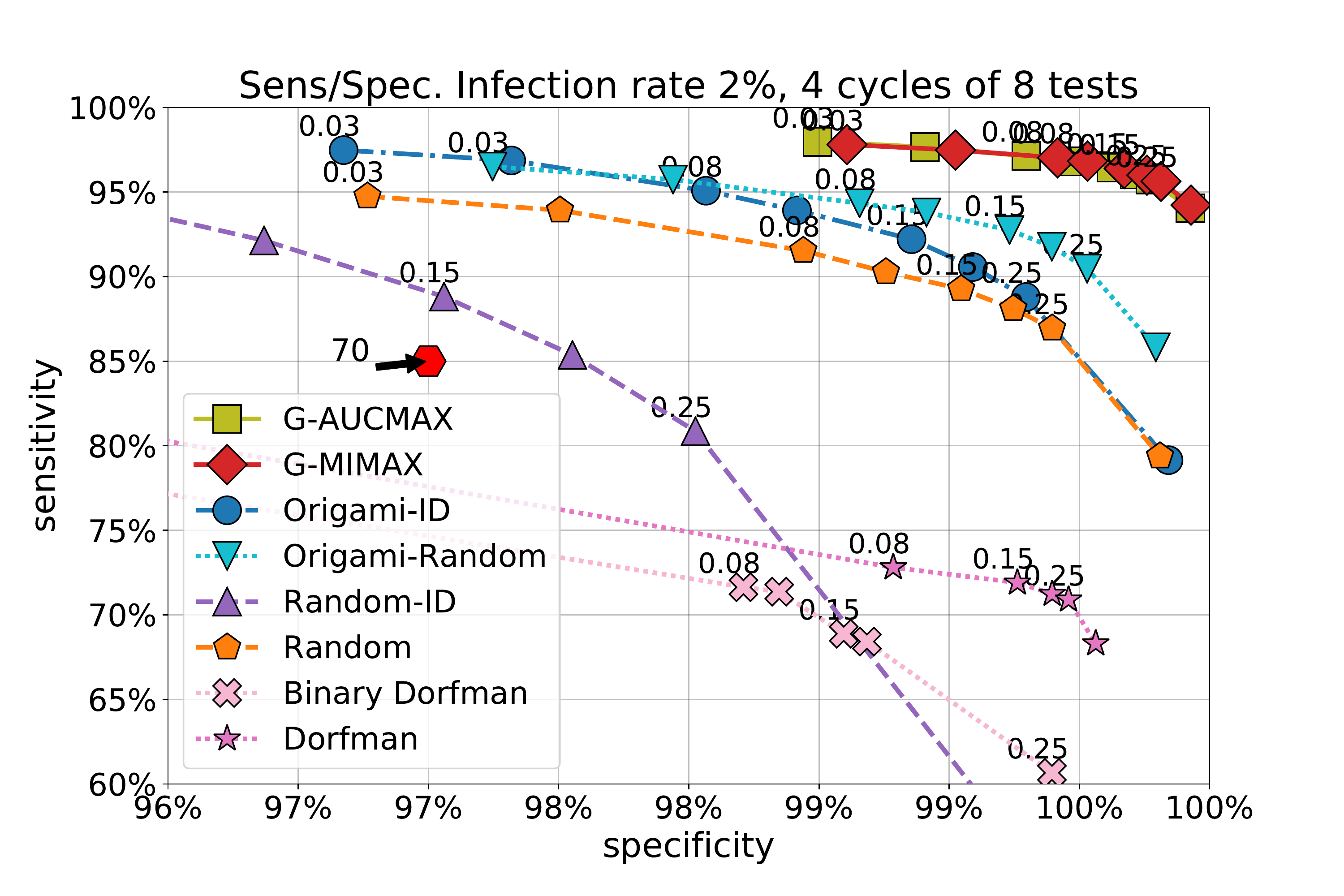}\\
    \includegraphics[width=.57\textwidth]{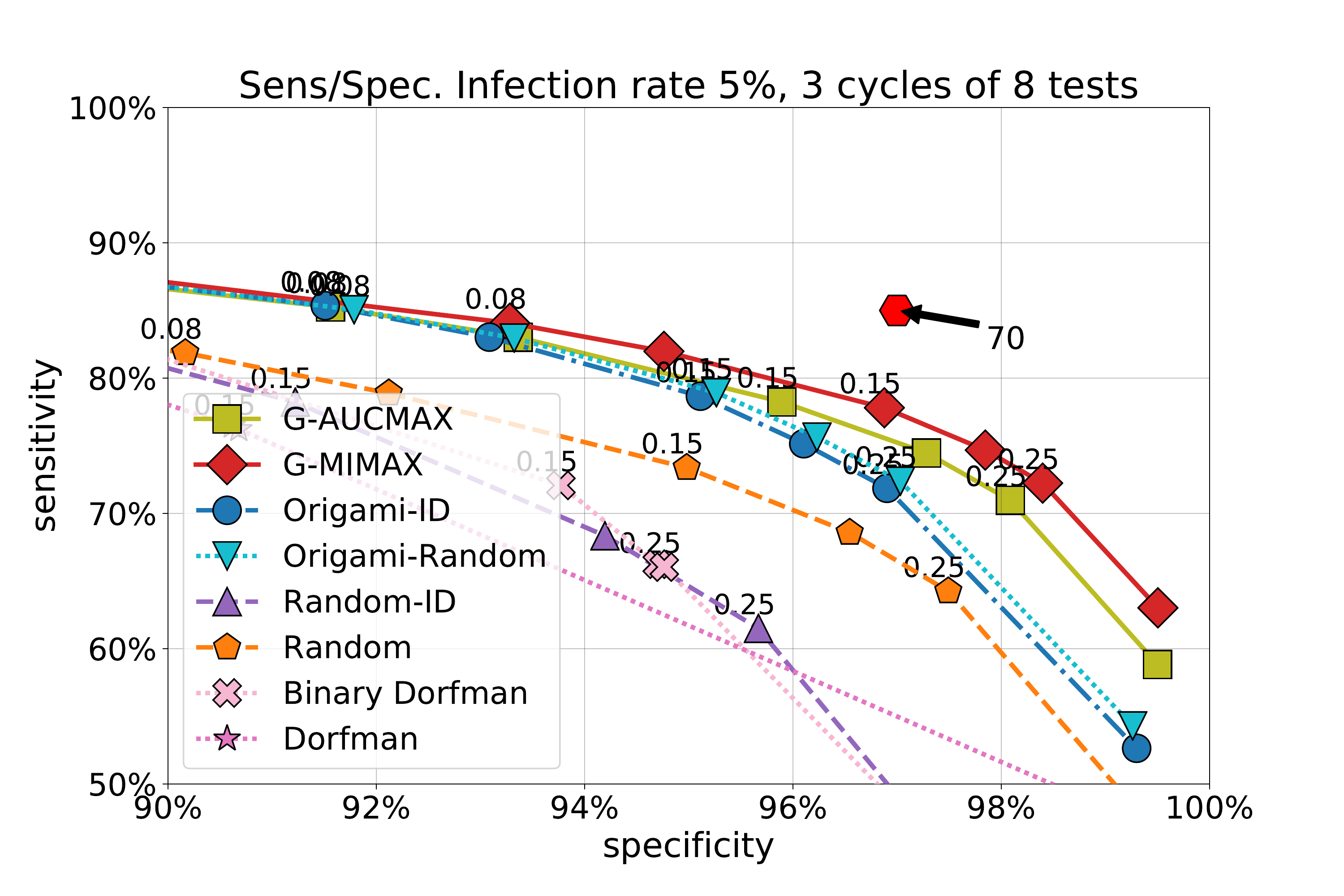}\hskip-.2cm
    \includegraphics[width=.57\textwidth]{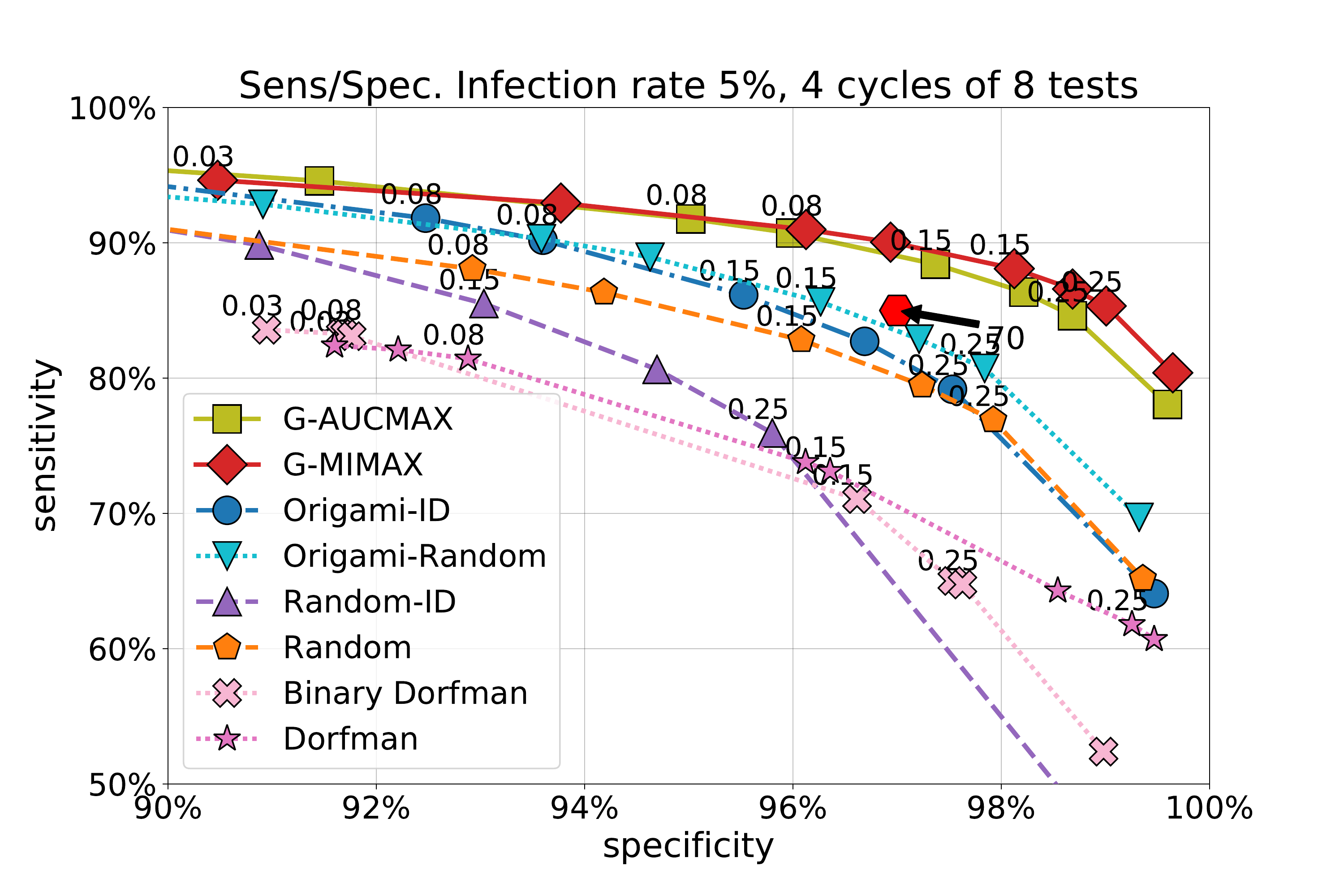}\\
    \includegraphics[width=.57\textwidth]{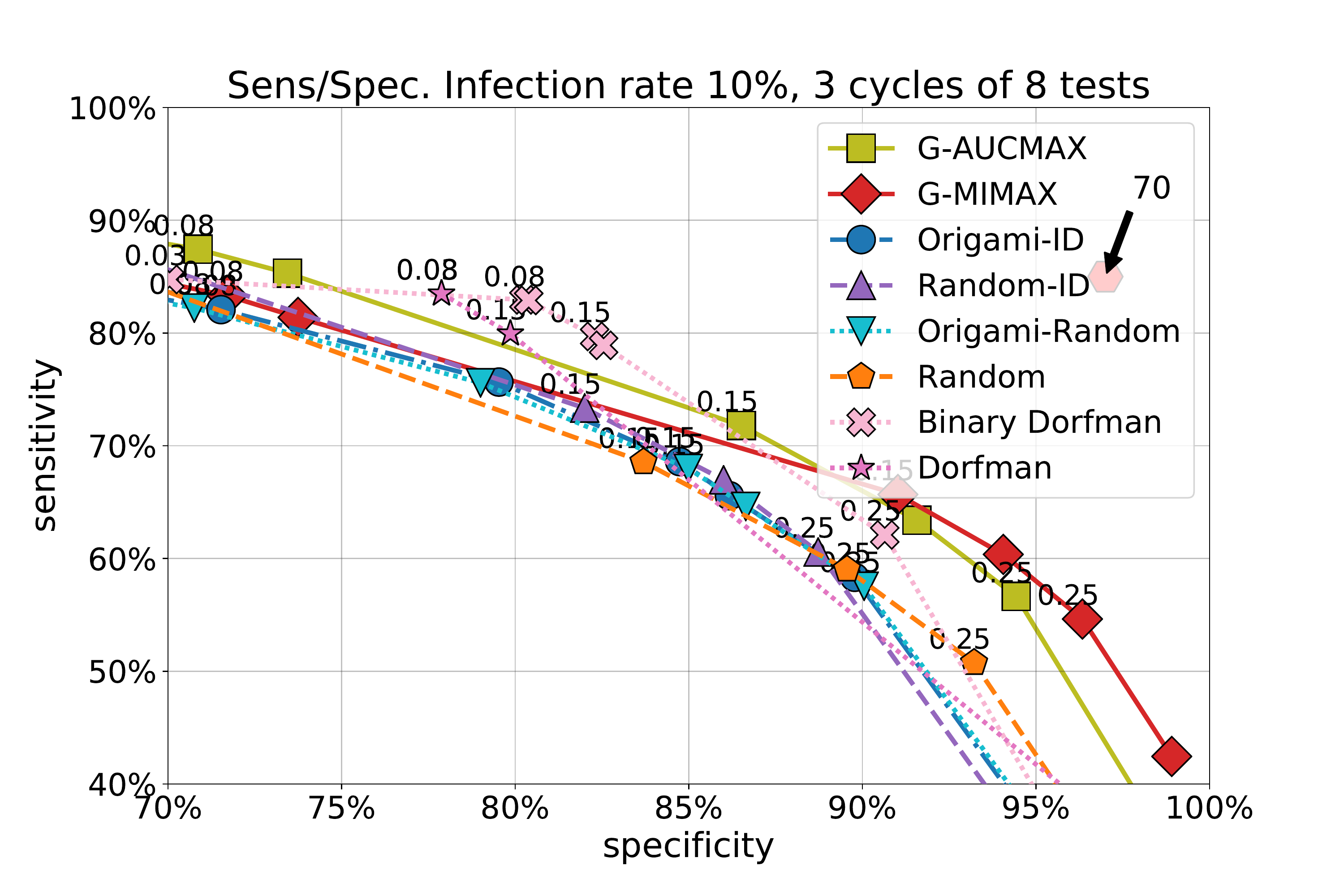}\hskip-.2cm
    \includegraphics[width=.57\textwidth]{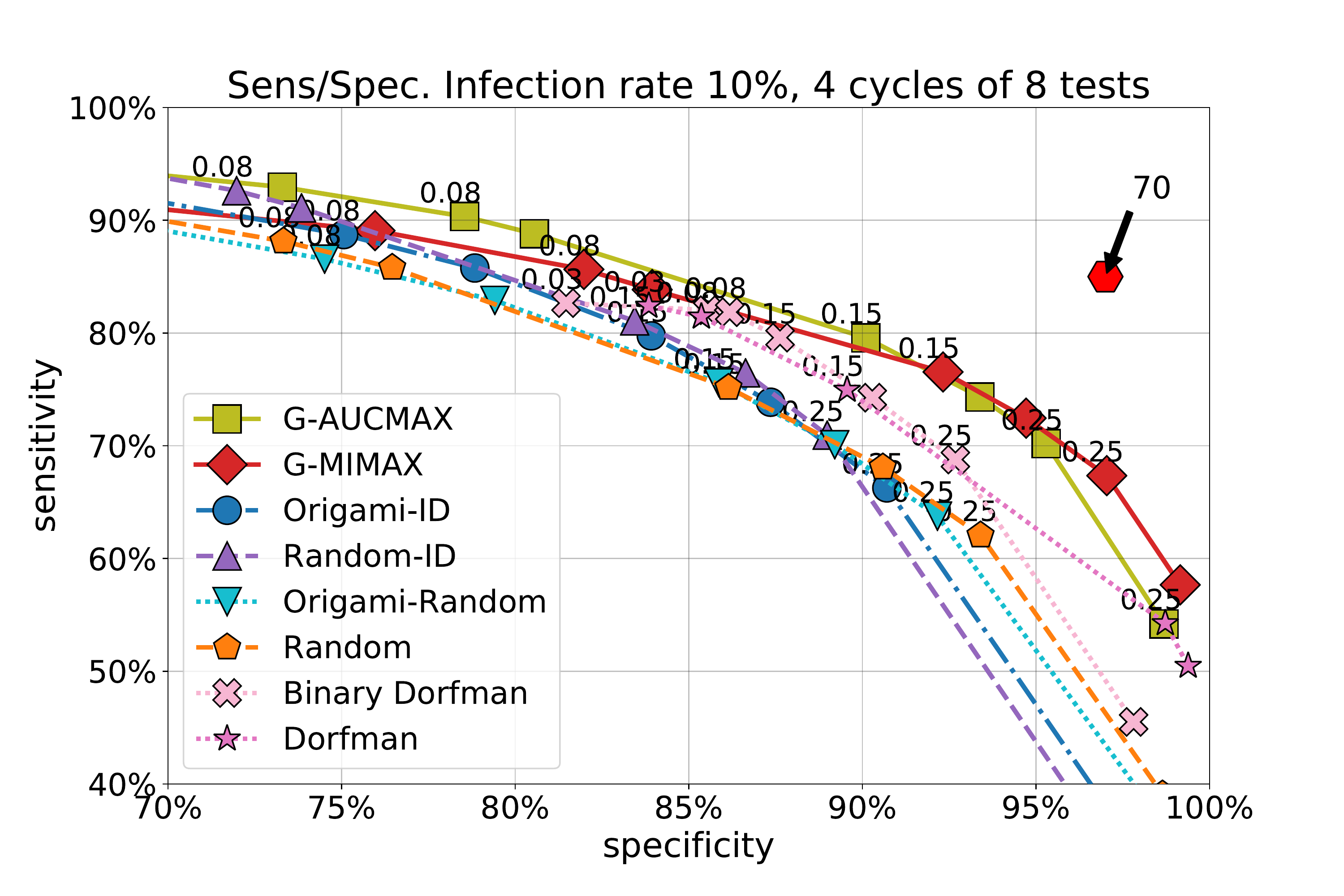}\\
\caption{Using the same setup as in Fig.~\ref{fig:perf_curve}, we report results for $3$ and $4$ testing cycles (corresponding therefore to 24 and 32 tests carried out in total). Note that, since the Origami assay only considers 22 tests,the slight difference in performance between Origami-ID and Origami-Random that arises on the left plots is only due to 2 tests, carried out using ID or randomly.}
    \label{fig:34cycles}
\end{figure}

\begin{figure}
    \includegraphics[width=.57\textwidth]{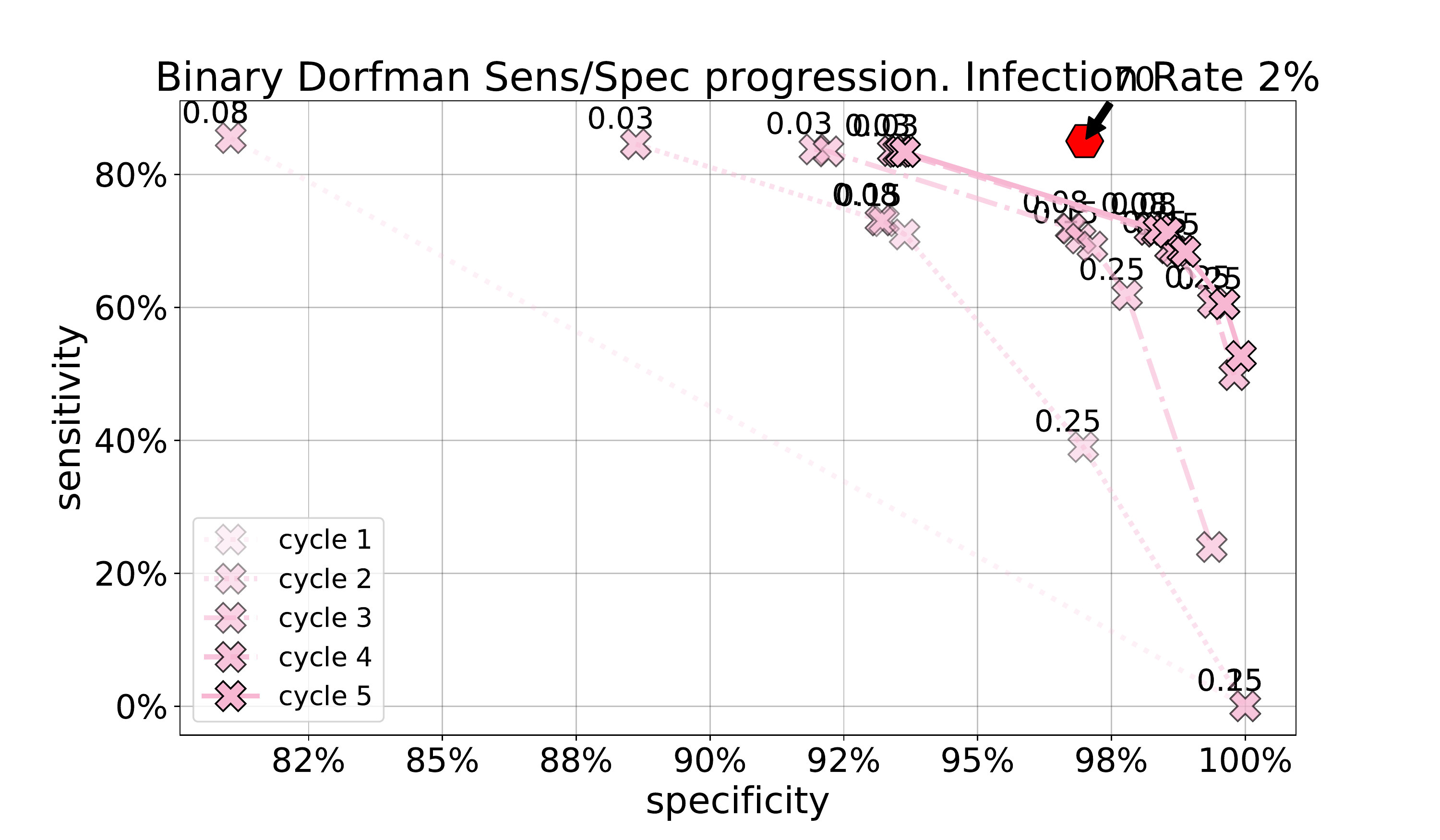}\hskip-.2cm
    \includegraphics[width=.57\textwidth]{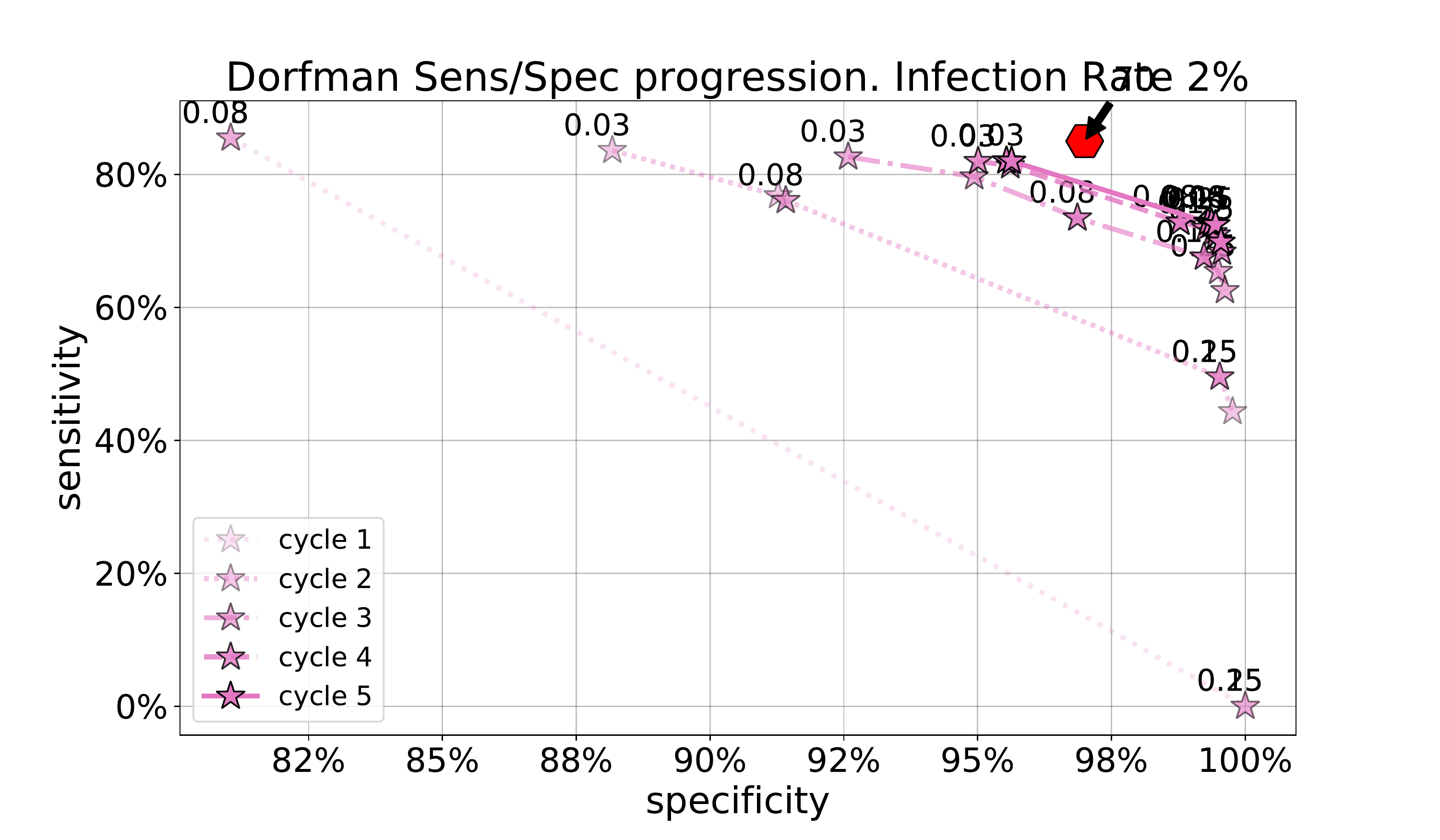}\\
    \hskip-.8cm
    \includegraphics[width=.57\textwidth]{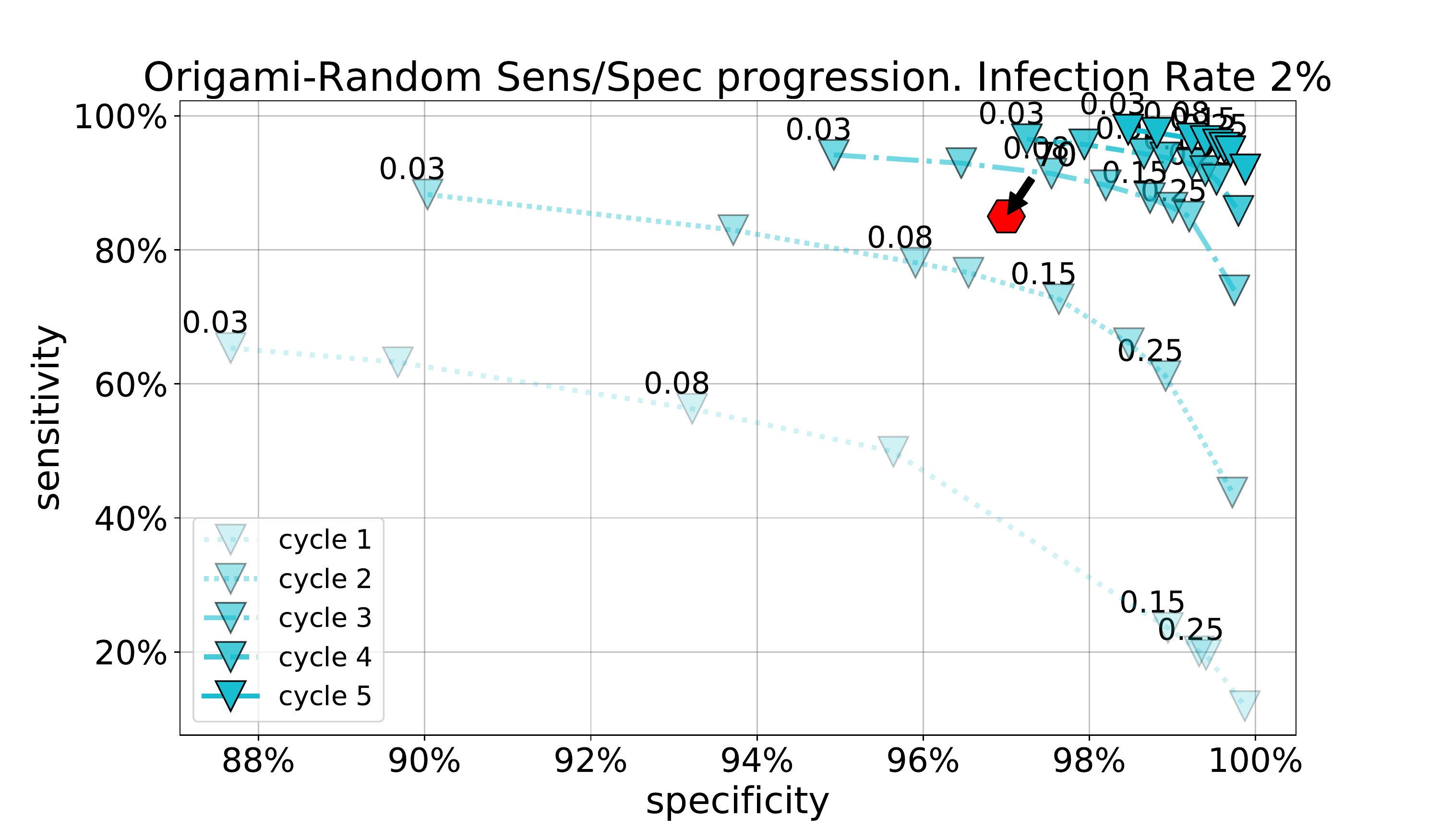}\hskip-.2cm
    \includegraphics[width=.57\textwidth]{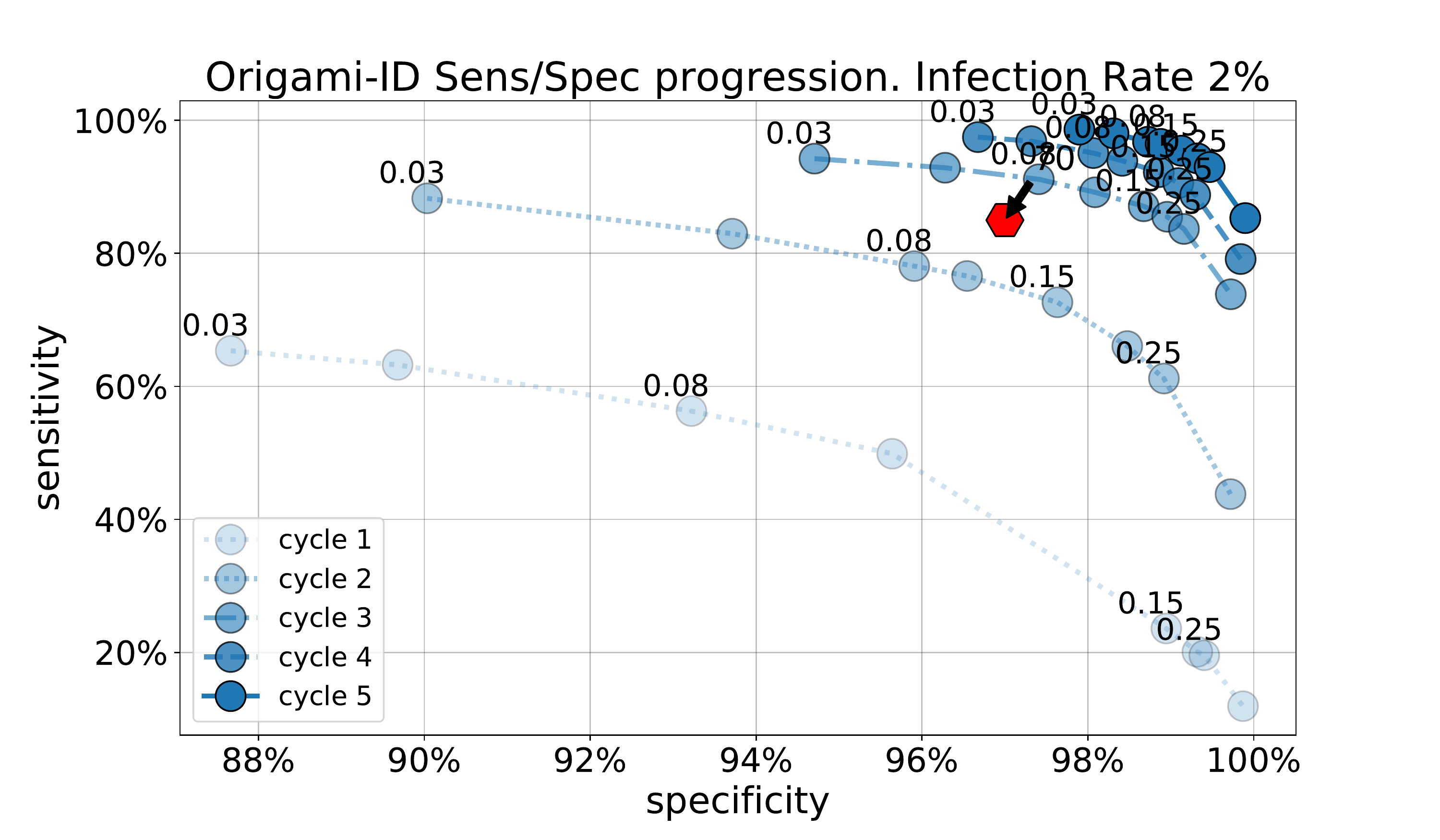}\\
    \hskip-.8cm
    \includegraphics[width=.57\textwidth]{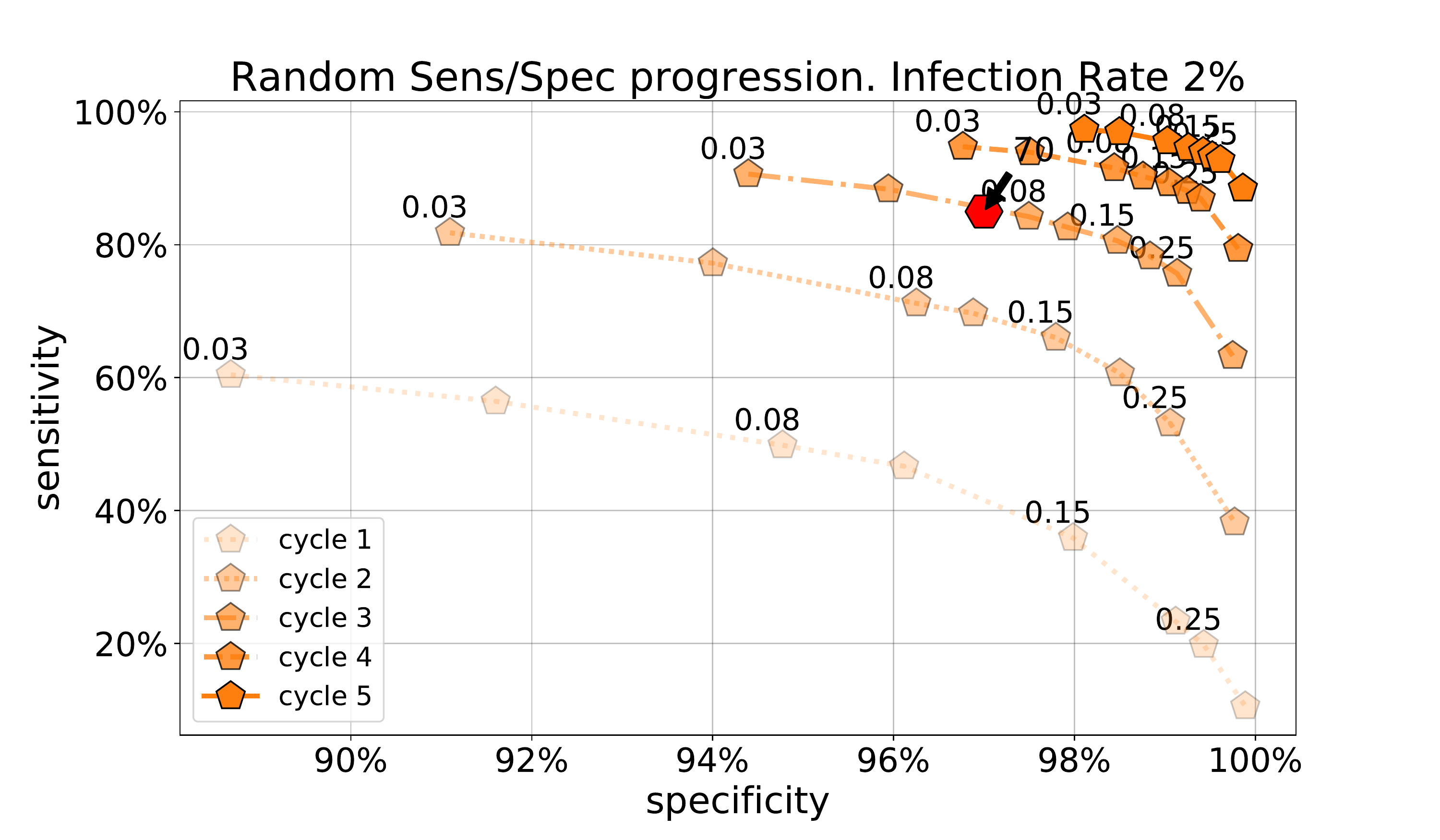}\hskip-.2cm
    \includegraphics[width=.57\textwidth]{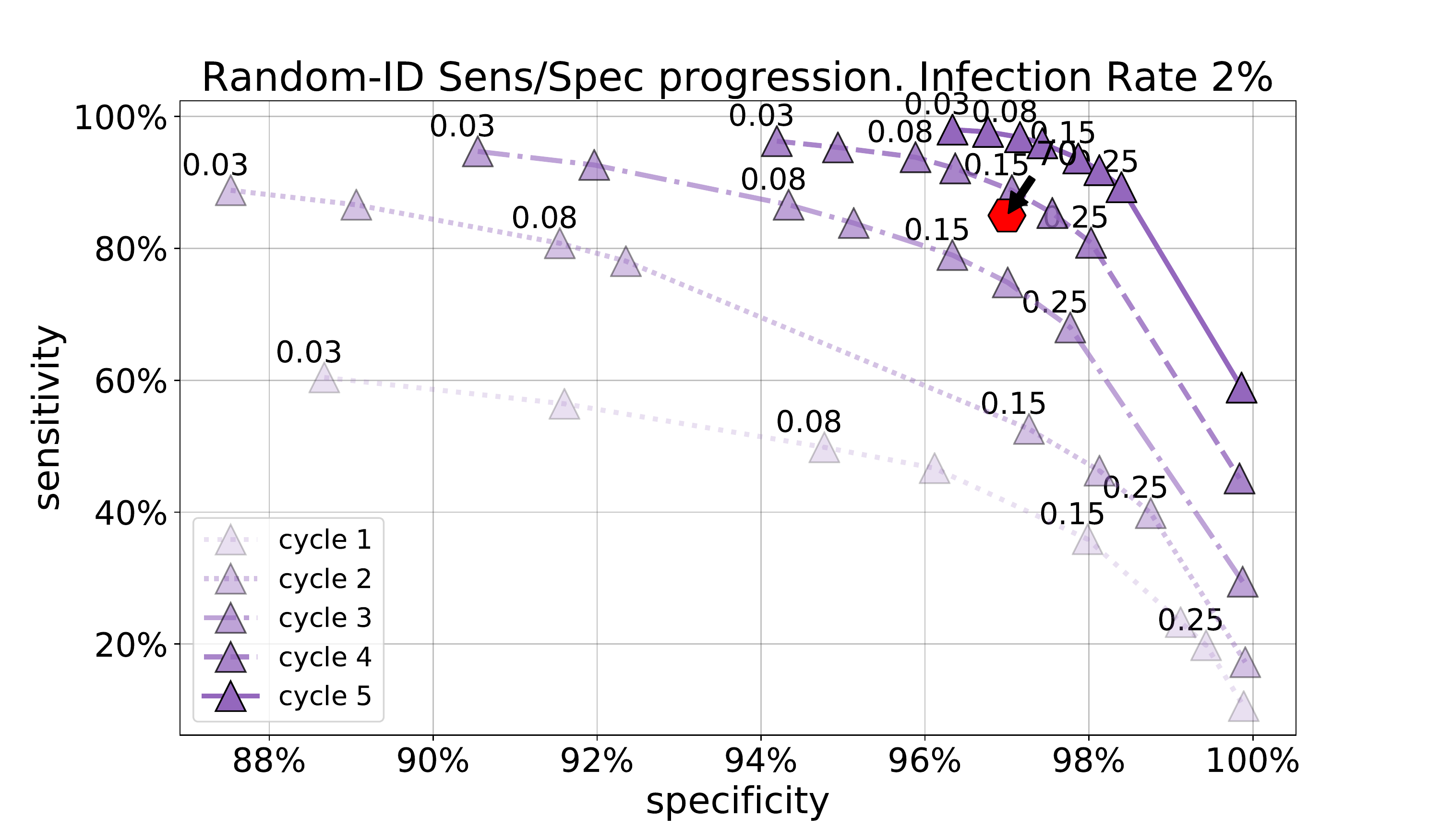}\\
    \hskip-.8cm
    \includegraphics[width=.57\textwidth]{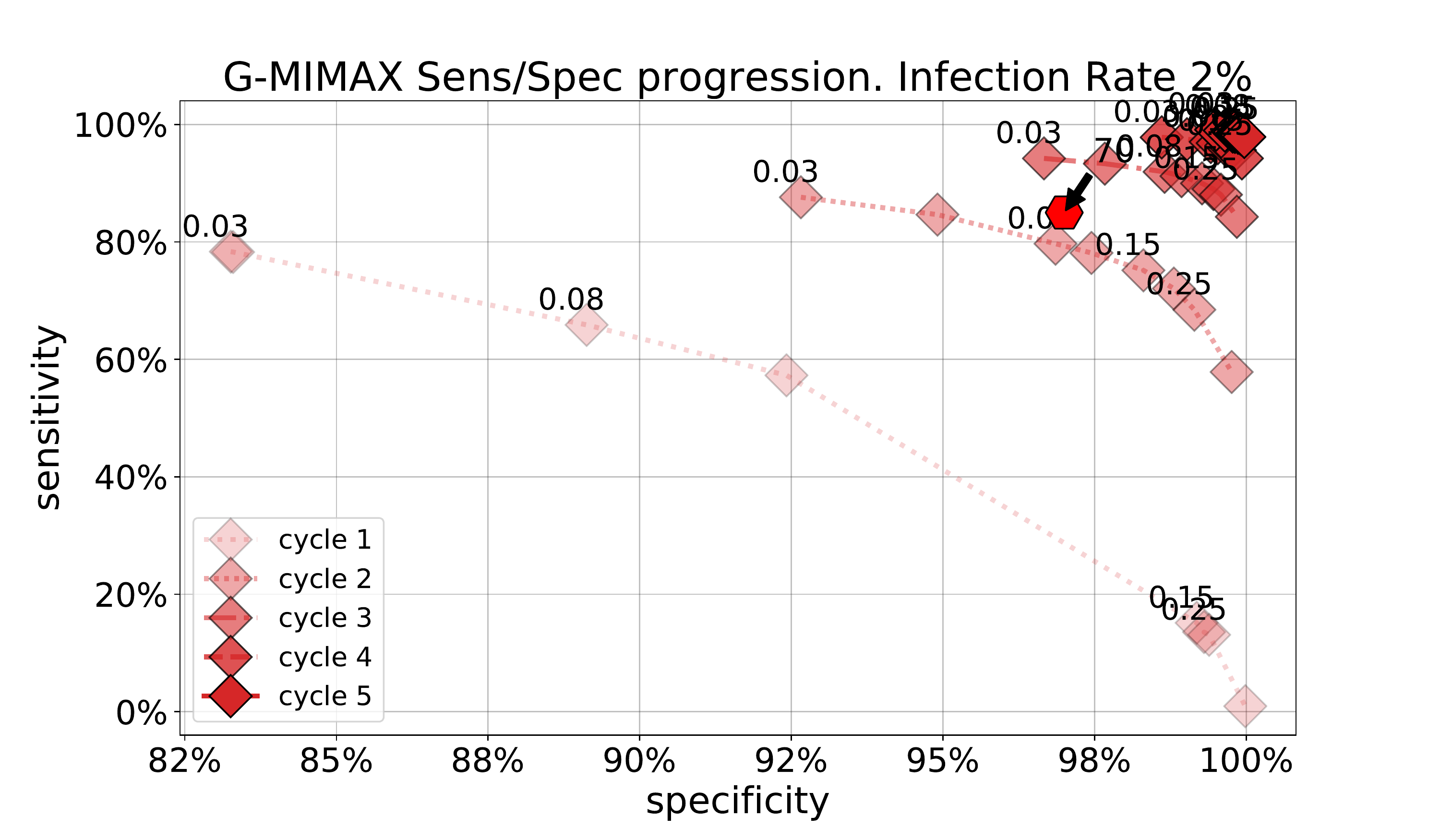}\hskip-.2cm
    \includegraphics[width=.57\textwidth]{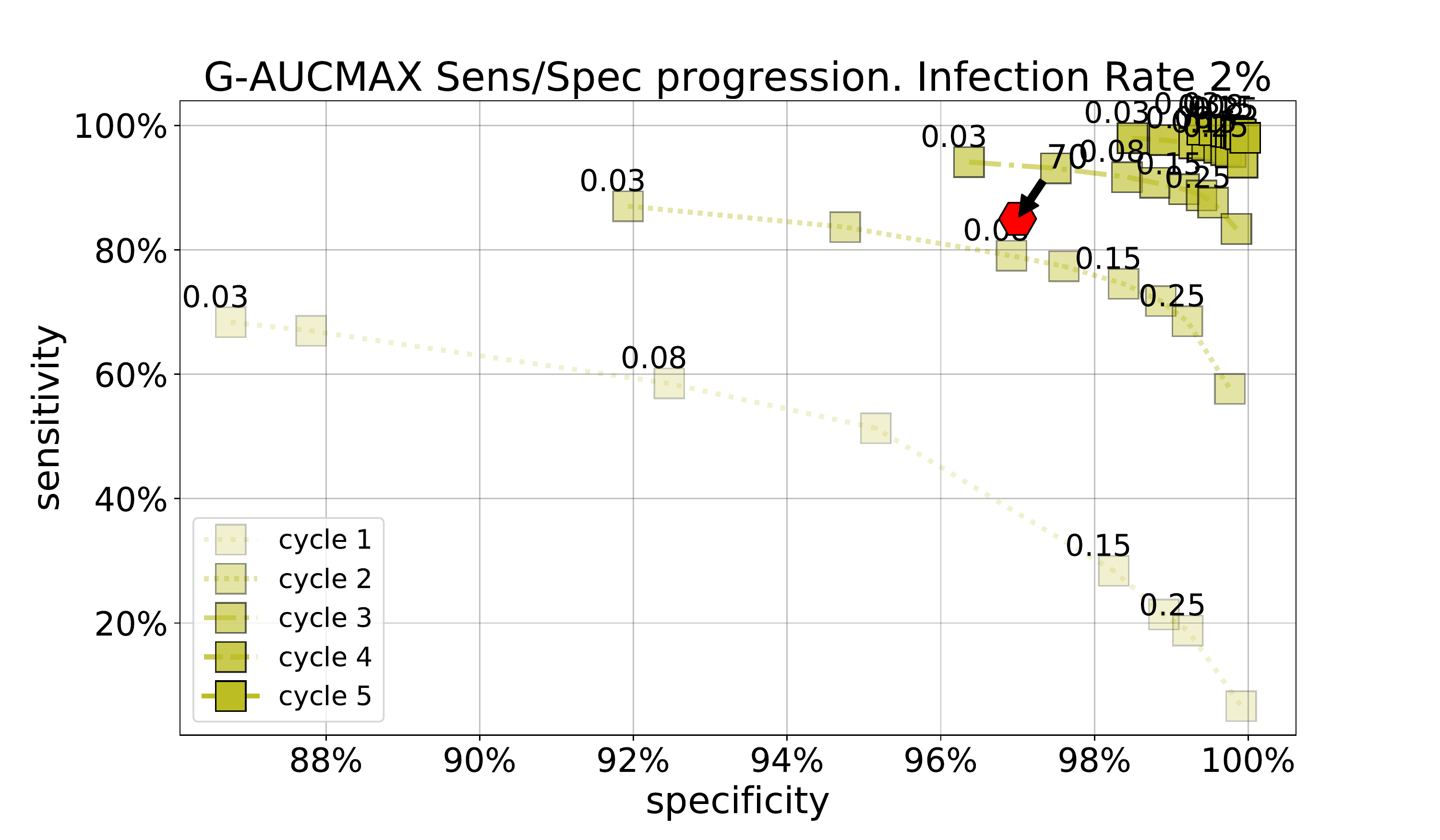}\\
\caption{Using the same setup as in Fig.~\ref{fig:perf_curve}, we report dynamic results for each policy, as the number of tests increases, for $q=2\%$. The number of tests is equal to $k$ (here 8) times the cycle number.}
    \label{fig:evolutionforall2}
\end{figure}

\begin{figure}
    \includegraphics[width=.57\textwidth]{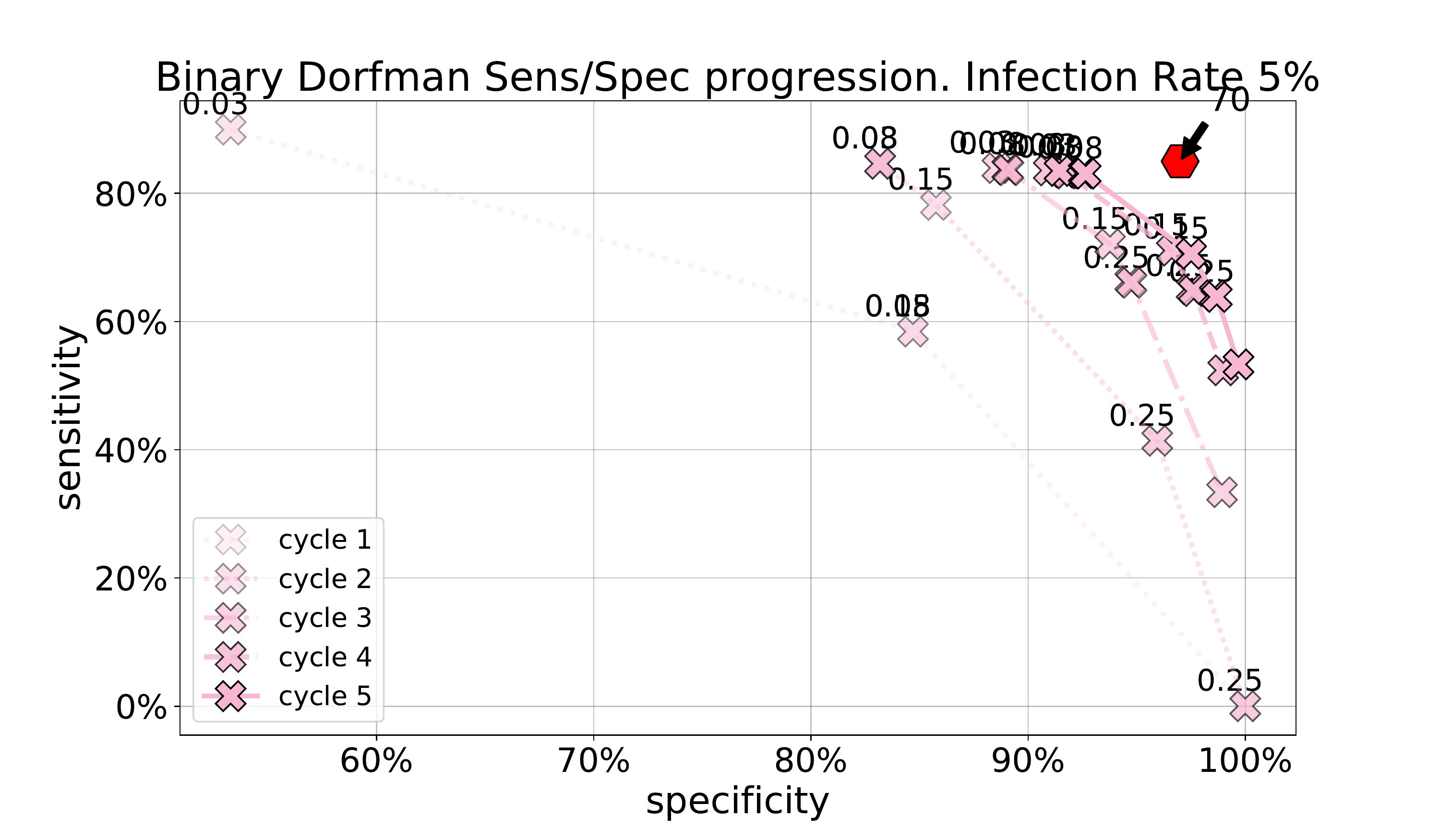}\hskip-.2cm
    \includegraphics[width=.57\textwidth]{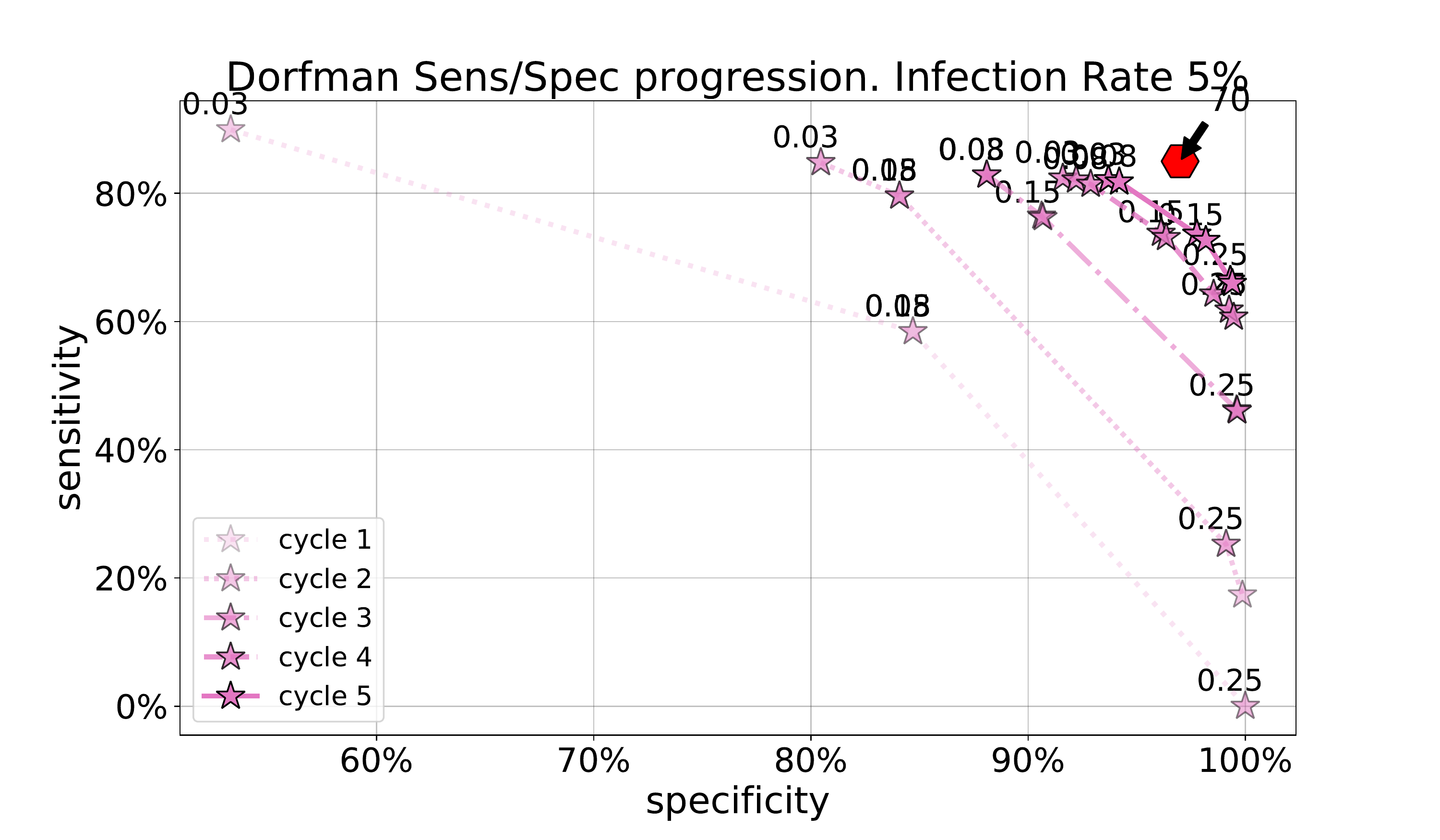}\\
    \hskip-.8cm
    \includegraphics[width=.57\textwidth]{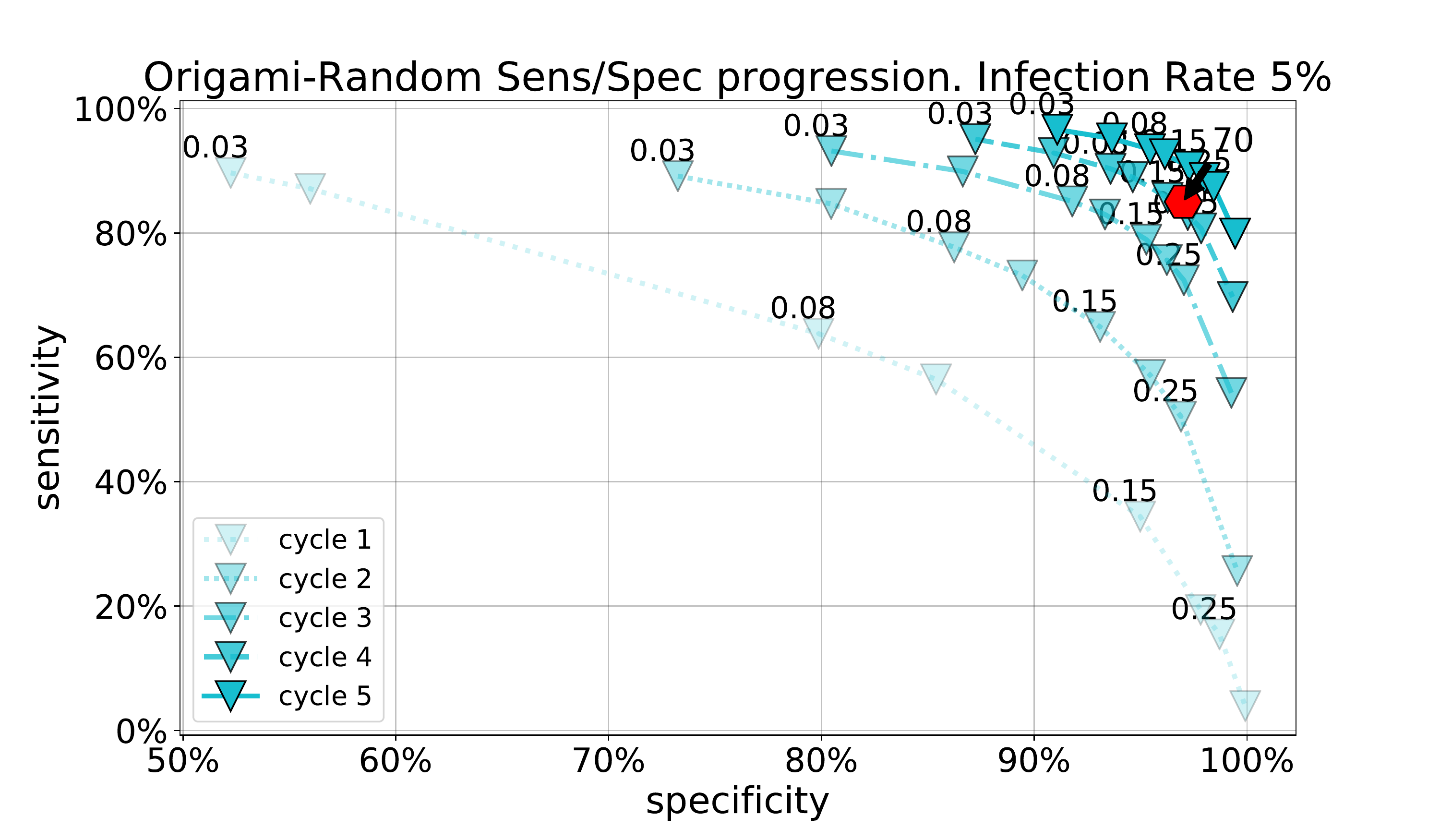}\hskip-.2cm
    \includegraphics[width=.57\textwidth]{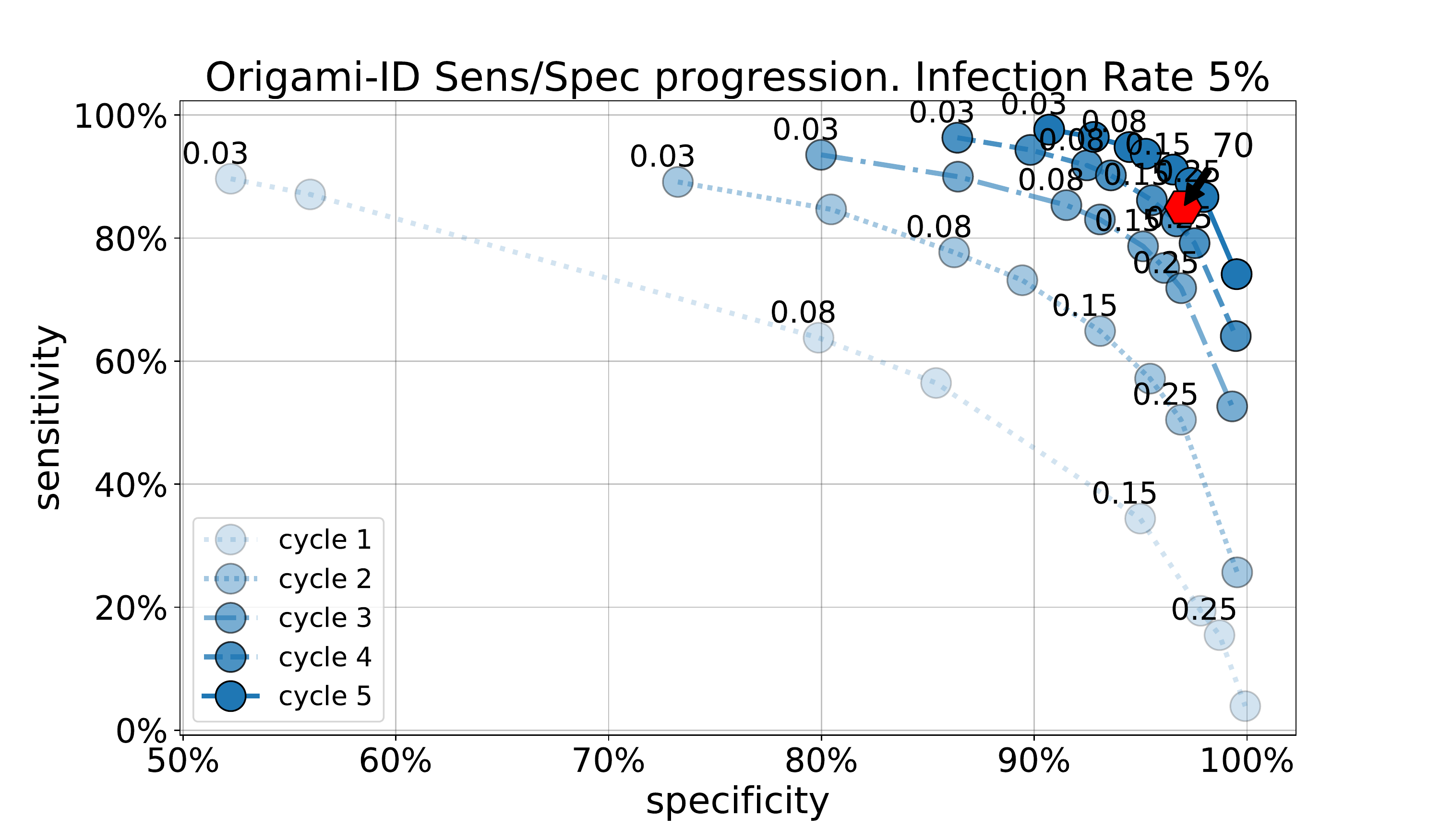}\\
    \hskip-.8cm
    \includegraphics[width=.57\textwidth]{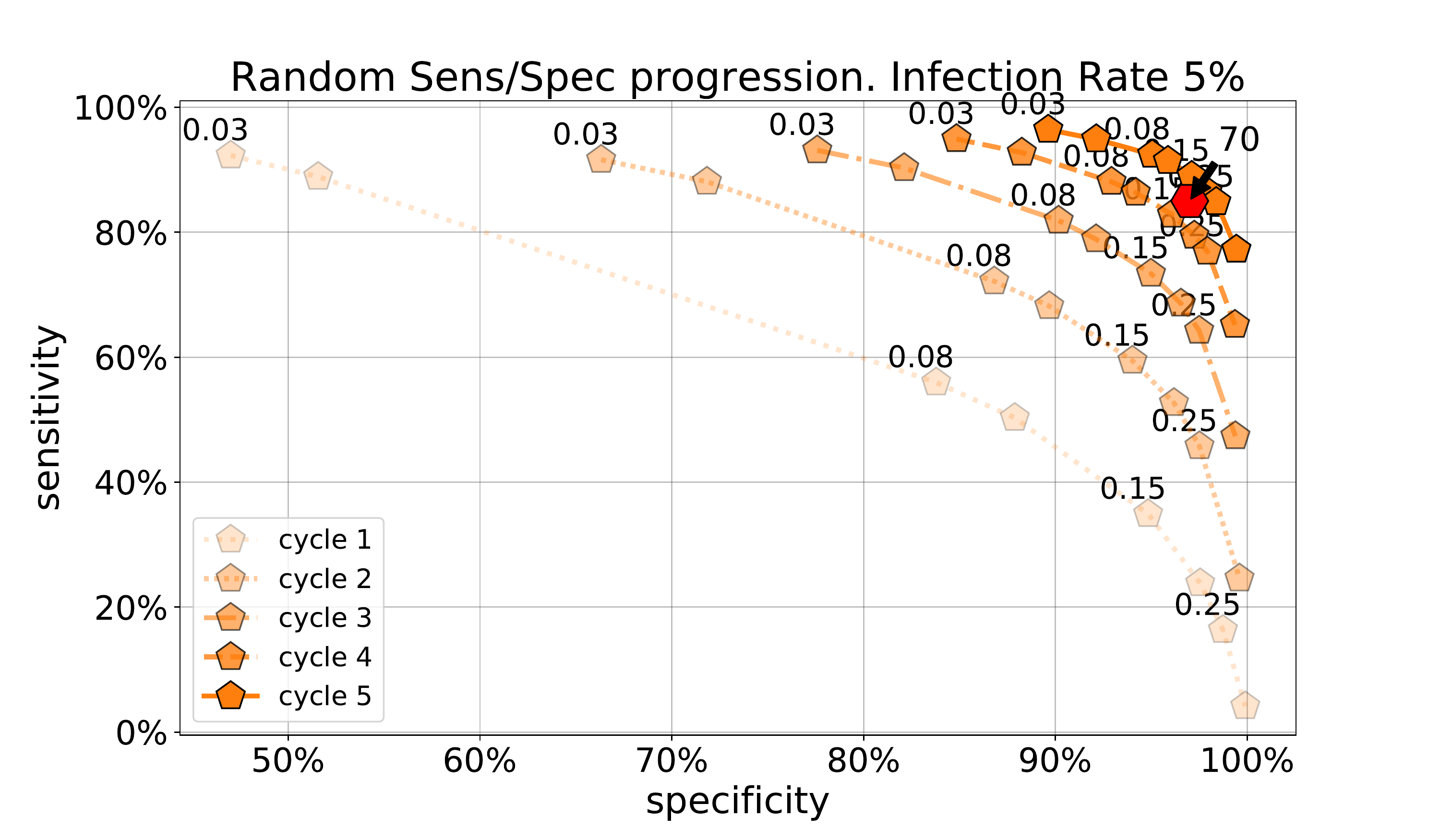}\hskip-.2cm
    \includegraphics[width=.57\textwidth]{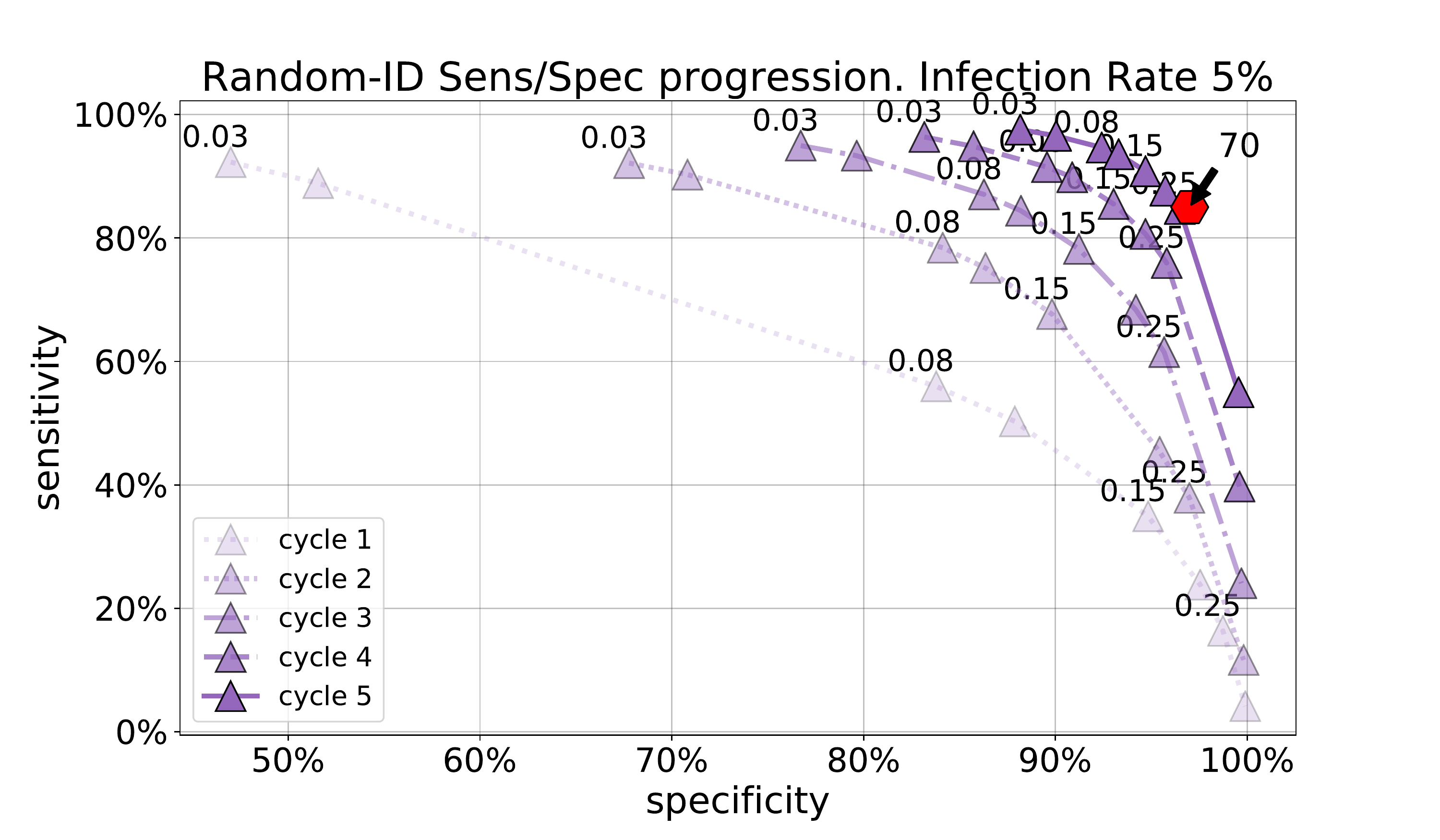}\\
    \hskip-.8cm
    \includegraphics[width=.57\textwidth]{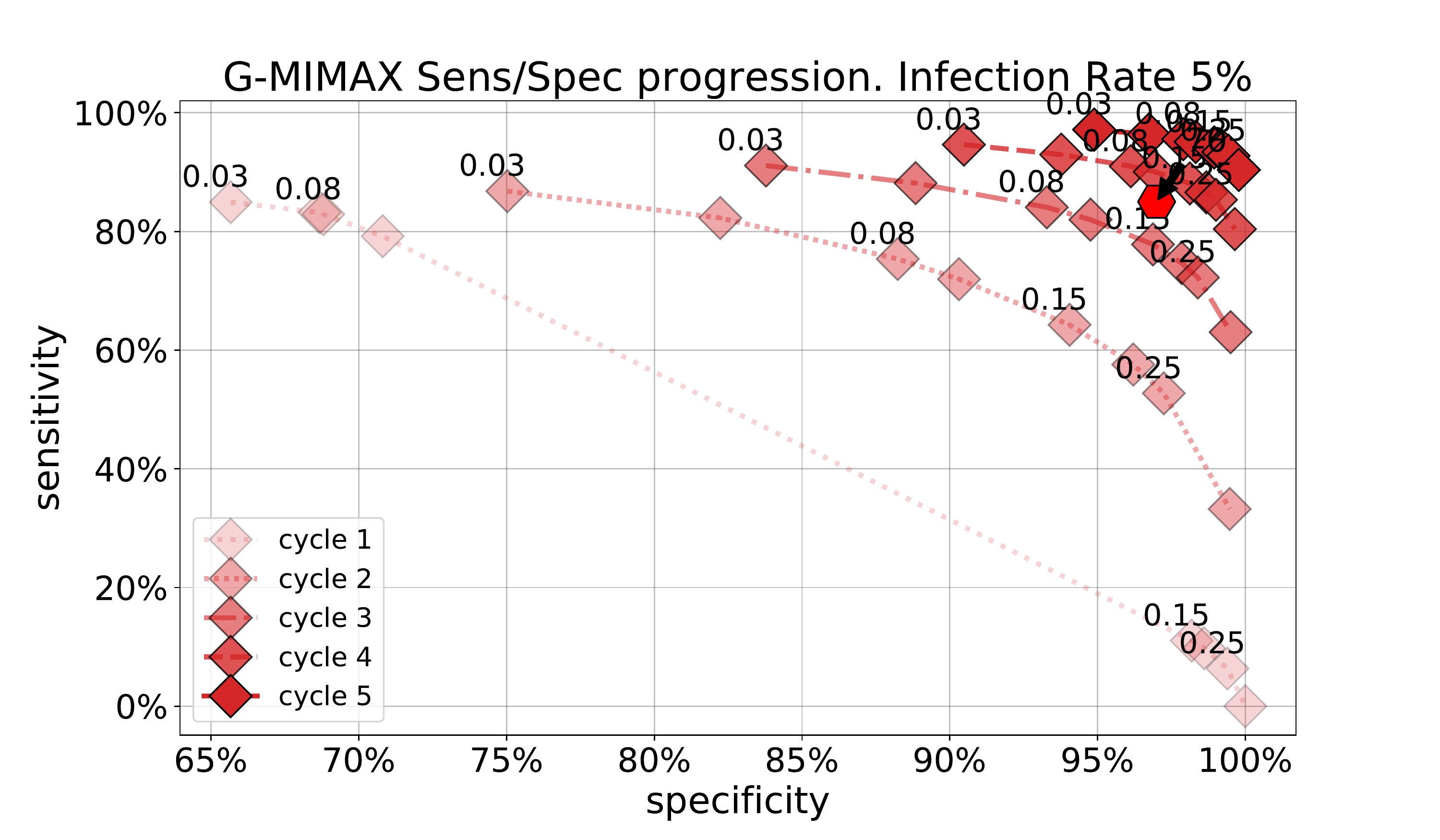}\hskip-.2cm
    \includegraphics[width=.57\textwidth]{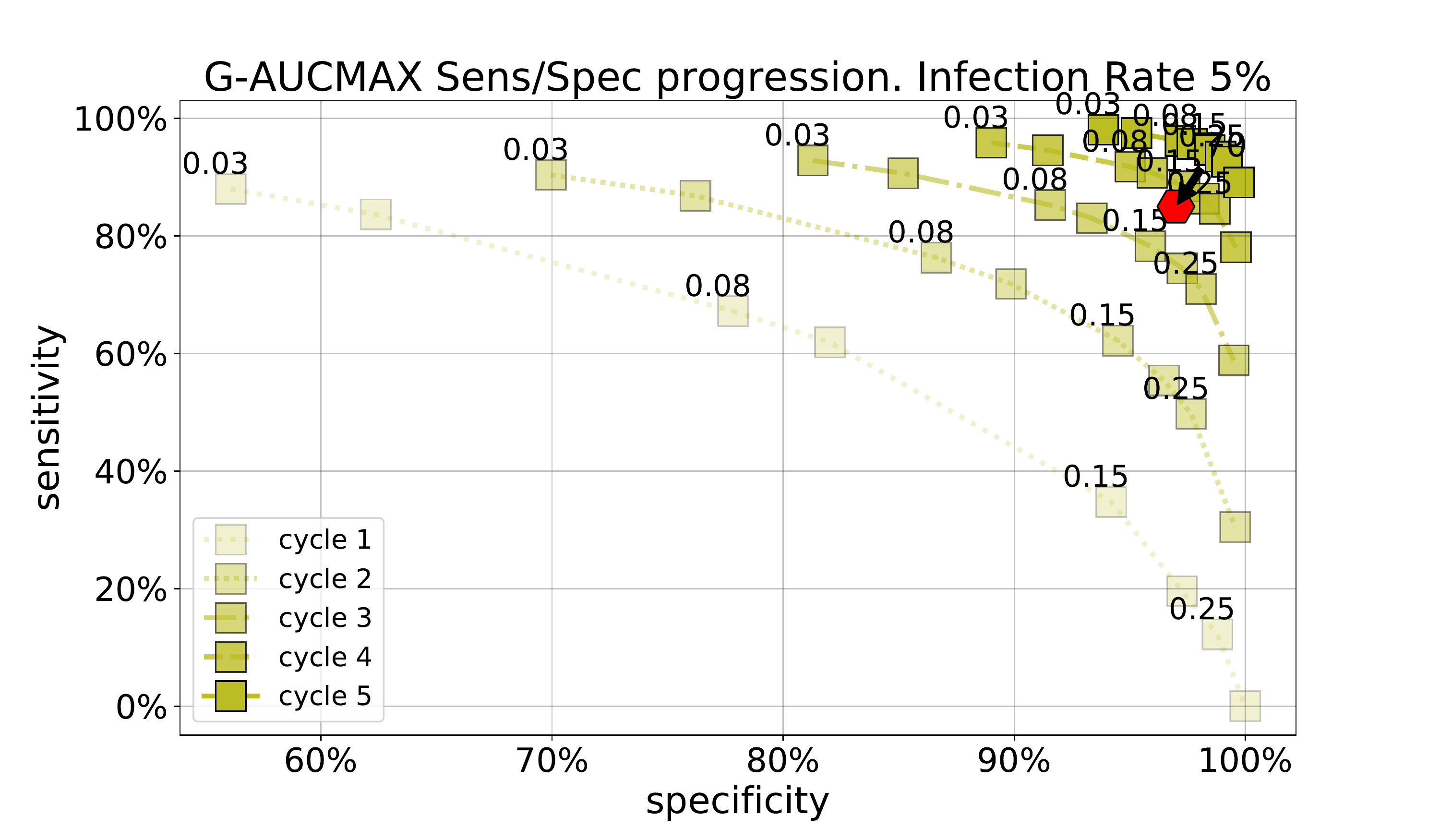}\\
\caption{Using the same setup as in Fig.~\ref{fig:perf_curve}, we report dynamic results for each policy, as the number of tests increases, for $q=5\%$. The number of tests is equal to $k$ (here 8) times the cycle number.}
    \label{fig:evolutionforall5}
\end{figure}

\begin{figure}
    \includegraphics[width=.57\textwidth]{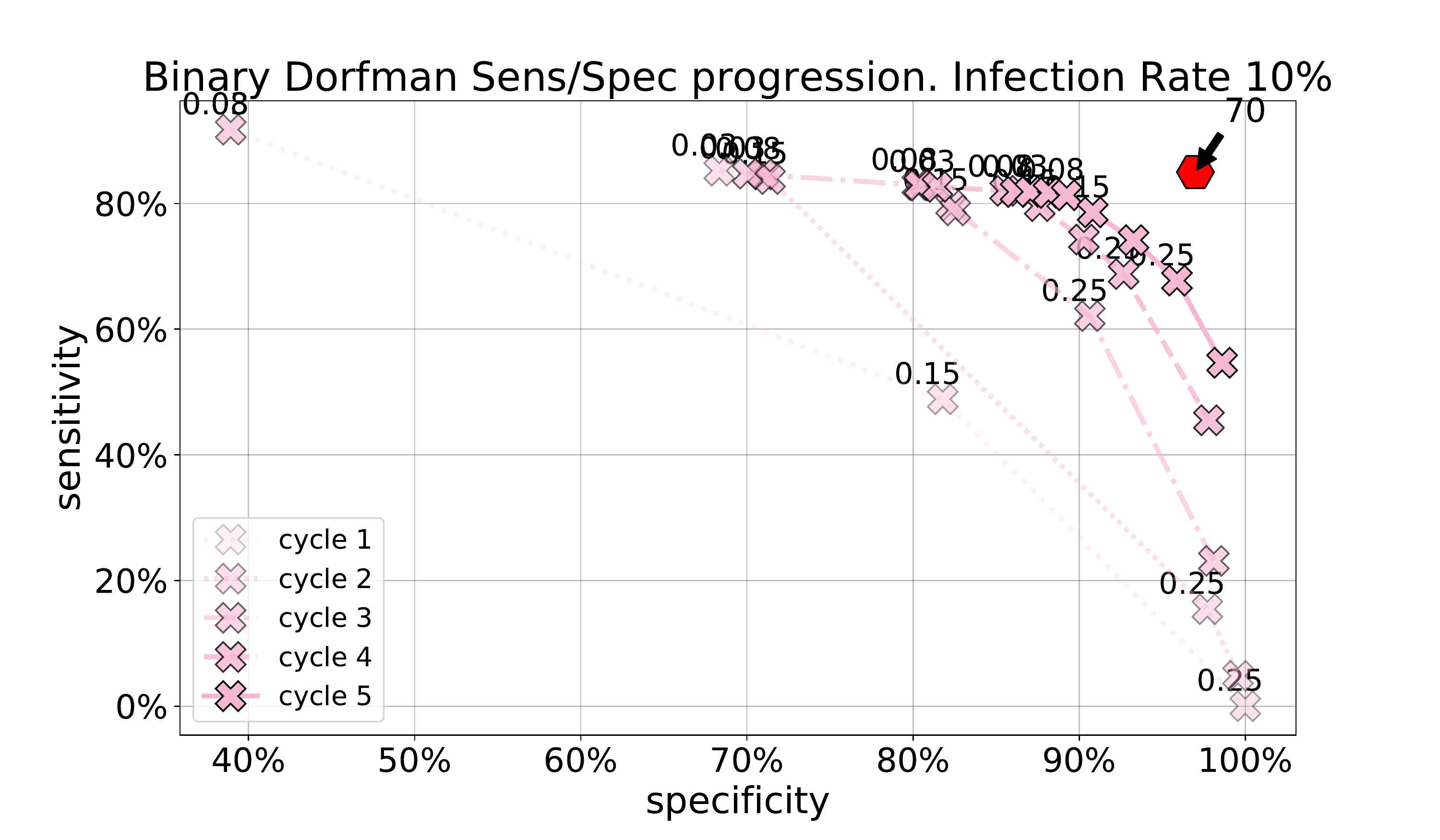}\hskip-.2cm
    \includegraphics[width=.57\textwidth]{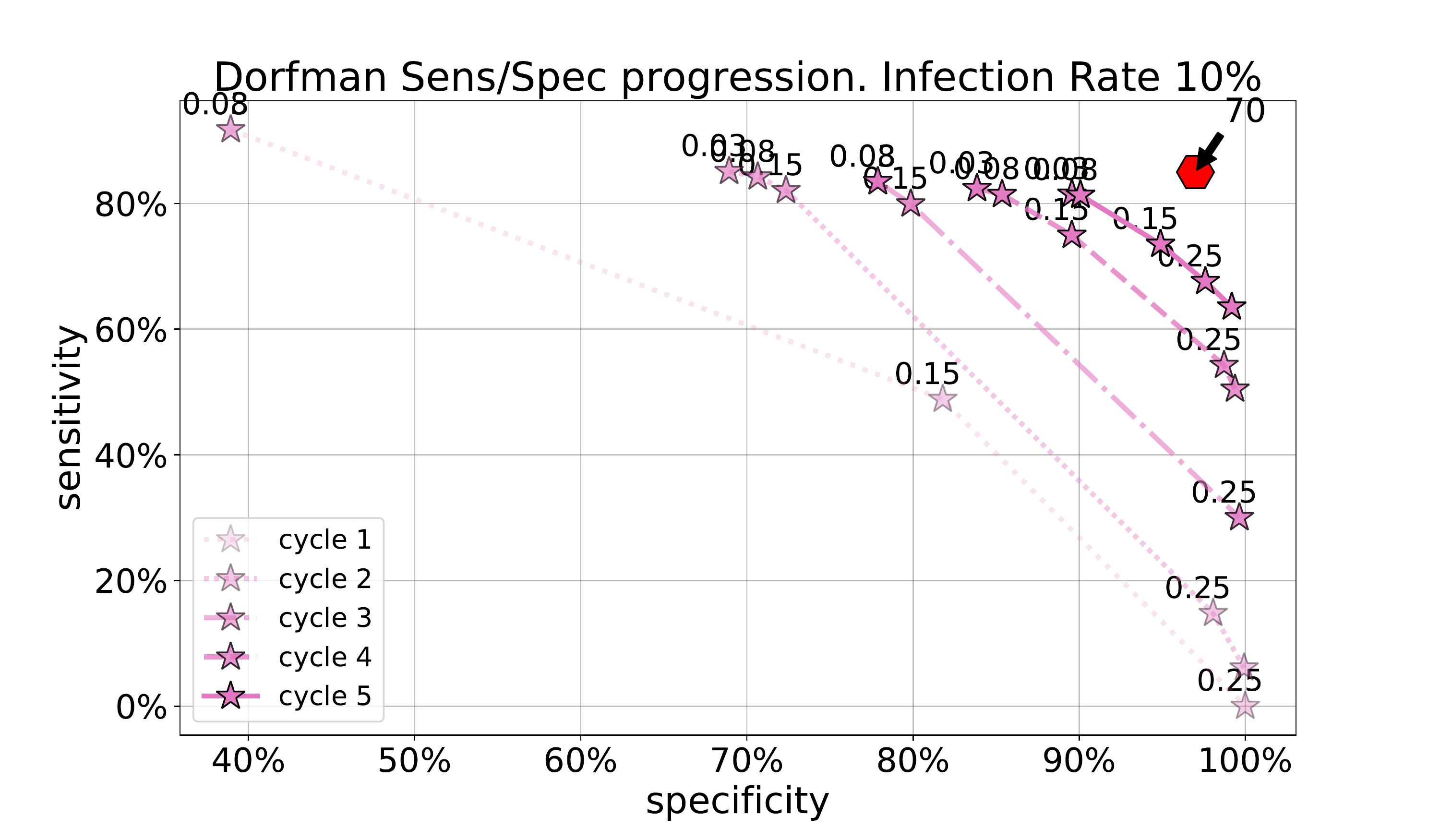}\\
    \hskip-.8cm
    \includegraphics[width=.57\textwidth]{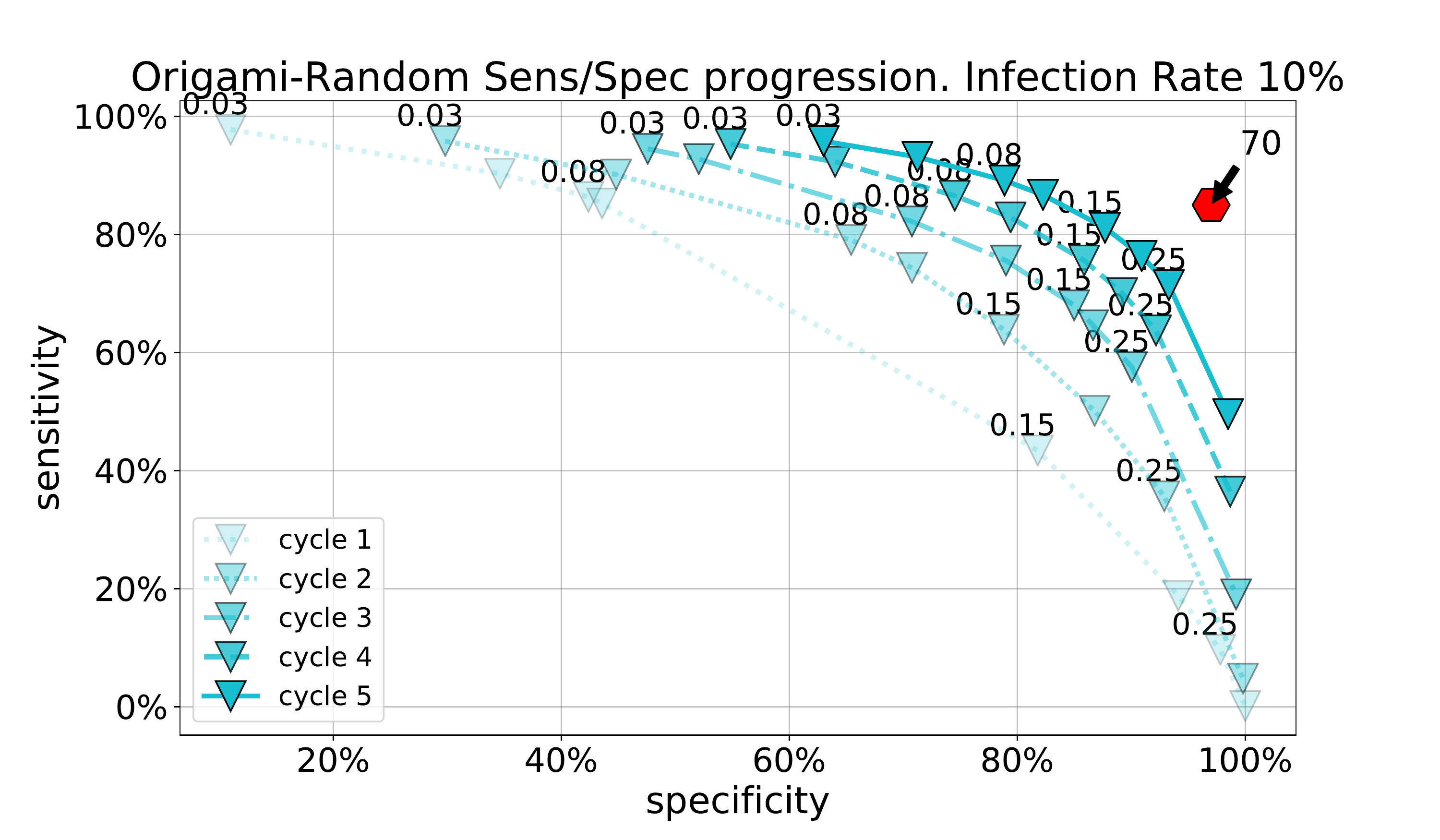}\hskip-.2cm
    \includegraphics[width=.57\textwidth]{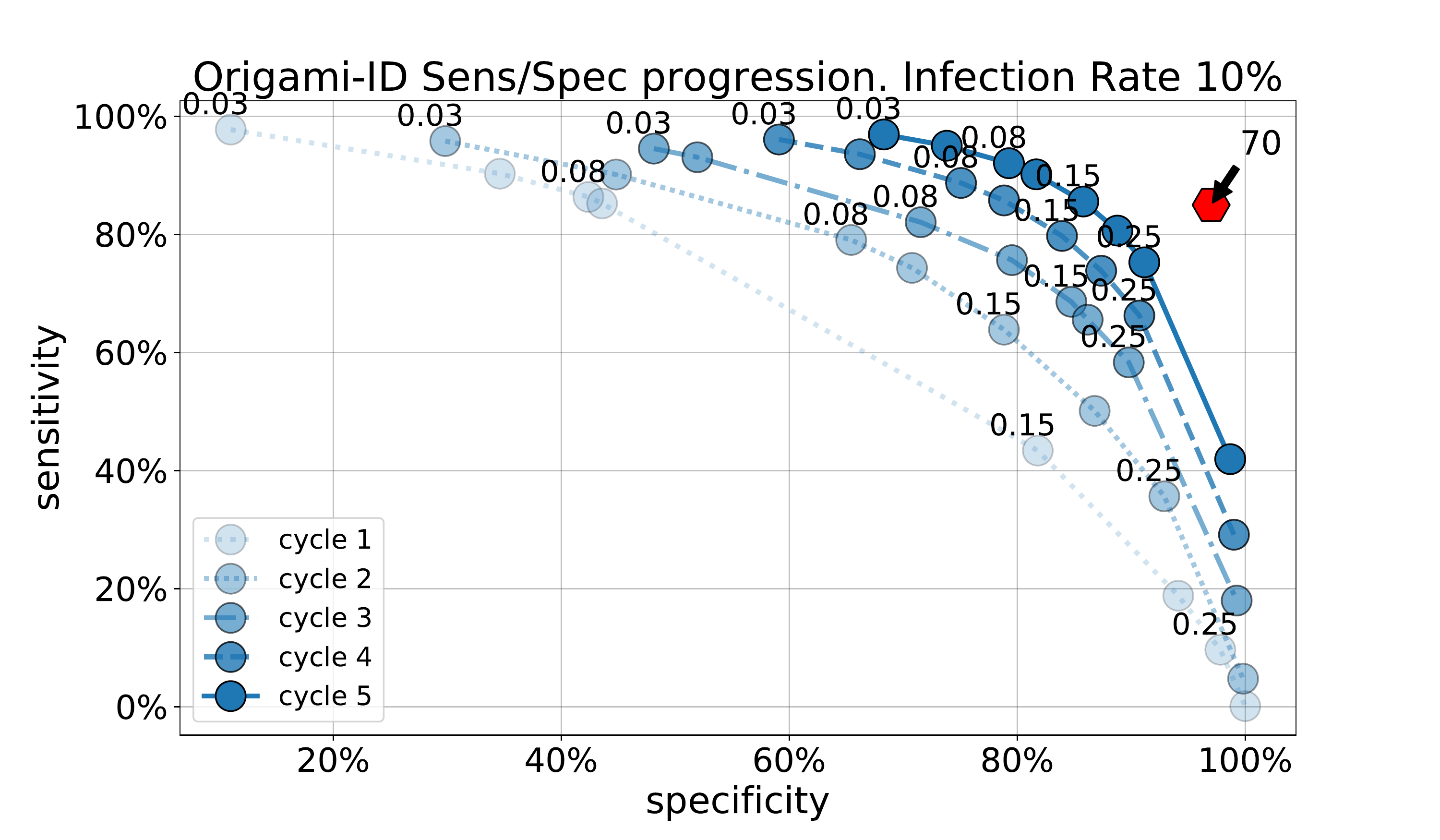}\\
    \hskip-.8cm
    \includegraphics[width=.57\textwidth]{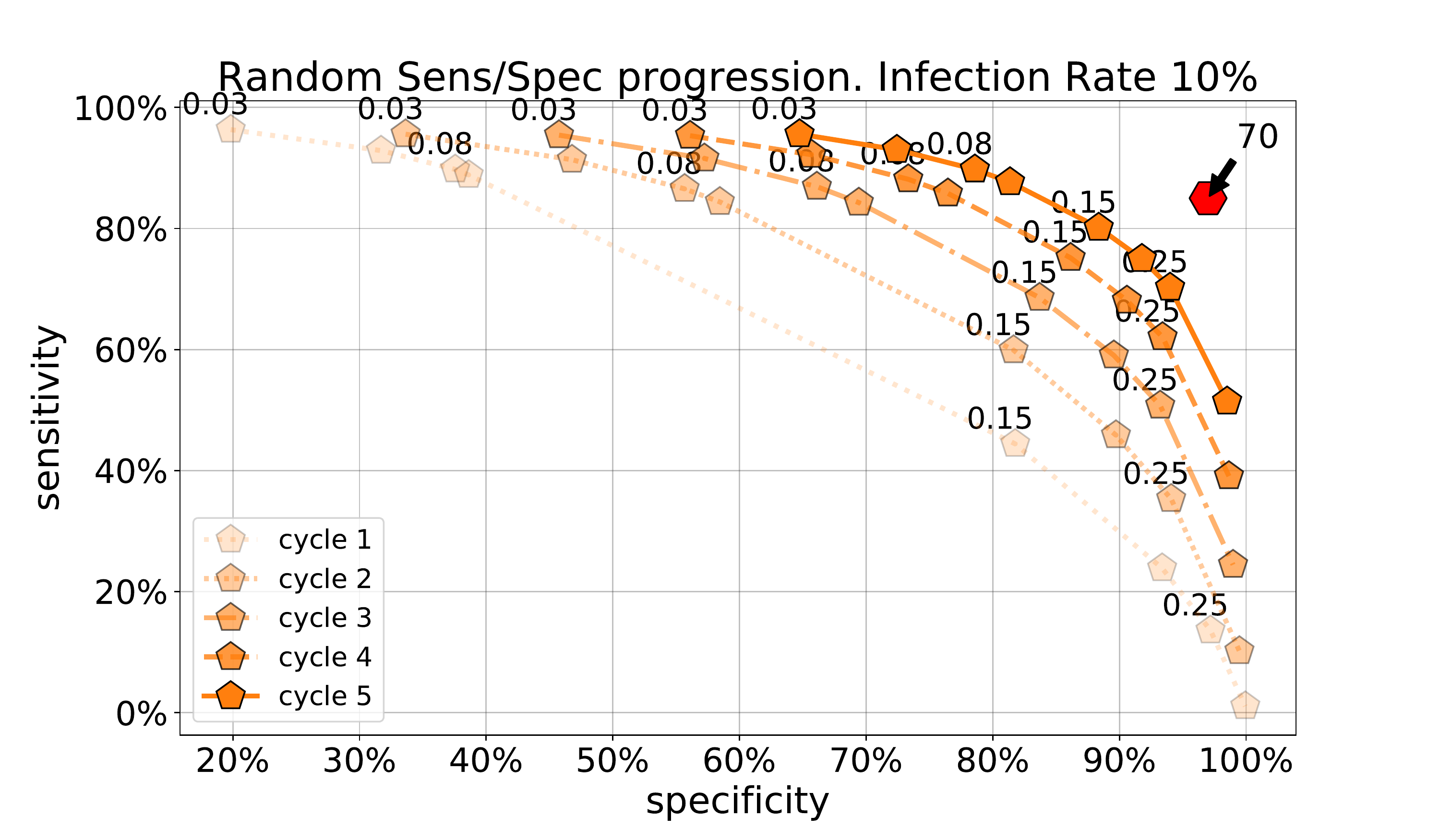}\hskip-.2cm
    \includegraphics[width=.57\textwidth]{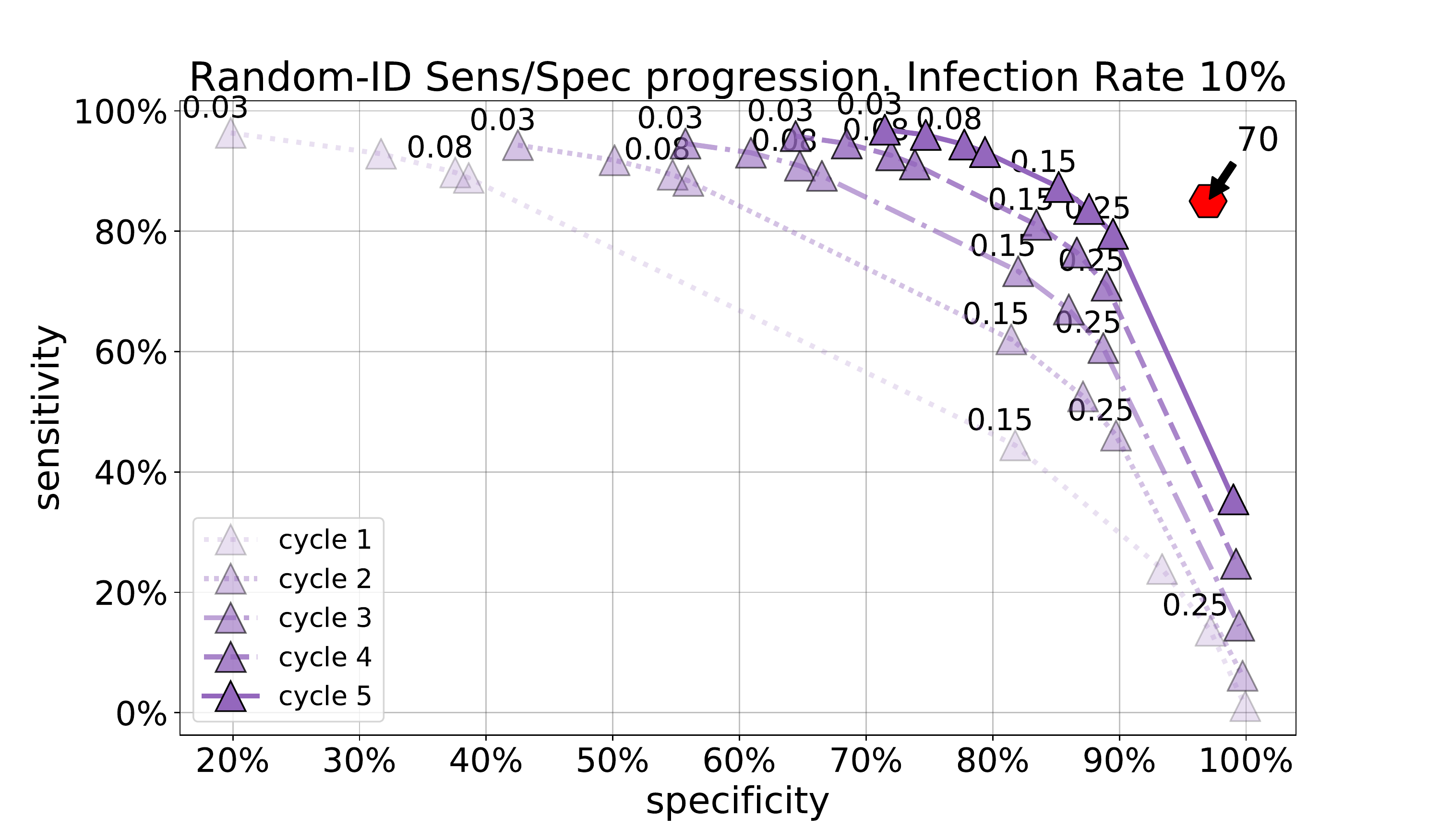}\\
    \hskip-.8cm
    \includegraphics[width=.57\textwidth]{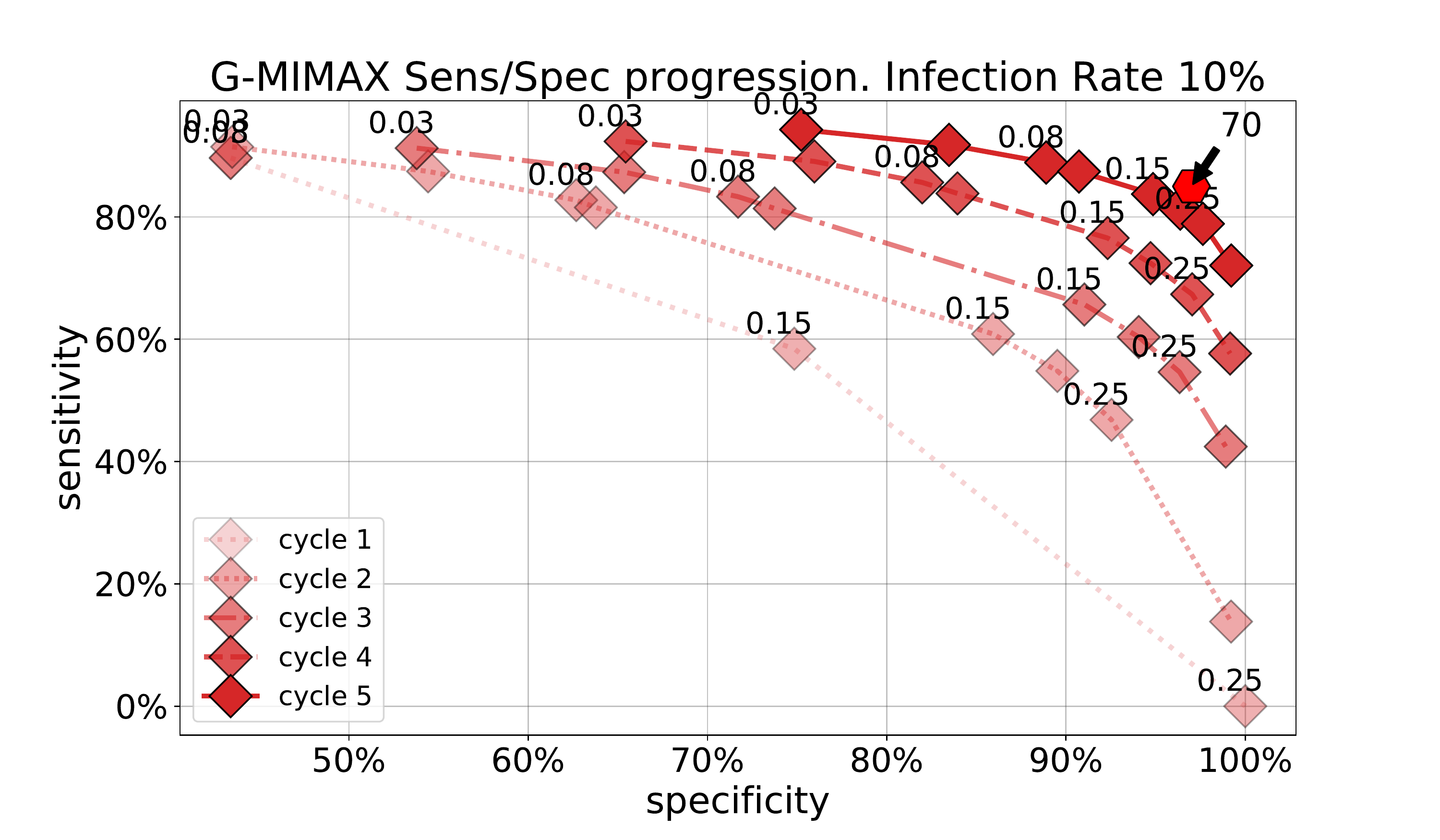}\hskip-.2cm
    \includegraphics[width=.57\textwidth]{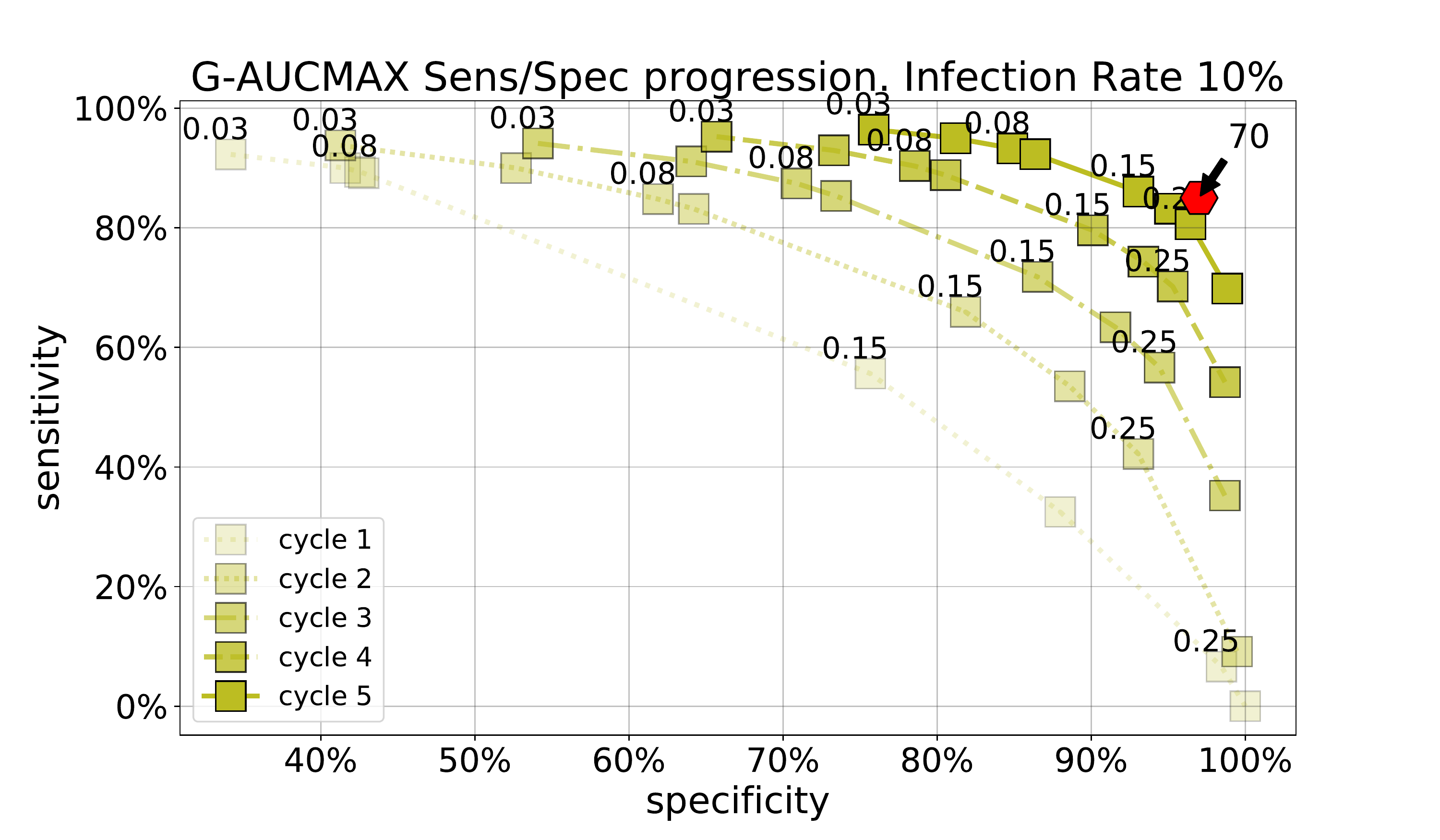}\\
\caption{Using the same setup as in Fig.~\ref{fig:perf_curve}, we report dynamic results for each policy, as the number of tests increases, for $q=10\%$. The number of tests is equal to $k$ (here 8) times the cycle number.}
    \label{fig:evolutionforall10}
\end{figure}

We provide in Fig.~\ref{fig:34cycles} additional plots that complement \ref{fig:perf_curve}, displaying how the specificity/sensitivity frontier evolves as the number of tests grows, for each of the policies we considered. We propose a more detailed view of that dynamic evoluation for each policy taken individually for all three base infection rates, in Fig.~\ref{fig:evolutionforall2}, ~\ref{fig:evolutionforall5} and ~\ref{fig:evolutionforall10}.

\subsection{Robustness to misspecification}\label{subsec:misspecification}
We study in this section robustness to misspecification of the policies we considered. Since all policies rely on a marginal decoder, their specificity / sensitivity hinges on the fact that the infection rate $q$ and testing device's noise parameters $s,\sigma$ both match with those used by the marginal decoder. We quantify how that performance varies under misspecification by considering the following perturbations: we use the setup from the right plot in Fig.~\ref{fig:perf_curve}, namely $q=5\%, s = 85\%, \sigma=97\%$, to generate the ground truth in our simulations, as well as to execute tests. On the other hand, the policies (along with their decoders), will be tested under 8 additional scenarios: $\hat{q} \in \{ 3\%, 5\%, 8\%\}$ and $\hat{s}\in \{ 78\%, 85\%, 92\%\}$. Naturally, when $\hat{q}=5\%$ and $\hat{s}=85\%$ we fall back on the well-specified scenario. To facilitate comparison, our two proposals \textbf{(G-AUCMAX)} and \textbf{(G-MIMAX)} are displayed 
\begin{figure}
    \includegraphics[width=.57\textwidth]{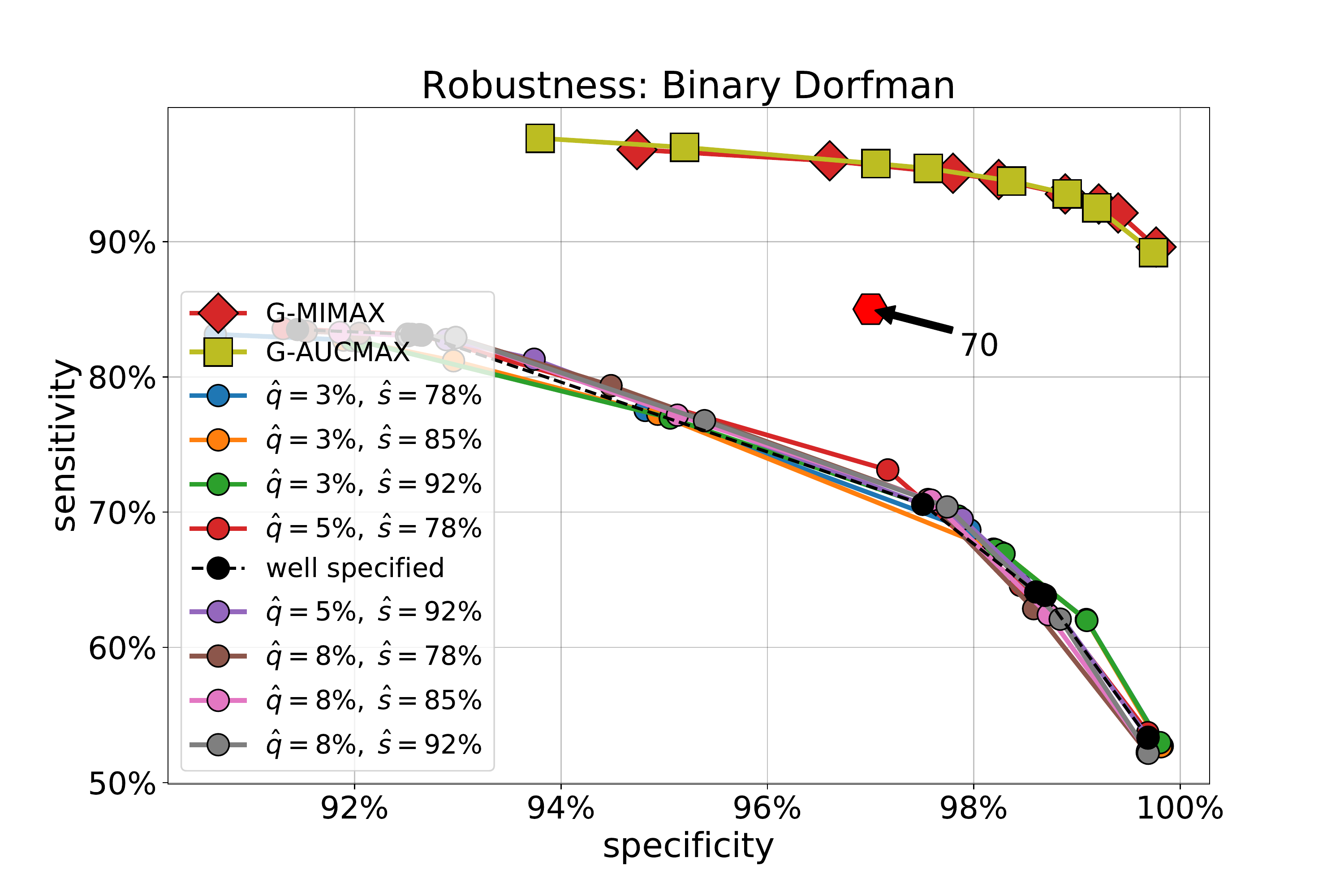}\hskip-.2cm
    \includegraphics[width=.57\textwidth]{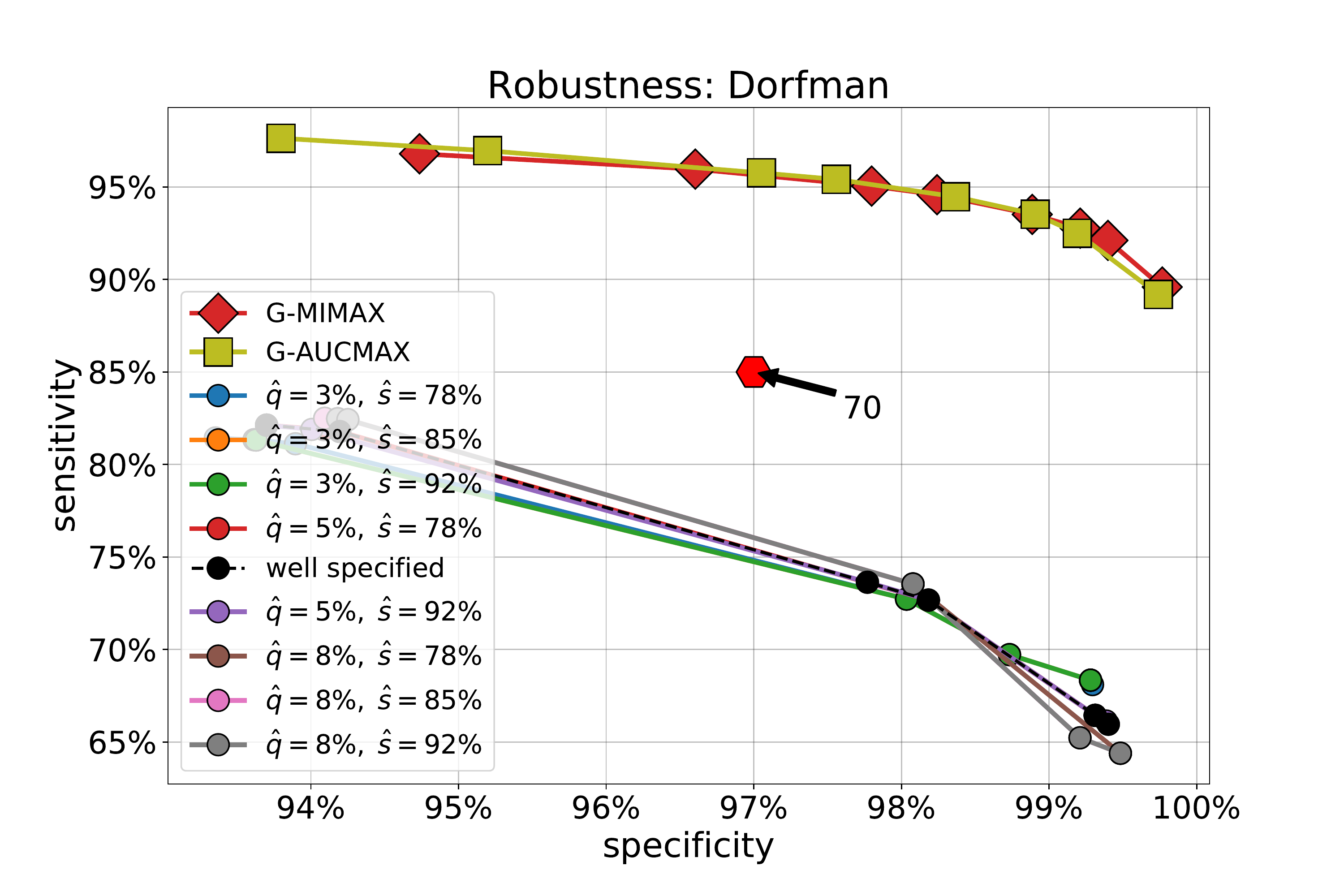}\\
    \hskip-.8cm
    \includegraphics[width=.57\textwidth]{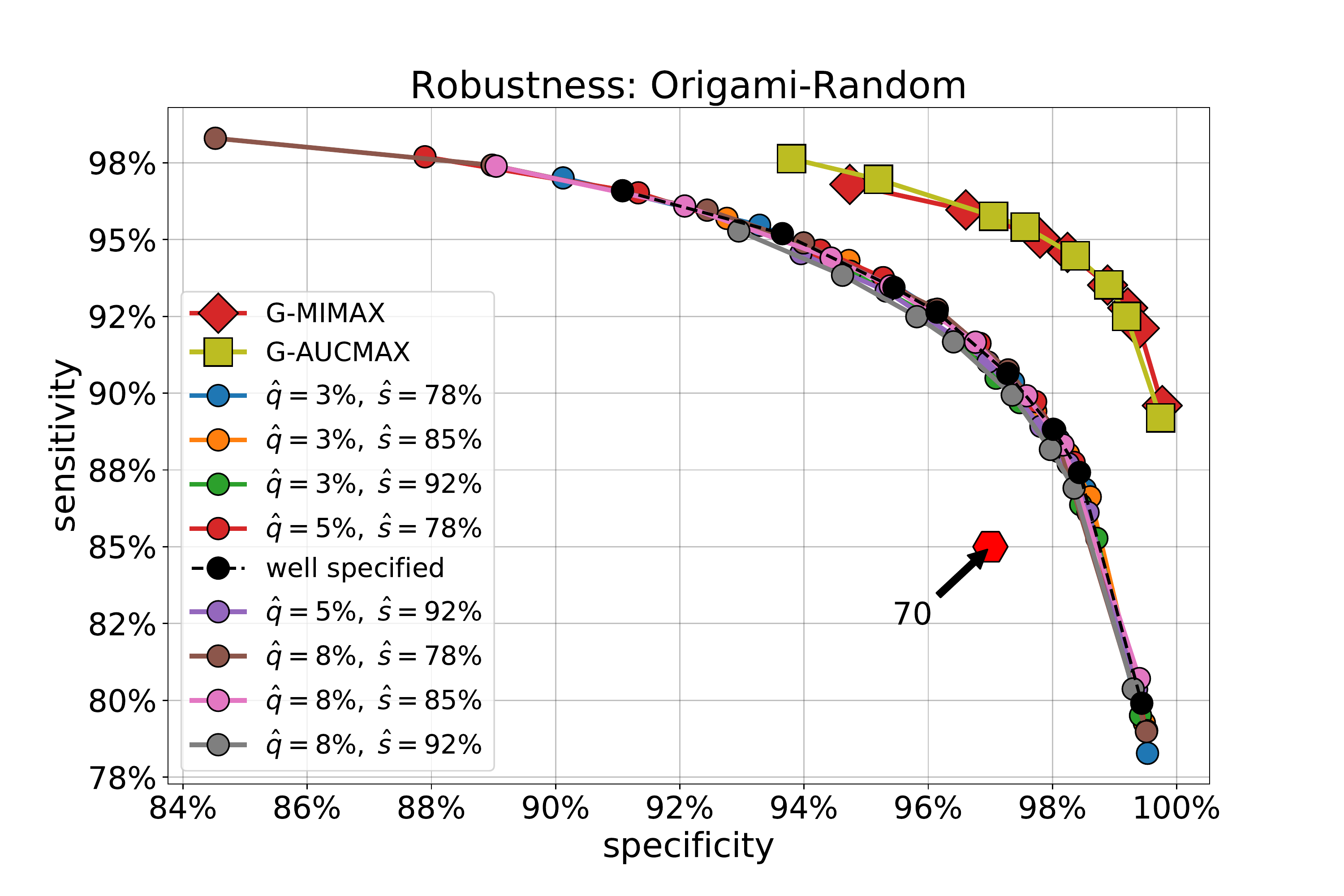}\hskip-.2cm
    \includegraphics[width=.57\textwidth]{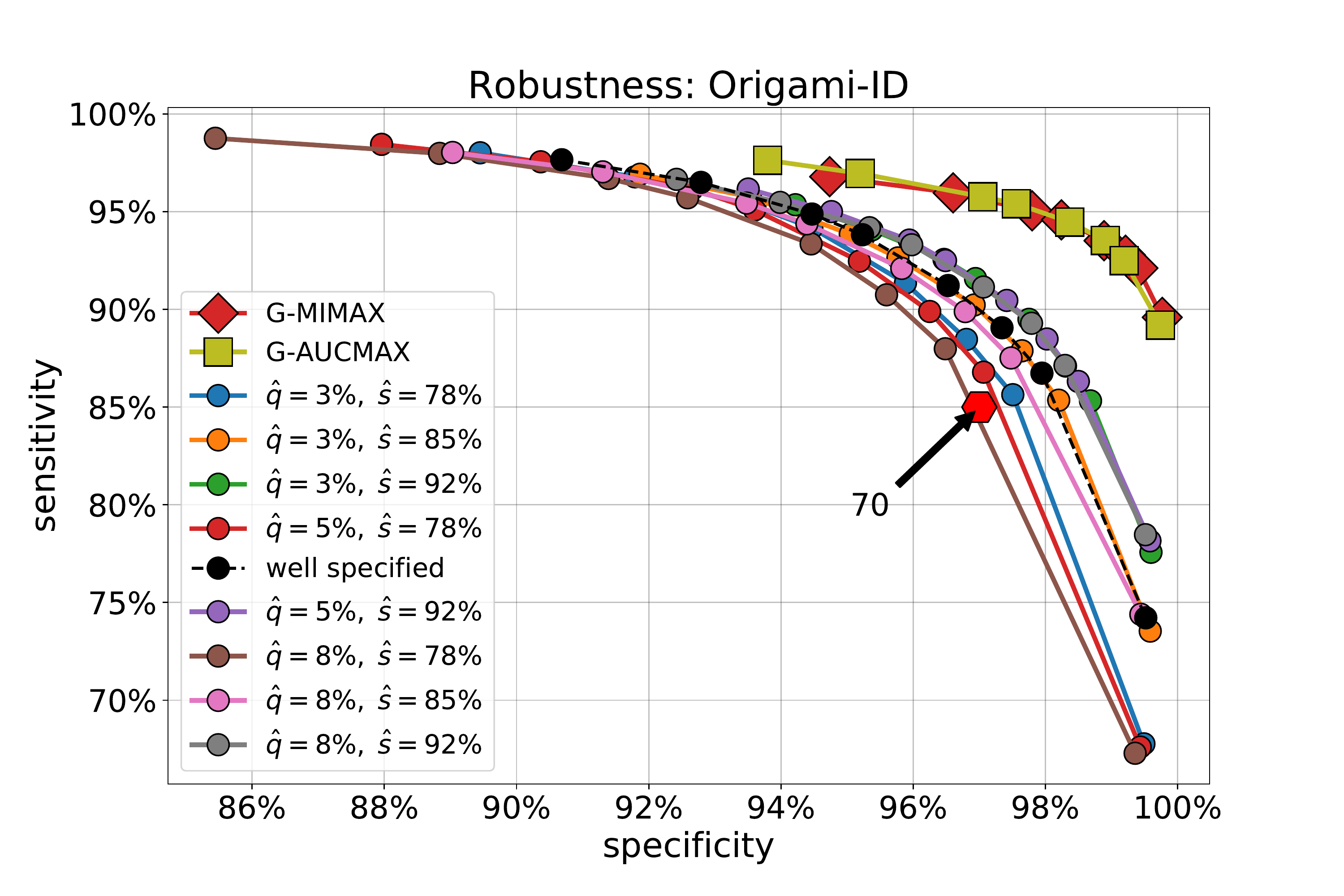}\\
    \hskip-.8cm
    \includegraphics[width=.57\textwidth]{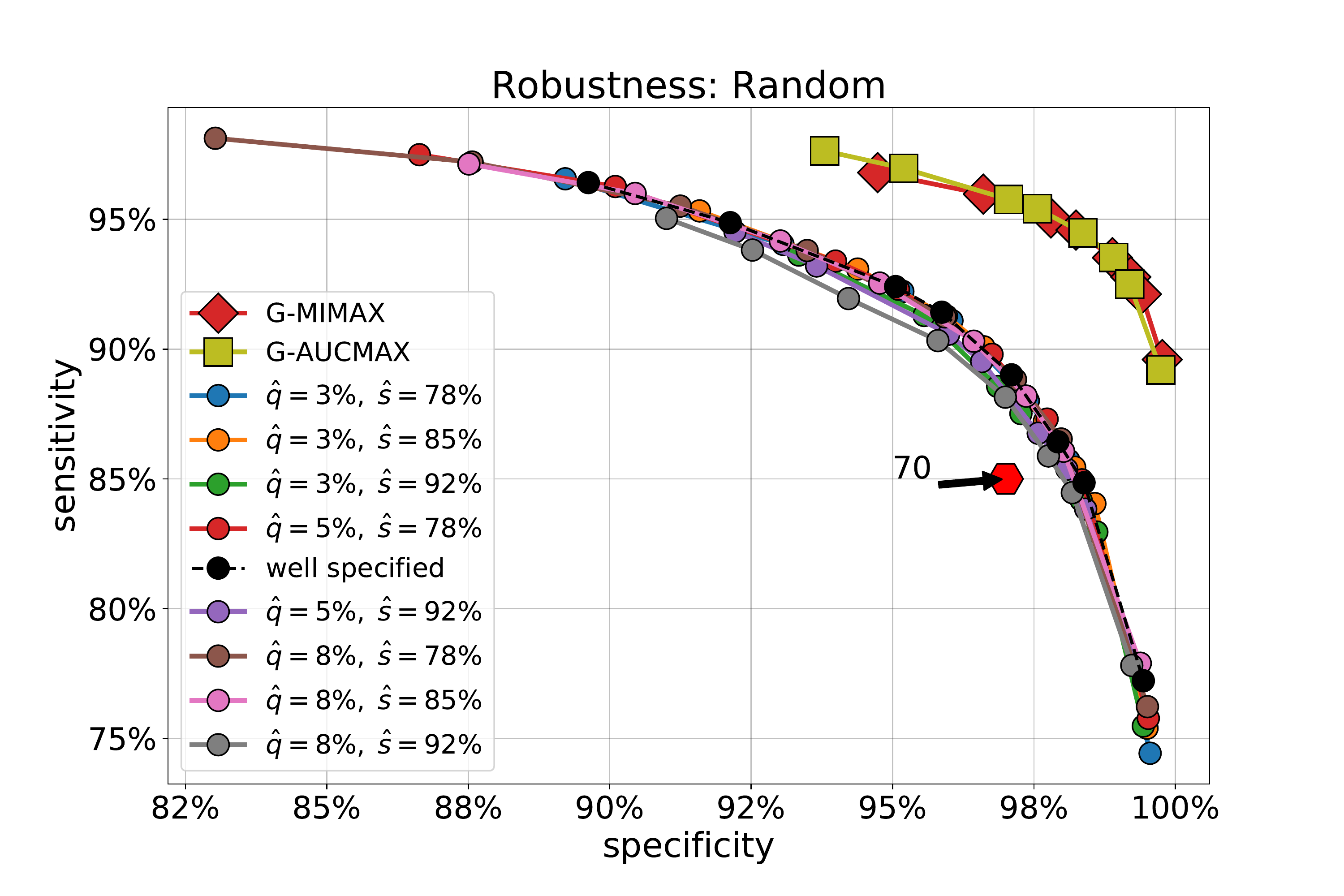}\hskip-.2cm
    \includegraphics[width=.57\textwidth]{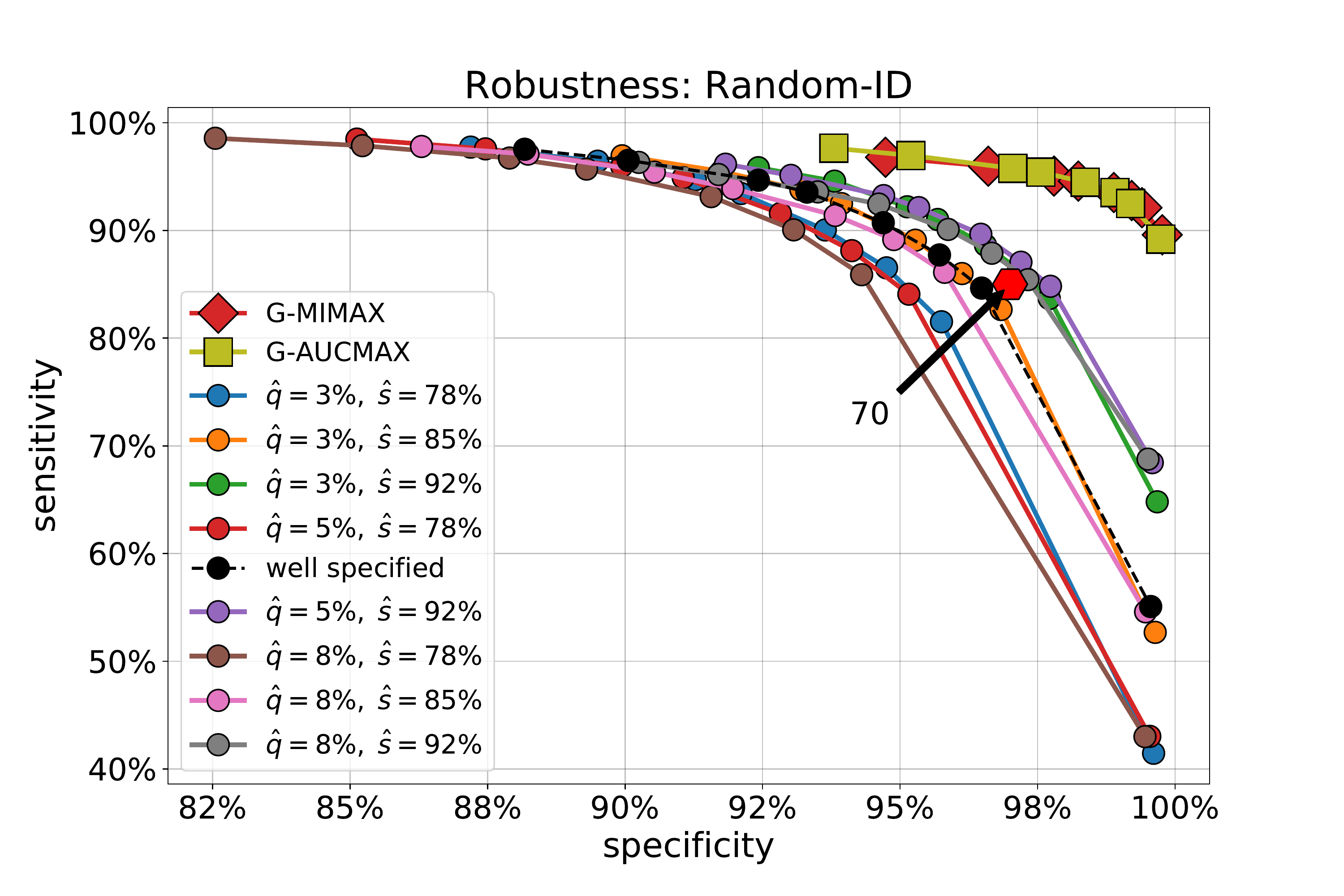}\\
    \hskip-.8cm
    \includegraphics[width=.57\textwidth]{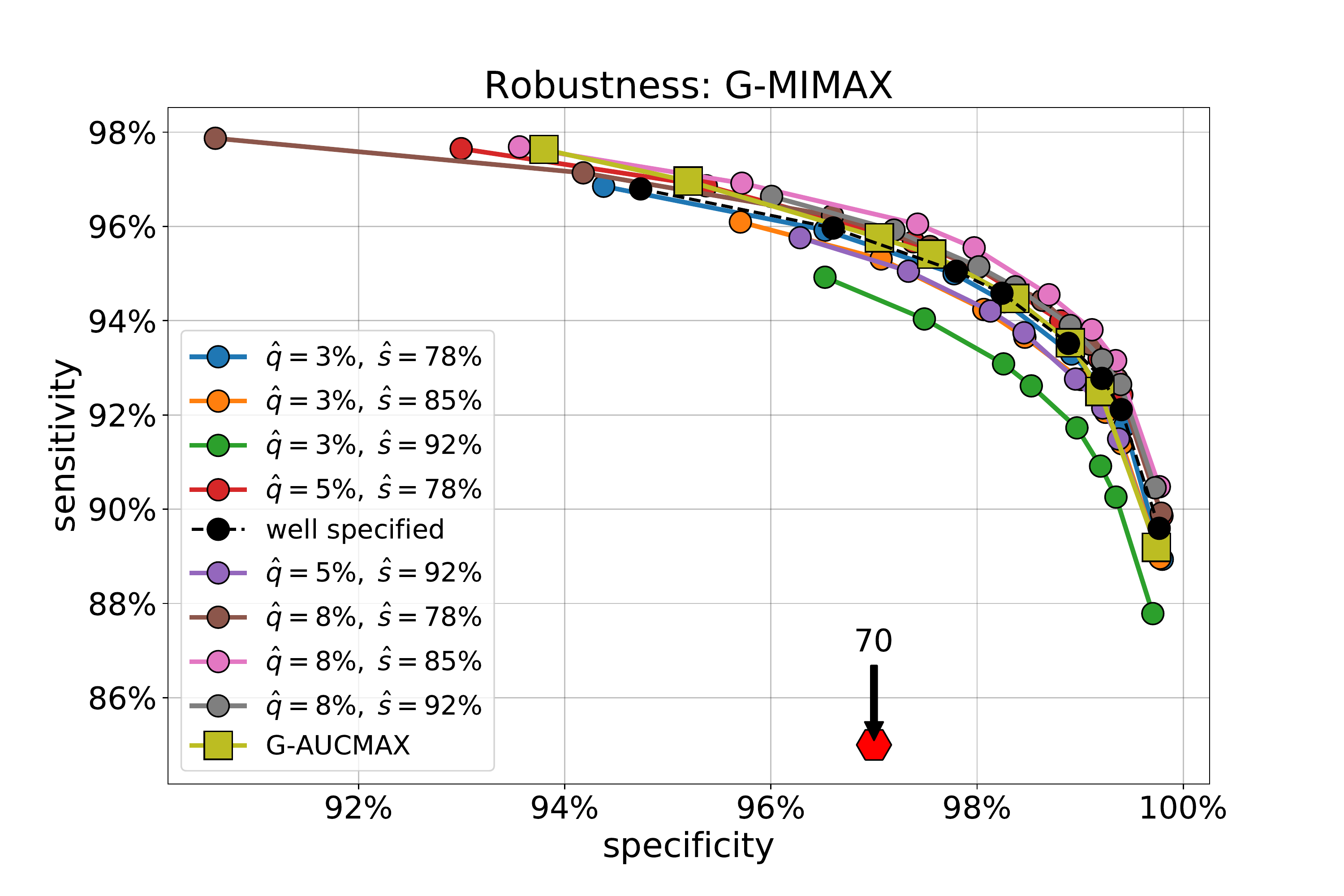}\hskip-.2cm
    \includegraphics[width=.57\textwidth]{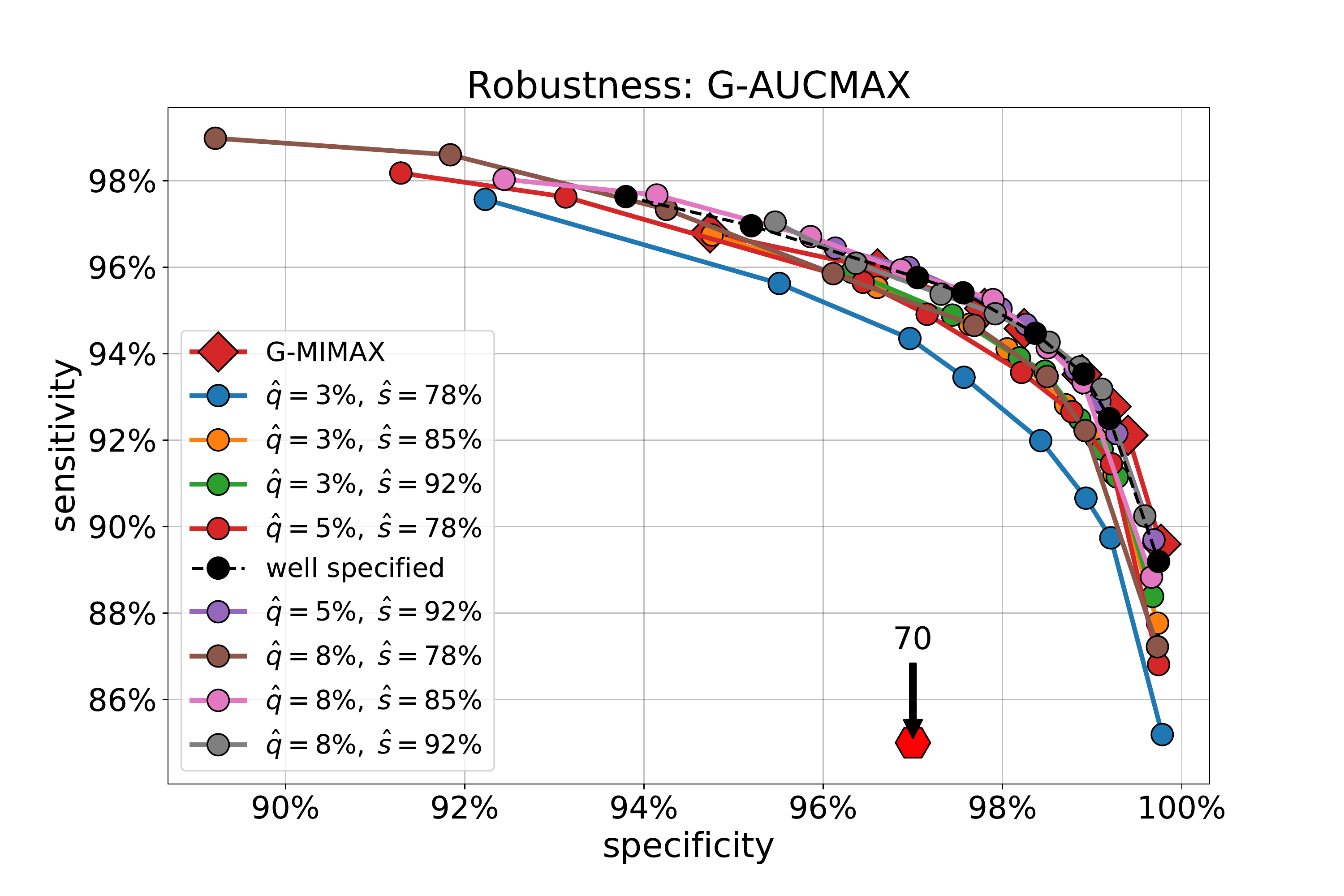}
\caption{Robustness to misspecification of two crucial parameters: prior infection rates and sensitivity $\hat{q}$ and $\hat{s}$, compared to ground truth parameters used to generate ground truth and tests $q=5\%$ and $s=85\%$. Specificity is well specified in all experiments, i.e. $\sigma=\hat{\sigma}=97\%$. Note that scales are relative to each plot, and highlight the robustness of our methods (bottom) to misspecification.}
    \label{fig:perf_robustness}
\end{figure}

\subsection{Experiments with $k=1$}\label{subsec:1cycle}
\begin{figure}
    \centering
    \includegraphics[width=.80\textwidth]{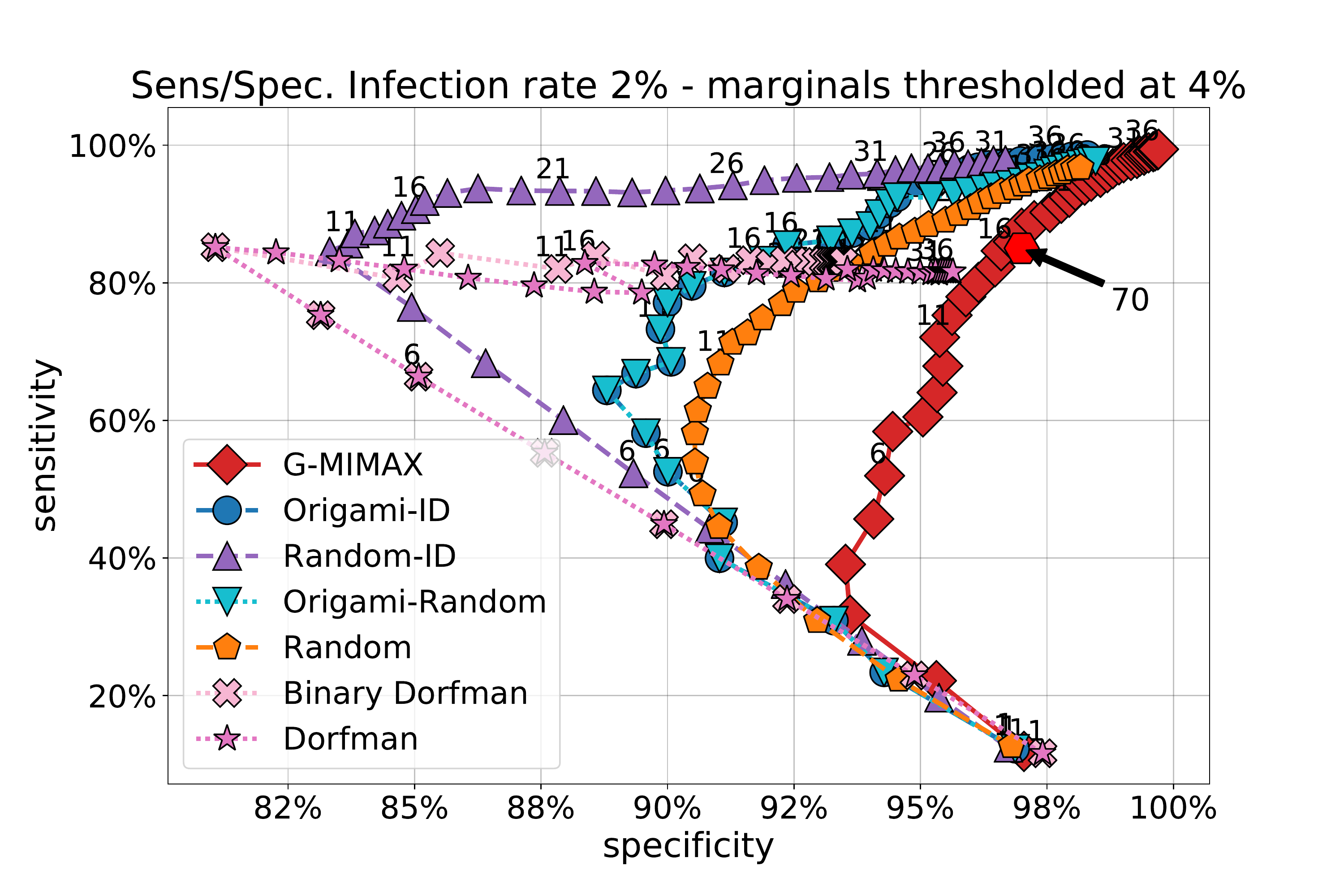}\\
    \includegraphics[width=.80\textwidth]{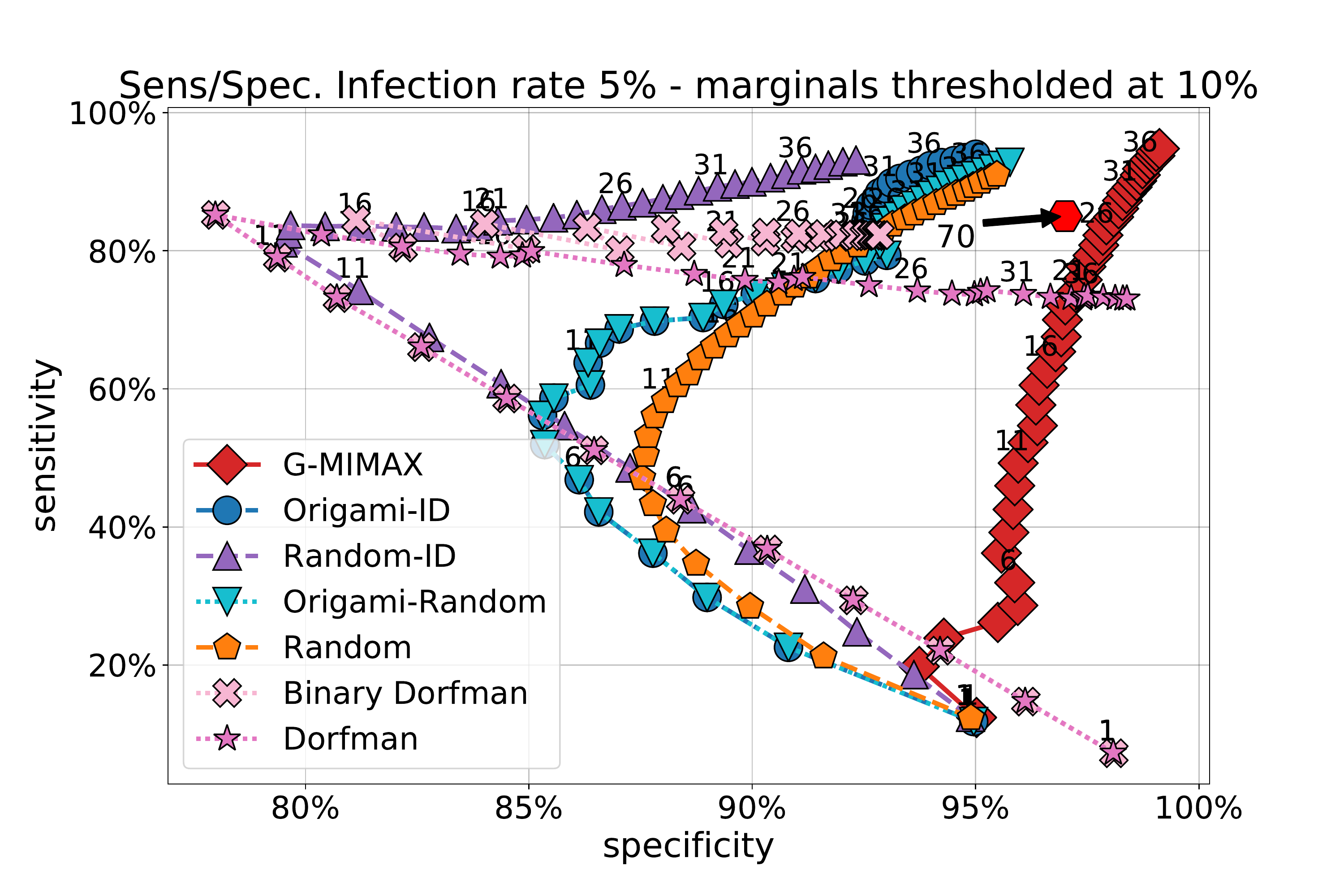}\\
    \includegraphics[width=.80\textwidth]{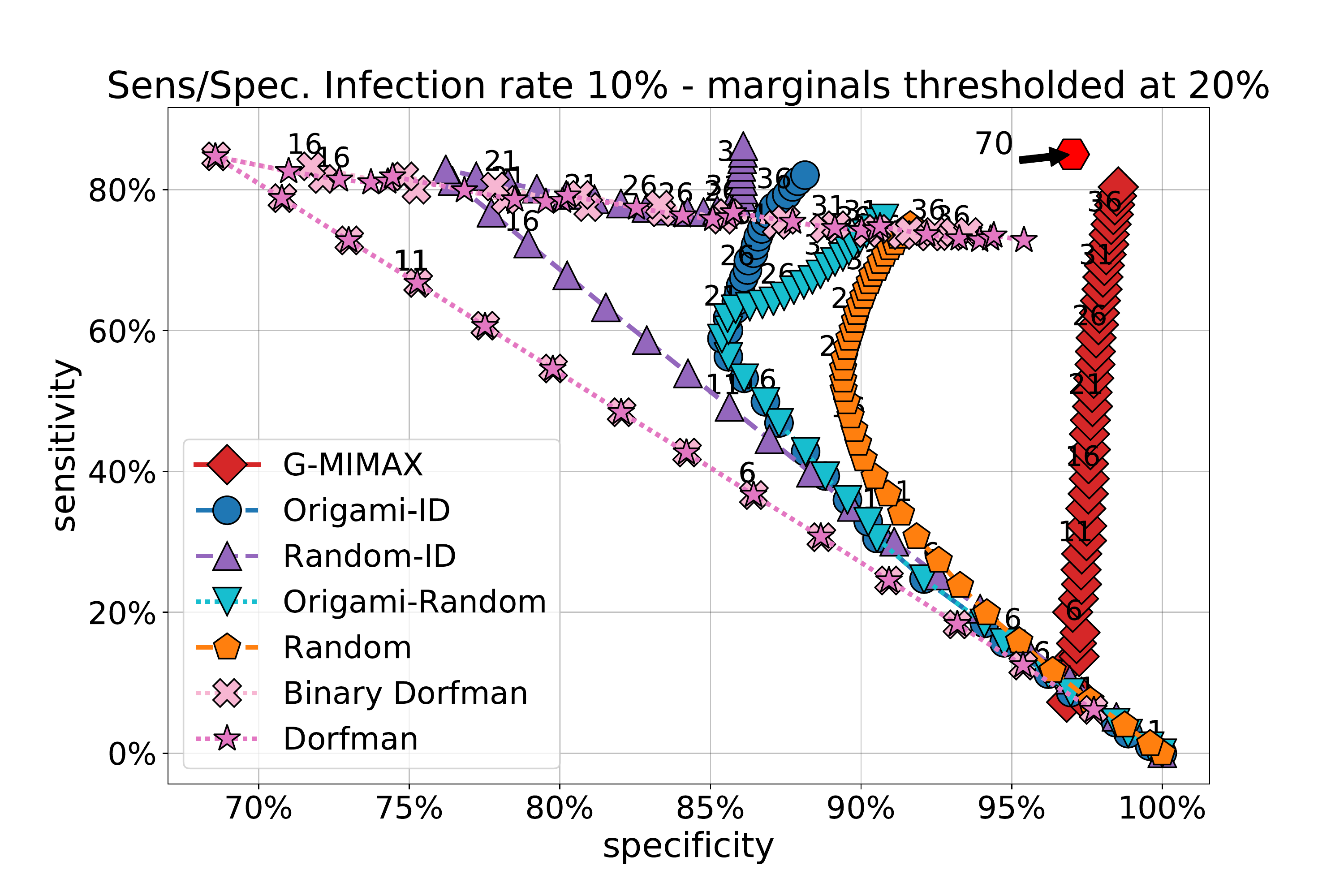}\\
\caption{Experiments in which only one test is carried out at a time before recomputing the marginal and deciding on the next test. Here $F=5, B=4$ for G-MIMAX (with $N=10000$) and all methods rely on a LBP decoder only.}
    \label{fig:k1}
\end{figure}

We consider now in Fig.~\ref{fig:k1} a setup where $k=1$. In that setting, we can have a fine grained picture of what each of the considered policies does when using the latest test result to produce a new group. The adaptiveness of \textbf{(G-MIMAX)} is showed case here, as we see the method maintain an acceptable specificity to highlight progressively positives while making few mistakes. This setting is particularly relevant to compare in an idealized setting our approach to the performance of Dorfman baselines.

\subsection{Varying sensitivity}\label{subsec:varying}
We explore an additional experimental setup that is a bit more ambitious in scale, since $n=96$, $\nmax=12$, $k=10$ and $T=4$. In that setting, we assume correct specification, with slightly different infection rates than those used before, $q\in\{2\%, 4\%, 7\%\}$ and a more reliable specificity $\Spe=0.99$, but factor in a decreasing sensitivity as the group size increases, $\Seg = (91 - g)\%$. We also consider in that setup various iterations for our greedy forward-backward approach, and $N\in\{10000, 20000\}$ total particles.

\begin{figure}
    \includegraphics[width=.57\textwidth]{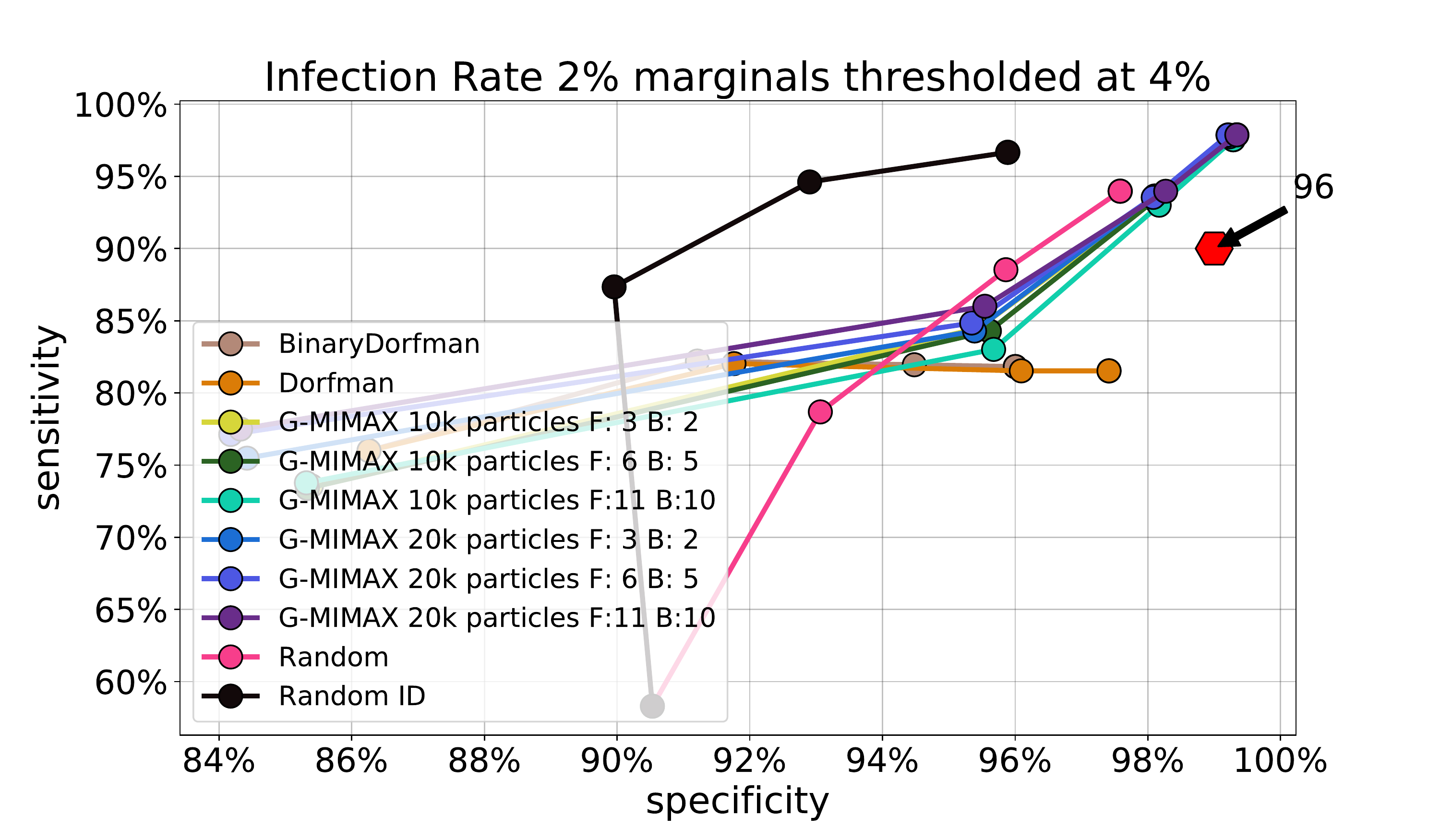}\hskip-.2cm
    \includegraphics[width=.57\textwidth]{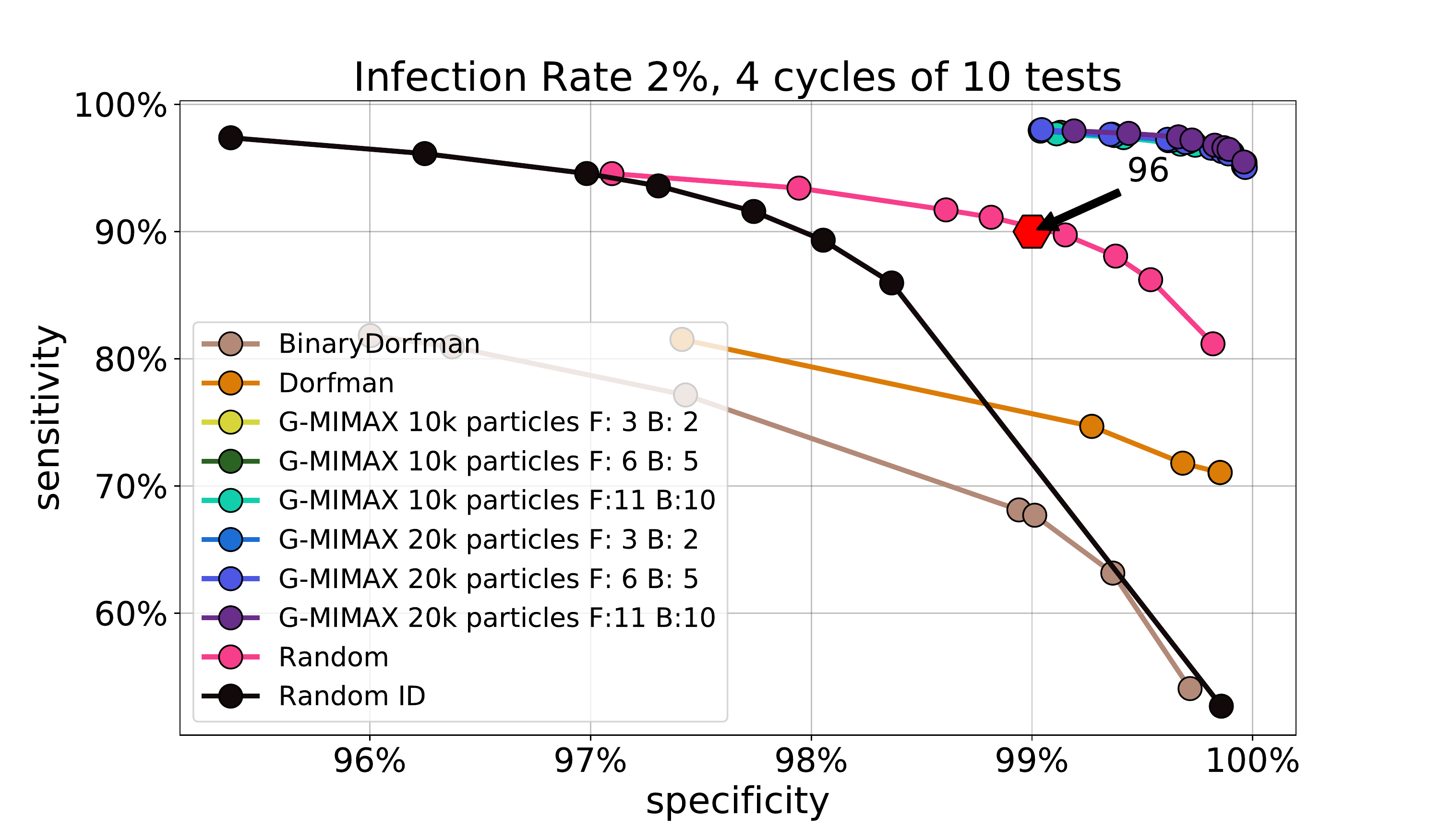}\\
    \hskip-.8cm
    \includegraphics[width=.57\textwidth]{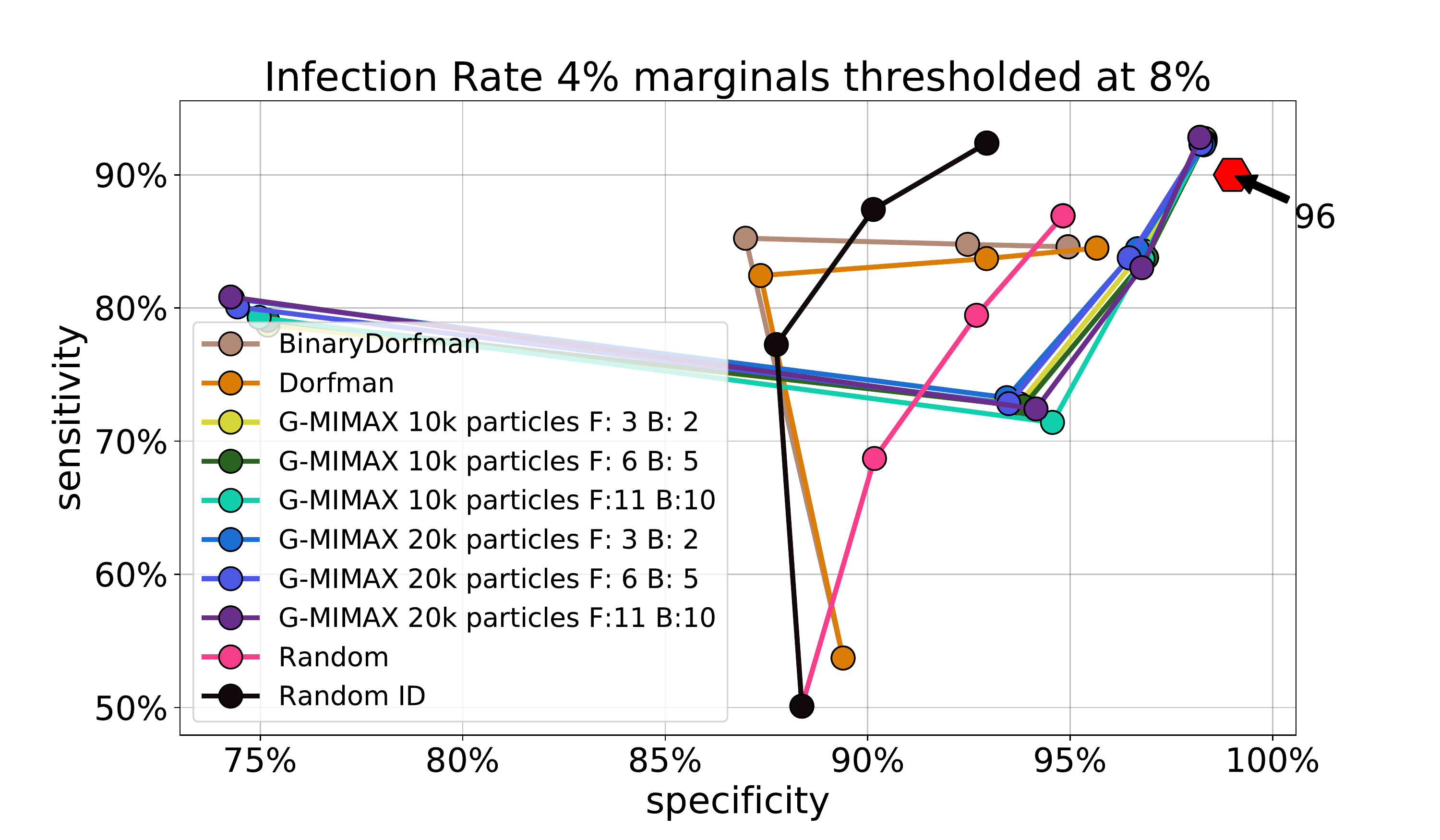}\hskip-.2cm
    \includegraphics[width=.57\textwidth]{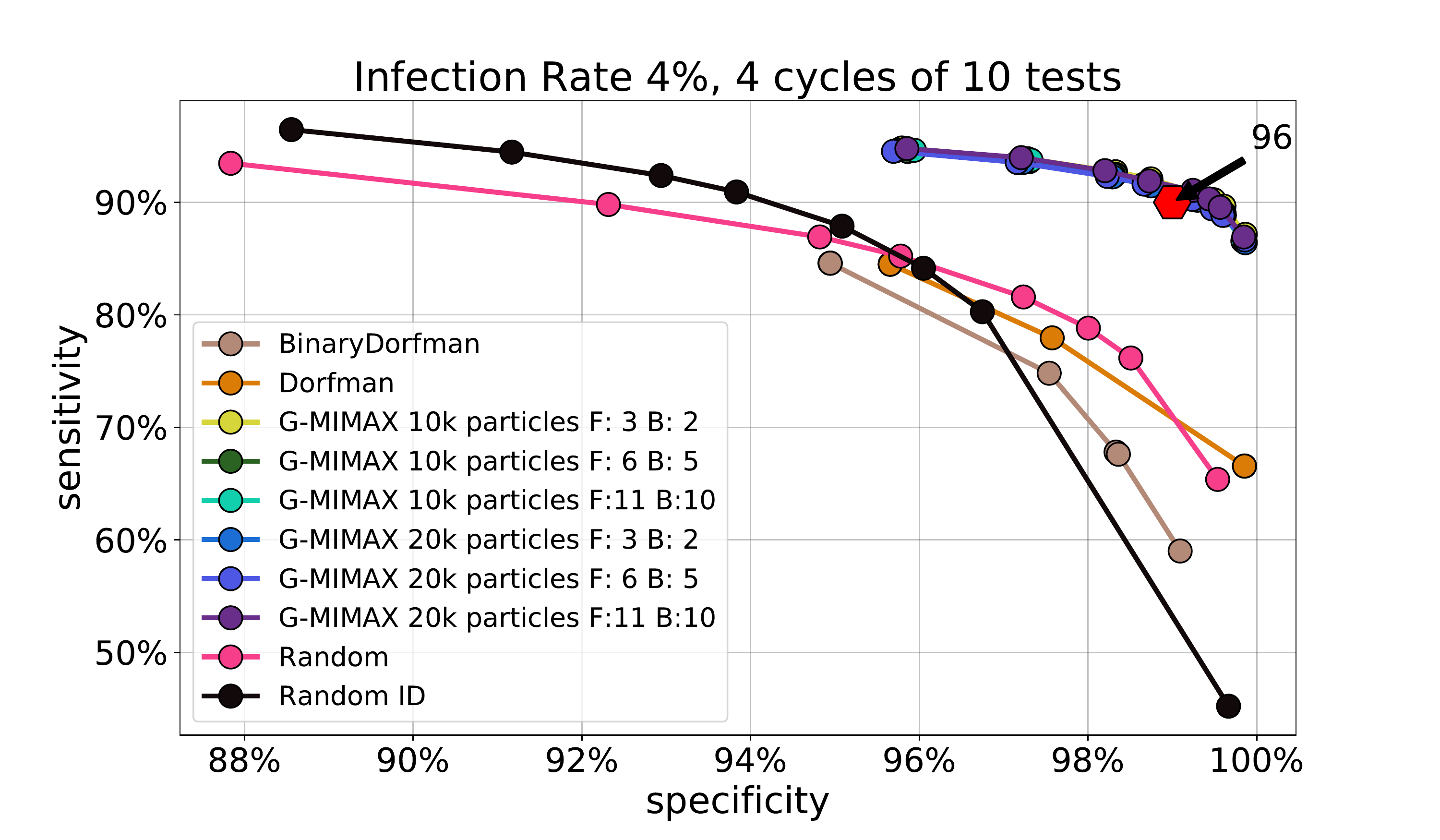}\\
    \hskip-.8cm
    \includegraphics[width=.57\textwidth]{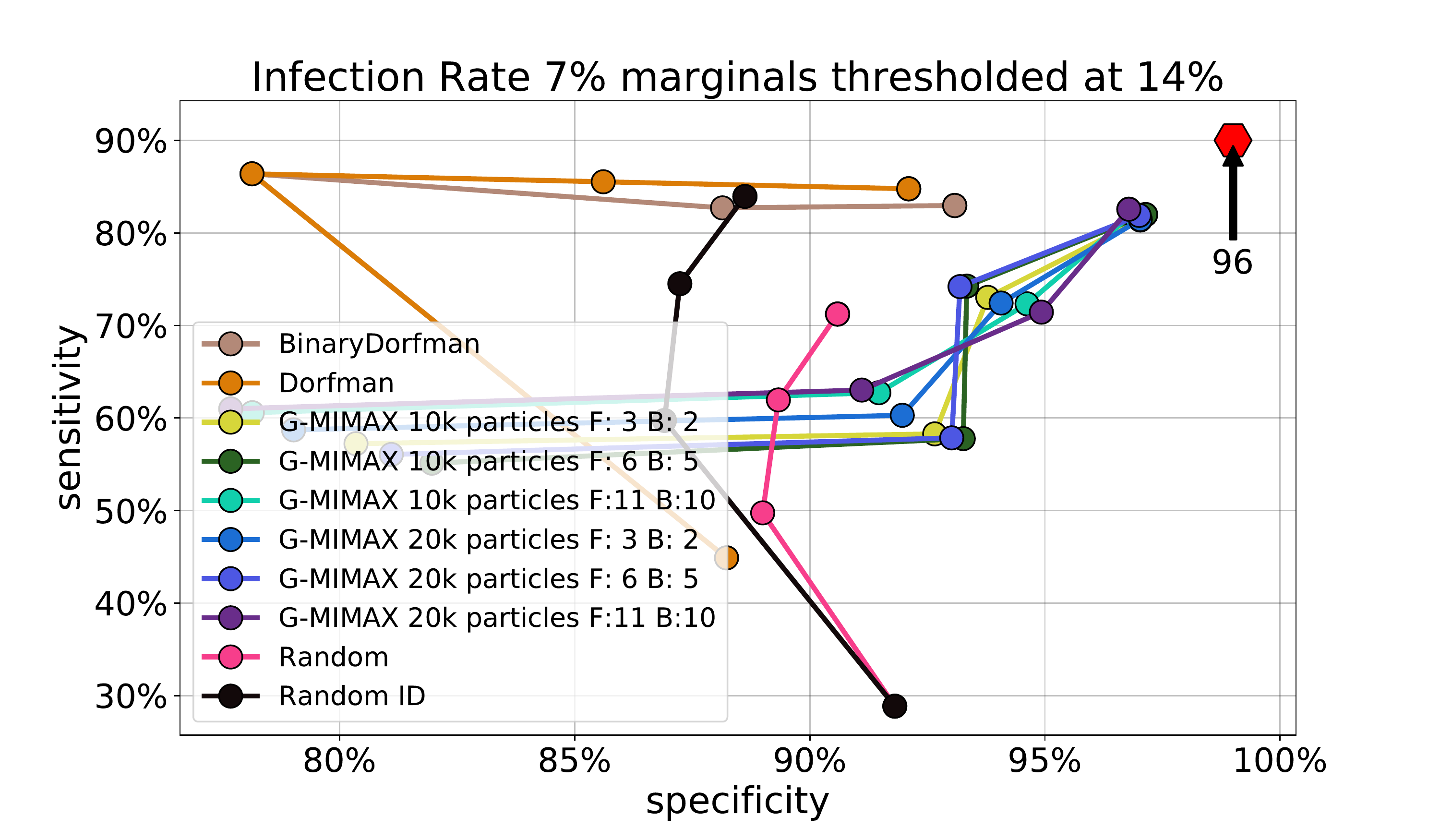}\hskip-.2cm
    \includegraphics[width=.57\textwidth]{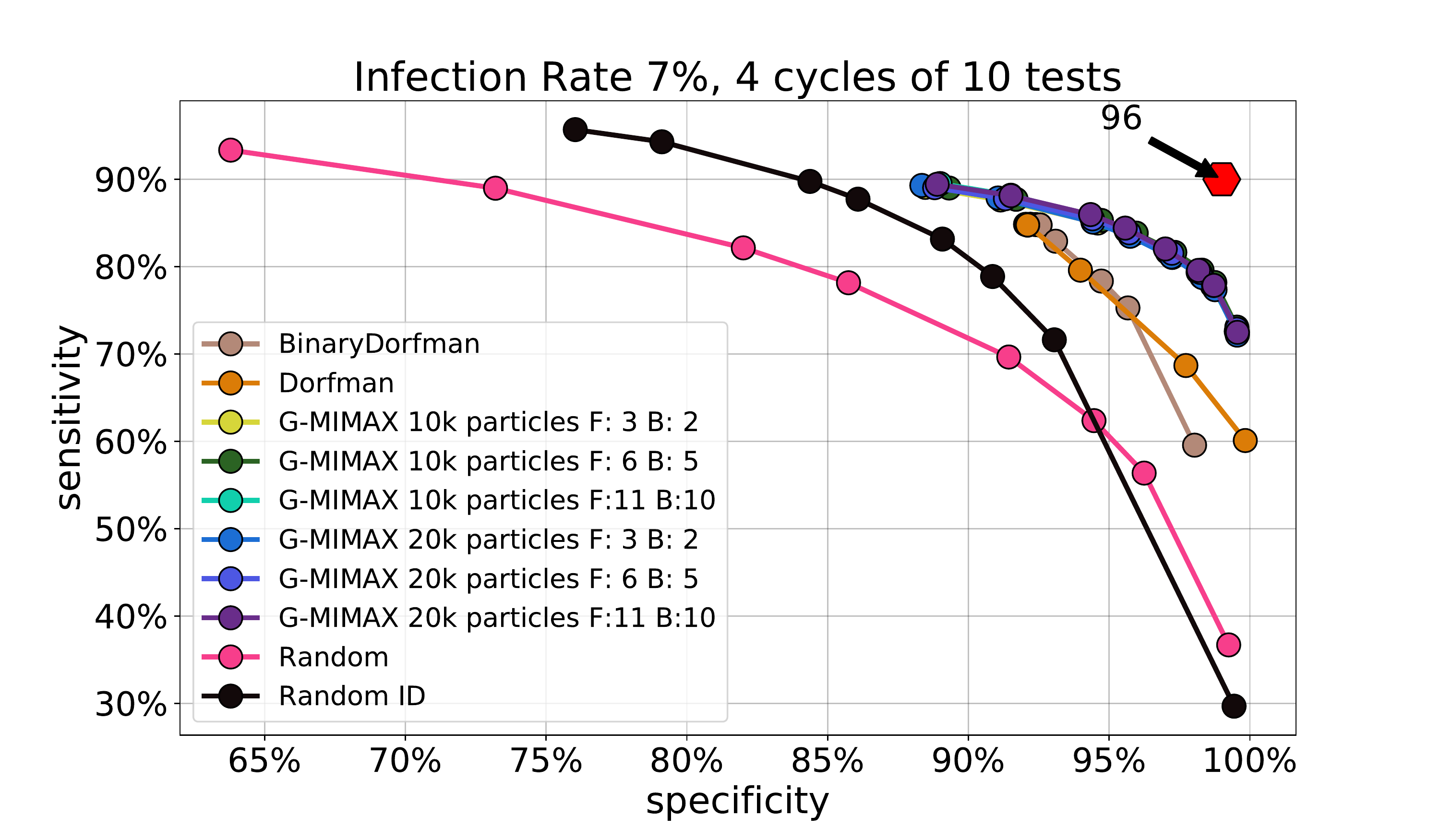}\\
    \hskip-.8cm
\caption{Experiments with a sensitivity that decreases with group size. Here we also show the relatively minor impact for the G-MIMAX strategy of choosing parameters such as $N$ and $F/B$. The red hexagonal dot stands for the sensisitivy/specificity of a single test, knowing that the sensitivity for groups decreases by 1\% every time the group size is increased by 1.}
    \label{fig:varying2}
\end{figure}

\end{document}